\documentclass[12pt,a4paper, twoside]{article}

\usepackage[english]{babel}
\usepackage[T1]{fontenc}
\usepackage{csquotes}
\usepackage[a4paper,top=3cm,bottom=2cm,left=3cm,right=3cm,marginparwidth=1.75cm]{geometry}
\usepackage{amsfonts}
\usepackage[style=numeric,sorting=none,giveninits=true]{biblatex}
\usepackage{amsmath,latexsym}
\usepackage{amssymb}
\usepackage{graphicx}
\usepackage{xcolor}
\usepackage{float}
\usepackage[colorlinks=true, allcolors=blue]{hyperref}
\usepackage[english]{babel}

\newcommand\calC{\mathcal{C}}

\begin{document}
\begin{titlepage}

\vfill

\begin{center}
   \baselineskip=16pt
   {\Large\bf Timelike boundary and corner terms in the causal set action}
  \vskip 1.5cm

Fay Dowker${}^{a,b}$, Roger Liu${}^{a,c}$, and Daniel Lloyd-Jones${}^{a,d}$ \\
     \vskip .6cm
            \begin{small}
      \textit{
      		${}^{a}${Blackett Laboratory, Imperial College, Prince Consort Road, London, SW7 2AZ, UK}\\ \vspace{5pt}
${}^{b}${Perimeter Institute, 31 Caroline Street North, Waterloo ON, N2L 2Y5, Canada}\\ 
 \vspace{5pt}
  ${}^{c}${Mathematical Institute, University of Oxford, Woodstock Road, Oxford, OX2 6GG, U.K.}\\ 
 \vspace{5pt}
 ${}^{d}${Department of Applied Mathematics and Theoretical Physics, University of Cambridge, Wilberforce Road, Cambridge CB3 0WA, UK}\\ 
 \vspace{5pt}
 email: f.dowker@imperial.ac.uk, xiaodi.liu@wadham.ox.ac.uk, dl738@cam.ac.uk
              }
              \end{small}                       \end{center}
\vskip 2cm
\begin{center}
\textbf{Abstract}
\end{center}
\begin{quote}
The causal set action of dimension $d$ is investigated for causal sets that are Poisson sprinklings into manifolds that are regions of $d$-dimensional Minkowski space. Evidence, both analytic and numerical, is provided for the conjecture that as the discreteness length $l$ tends to zero,
the mean of the causal set action over Poisson sprinklings into a manifold with a timelike boundary, is dominated by a term proportional to the volume of the timelike boundary and diverges like $l^{-1}$. A novel conjecture for the contribution to the causal set action from co-dimension two corners, also known as joints, is proposed and justified.  
\end{quote}

\vfill

\end{titlepage}
\pagenumbering{roman}
\tableofcontents
\newpage
\pagenumbering{arabic}

\section{Introduction}
In the causal set approach to quantum gravity, a major direction in constructing a quantum dynamics for causal sets is what might be called \textit{state sum models} in which the heuristic of the continuum gravitational path integral over manifolds and geometries, 
\begin{equation}\label{eq:continuumsoh}
    Z(V) = \sum_{M\in \cal{M}} \int_{g \in \mathcal{G}} [dg] e^{iS(g)/\hbar}\,,
\end{equation}
is interpreted as being fundamentally a path \textit{sum}, $Z$, given by:
\begin{equation}\label{eq:causetsoh}
    Z(n) = \sum_{\mathcal{C}\in \Omega_n} e^{iS(\mathcal{C})/\hbar}
\end{equation}
where $\Omega_n$ is some collection of (order isomorphism classes of) causal sets of cardinality $n$  \cite{Surya:2011du,Glaser:2017sbe}. The model is defined by $\Omega_n$ and by the action $S(\mathcal{C})$ and there may be a symmetry factor such as $1/|Aut(\mathcal{C})|$ where $Aut(\mathcal{C})$ is the group of automorphisms of $\mathcal{C}$.\footnote{Another direction in the quest for a quantum dynamics for causal sets is to generalise the Classical Sequential Growth models of Rideout and Sorkin \cite{Rideout:1999ub} and to construct the decoherence functional or double path integral of a quantum measure theory for a growing causal set \cite{Dowker:2010qh,Surya:2020cfm}.}  \par

The best studied candidates for the action of a causal set are the members of the Benincasa-Dowker-Glaser (BDG) family of actions \cite{Benincasa:2010ac,Benincasa:thesis,
Dowker:2013vba,Glaser:2013xha}. These causal set actions and their smeared versions will be defined in the next section. The discreteness of causal sets not only means that the gravitational path integral becomes a  path sum but, when the sum is restricted to causal sets of a fixed cardinality as above, the sum is actually finite and concerns about mathematical existence and convergence of the path integral are alleviated. This finiteness is one of the original motivations for causal set theory \cite{Bombelli:1987aa}. \par

In order to recover physics as we currently understand it, the path sum would have to exhibit a continuum regime in which some version of General Relativity (GR) emerges. 
A necessary condition for this to happen is that the non-manifoldlike causal sets -- i.e. causal sets that are not approximated by any continuum Lorentzian spacetime -- should be suppressed in the sum over causal sets. There is mounting evidence that the BDG action can do the job of suppressing certain classes of non-manifoldlike causal sets in the sum. One important such class of non-manifoldlike causal sets is the Kleitman-Rothschild (KR) orders. These causal sets only have 3 layers, representing universes that, roughly speaking, last only 3 Planck times  and so are clearly non-manifoldlike. However, the KR orders are the most numerous causal sets: the  ratio between the number of  KR orders of cardinality $n$ and the number of  causal sets of cardinality $n$ tends to 1 as $n$ tends to infinity \cite{Kleitman:1970,Kleitman:1975}. However, this entropic catastrophe is averted for the KR orders which are shown to be strongly suppressed in the path sum by the BDG action \cite{Loomis:2017jhn} and further work extends this to other actions and 
other classes of non-manifoldlike causal sets \cite{Mathur:2009,Cunningham:2019rob,CarlipCarlipSurya2024}.

A second necessary condition for GR to be recovered from the path sum is that causal sets that are well-approximated by continuum Lorentzian spacetimes that are not ``GR solutions'' should be suppressed in the sum, relative to those that are approximated by GR solutions. This condition has  at least  two different aspects. One aspect is the suppression of causal sets corresponding to non-solutions of the Einstein equations relative to solutions of the Einstein equations, for which reason the action of a manifoldlike causal set should approximate the Einstein Hilbert action so that the usual Feynmannian stationary phase argument may be invoked \cite{feynman}:
\begin{quote}
    “Suppose that for all paths, $S$ is very large compared to $\hbar$. One path contributes a certain amplitude. For a nearby path, the phase is quite different, because with an enormous $S$ even a small change in S means a completely different phase - because $\hbar$ is so tiny. So nearby paths will normally cancel their effects out in taking the sum - except for one region, and that is when a path and a nearby path all give the same phase in the first approximation (more precisely, the same action within $\hbar$). Only those paths will be the important ones.”
\end{quote}

The second aspect is the related question of what extra conditions, in addition to being a solution of the Einstein equations, a Lorentzian geometry should satisfy to be considered a ``GR solution''. \par

For example, a broad class of possible ``extra conditions'' are the various causality conditions on spacetime such as strong causality or global hyperbolicity. We expect that quantum gravity when we have it will tell us  what the physical axioms of General Relativity actually are. Quantum gravity may rule out certain classes of spacetimes even though they are, formally, solutions of the Einstein equations. An immediate example of this is the ruling out of spacetimes with Closed Timelike Curves (CTCs) from Causal Set quantum gravity: a Lorentzian geometry with CTCs cannot approximate any causal set and so cannot emerge as a prediction from a sum-over-causal-sets. \par

In this work we investigate properties of the BDG causal set action with an eye to the question of which solutions of the vacuum Einstein equations may be favoured over which other solutions of the vacuum Einstein equations by a causal set sum-over-histories. We consider only finite volume flat spacetimes and so they are all, formally, solutions of the vacuum Einstein equations but they will differ in their shapes, causal properties and boundary properties.

\section{The Causal Set Action}

\subsection{The discrete-continuum correspondence for causal sets}

A causal set $(\calC, \prec)$ is well-approximated by a $d$-dimensional Lorentzian manifold $(M, g)$ at length scale $l$  if there is a  \textit{faithful embedding} of $(\calC, \prec)$ in  $(M, g)$ at density $\rho = l^{-d}$\cite{Bombelli:1987aa}. For more details about this criterion see \cite{Bombelli:1987aa} and section 3.2 of  \cite{dowker2021recovering_published}. In the causal set literature it is widely assumed that a causal set $(\calC, \prec)$ is  faithfully embeddable in Lorentzian manifold $(M, g)$ if and only if $(C, \prec)$ is a \textit{typical outcome} of a Poisson point process in $(M, g)$ of density $\rho = l^{-1}$
 \cite{Bombelli:1987aa}. This process is known as sprinkling and an outcome of the process is referred to as a sprinkling or as a sprinkled causal set \footnote{Strictly, the outcome of the sprinkling process is a causal set $(\calC, \prec)$ \textit{together with} its embedding in $(M, g)$. If should be clear in context whether ``sprinkled causal set'' refers to the causal set only or refers to the set and its embedding.}. For the purposes of this paper we will follow the literature that assumes manifoldlikeness for a causal set is equivalent to the causal set being a typical Poisson sprinkling.\footnote{For 
 discussion about the relationship between faithful embedding and Poisson sprinkling see \cite{Saravani:2014gza}.}\par

There is growing evidence that the discrete-continuum correspondence described above works, namely that a typical sprinkled causal set is indeed rich enough to encode all the geometric information of the Lorentzian geometry on scales large compared to the discreteness scale set by the density. 
Much of the evidence is of the following form. Let  $\mathcal{G}(M, g)$ be a geometric or topological property of the Lorentzian manifold $(M,g)$ for which we are able, somehow or other, to glean a causal set analogue $G(\calC, \prec)$ which is a function of a causal set (any causal set) $(\calC, \prec)$. By Poisson sprinkling the manifold at density $\rho$ and evaluating  $G$ on the random sprinkled causal set, $G$ becomes a random variable $\boldsymbol{G}(M, g, \rho)$ depending on the Lorentzian geometry and the density. If a typical value of $\boldsymbol{G}$ is close to the continuum geometric quantity $\mathcal{G}$ then a typical sprinkled causal set indeed encodes that geometric quantity. 
\par
Therefore, with a proposed causal set geometric property $G$ in hand, one first computes the mean, $\langle \boldsymbol{G} \rangle $ of $\boldsymbol{G}$ in the sprinkling process in the limit as $\rho$ the sprinkling density tends to infinity. If this gives the correct continuum quantity $\mathcal{G}(M)$ then we're in business and the next task becomes one of working out i) the finite $\rho$ correction to the limiting value to show it is small for large enough $\rho$ and ii) the fluctuations around the mean to show they are small so that the value of $G$ for a typical sprinkling at Planckian density is close to the continuum value. \par

There is a growing list of proposals for causal set analogues of continuum geometric quantities \cite{Surya_2019,dowker2021recovering}. Some proposals are only verified to work in regions of Minkowski space and some are verified to have the correct mean over sprinklings but yet lack a proper investigation of the fluctuations around the mean. 

\subsection{The BDG Action}

The Ricci scalar curvature is a geometric quantity with a causal set analogue and summing the scalar curvature analogue over the whole causal set gives the Benincasa-Dowker-Glaser 1-parameter family of actions for a causal set $\mathcal{C}$ \cite{Benincasa:2010ac,dalembertians_various_dimensions,action_constants,Benincasa:thesis}:
\begin{equation} \label{action}
\frac{1}{\hbar}S^{(d)}(\mathcal{C}) = -\alpha_{d} \left(\frac{l}{l_{p}}\right)^{d-2}\left(N + \frac{\beta_{d}}{\alpha_{d}}
\sum_{i=1}^{n_{d}} C^{(d)}_{i} N_{i}\right), 
\end{equation}
where $l_{p}:=(8\pi G \hbar)^{1/{d-2}}$ is the Planck length in $d$ dimensions, $\alpha_{d}$, $\beta_{d}$ and $C^{(d)}_{i}$ are known, dimension dependent constants of order one  and $n_d = \lfloor{d/2}\rfloor + 2$. For fixed $d$, the $C^{(d)}_{i}$ constants alternate in sign with $i$. The values of these constants are given in \cite{action_constants}. 
$l$ is the \textit{fundamental length} in the theory and it is expected to be of order the Planck length so that the  factor $(l/l_p)^{d-2}$ in the action is a dimensionless constant of order 1. $N$ is the cardinality of the causal set $\mathcal{C}$ and $N_{i}$ is the number of pairs of elements $a$, $b$ such that $|[a,b]|=i-1$, where $[a,b]=\{c \in \mathcal{C} | a \prec c \prec b \}$, in other words, $N_i$ is the number of exclusive order intervals of cardinality $i-1$ \cite{10.1088/1361-6382/abc2fd}. It is important to notice that $N_i$ is a "bilocal" quantity:  we can write it explicitly in the form
\begin{equation}
    N_i=\sum_{x\in \mathcal{C}}\sum_{y\in \mathcal{C}} \chi(x,y)
\end{equation}
where $\chi(x,y)$ is a characteristic function  depending on 2 elements of the causal set,
\begin{equation}
        \chi(x,y)=
        \begin{cases}
            1,& |[x,y]|=i-1\\
            0,& \text{else}.
        \end{cases}
\end{equation}
This bilocal nature of the causal set action implies that, for subcausets $A$ and $B$ in a causal set $\mathcal{C}$ we have  $S^{(d)}(A\cup B)\ne S^{(d)}(A)+S^{(d)}(B)-S^{(d)}(A\cap B)$ because of bilocal contributions from order intervals that begin in one subcauset and end in the other. For subcausets $A$ and $B$ of a causal set $\mathcal{C}$, we can define a more general ``bi-action'' \cite{Benincasa:thesis}:
\begin{equation} \label{eq:biaction}
\frac{1}{\hbar}S^{(d)}(\mathcal{C};A,B) = -\alpha_{d} \left(\frac{l}{l_{p}}\right)^{d-2}\left(N[A,B] + \frac{\beta_{d}}{\alpha_{d}}
\sum_{i=1}^{n_{d}} C^{(d)}_{i} N_{i}[A,B]\right), 
\end{equation}
where $N[A,B] = |A\cap B|$, and $N_{i}[A,B]$ is the number of pairs of elements $a$, $b$ such that $a\in A$ and $b\in B$ and $|[b,a]|=i-1$. In other words $N_{i}[A,B]$ is the number of order intervals of cardinality $i-1$ such that the minimal element is in $B$ and the maximal element is in $A$.  
 We note that $S(\calC; \calC, \calC) = S(\calC)$.
\par
The explicit forms of equation \eqref{action} for $d=2,3,4$ are listed in Appendix \ref{App:A}.
\subsection{Mean of the Action}
Henceforth, for brevity we will refer to a Lorentzian manifold $(M,g)$ just as $M$ and a causal set $(\calC, \prec)$ as $\calC$ and will consider only sprinkled causal sets. We will assume that $M$ is of dimension $d$ that matches the action we calculate. \par

As described above, evaluating the $d$-action (\ref{action}) of a causal set sprinkled in a manifold $M$ of dimension $d$ and finite volume $V$  turns $N$, $N_i$ and the action $S$ itself into random variables, $\mathbf{N}$, $\mathbf{N}_i$ and $\mathbf{S}^{(d)}$ that each depend on $M$ and on the density of sprinkling $\rho$. For example $\langle \mathbf{N} \rangle = \rho V$. We will refer to $\mathbf{S}^{(d)}$ as the random action of $M$ and taking the mean of the random action over sprinklings gives
\cite{10.1088/1361-6382/abc2fd}
\begin{equation} \label{mean_action}
\begin{split}
    &\frac{1}{\hbar}\langle\boldsymbol{S}_{\rho}^{(d)}(M)\rangle= -\alpha_{d} \left(\frac{l}{l_{p}}\right)^{d-2}\left(\langle\boldsymbol{N}_{\rho}^{(d)}(M)\rangle + \frac{\beta_{d}}{\alpha_{d}}
    \sum_{i=1}^{n_{d}} C^{(d)}_{i} \langle\boldsymbol{N}_{i,\rho}^{(d)}(M)\rangle\right)\\
    &= -\alpha_{d} \left(\frac{l}{l_{p}}\right)^{d-2}\Bigg( \rho V 
    + \frac{\beta_{d}}{\alpha_{d}}
    \sum_{i=1}^{n_{d}} C^{(d)}_{i} \rho^{2} \int_{M} \,dV_x \int_{M \cap J^{+}(x)} \,dV_y  \text{ } \frac{(\rho V_{xy})^{i-1} e^{-\rho V_{xy}}}{(i-1)!}\Bigg)\\
\end{split}
\end{equation}
where $V_{xy}$ is the volume of the causal interval between points $x$ and $y$. $dV_x=d^{d}x \sqrt{-g(x)}$ and likewise for $dV_y$. The origin of the integral expression in the second line is  the Poissonian probability of exactly $i-1$ points being sprinkled in the causal interval  of volume $V_{xy}$ between $x$ and $y$. This expression can be written more compactly in terms of a differential operator $\hat{\mathcal O}_d$ \cite{10.1088/1361-6382/abc2fd}:
\begin{equation} \label{mean_action_operator}
\begin{split}
    &\frac{1}{\hbar}\langle \boldsymbol{S}_{\rho}^{(d)}(M)\rangle  = -\alpha_{d} \left(\frac{l}{l_{p}}\right)^{d-2}\left( \rho V 
    + \frac{\beta_{d}}{\alpha_{d}}
    \rho^{2} \hat{\mathcal O}_d X_{\rho}\right)\\
    &X_{\rho} = \int_{M} \,dV_x \int_{M \cap J^{+}(x)} \,dV_y  \text{ } e^{-\rho V_{xy}},\\
    &\hat{\mathcal O}_d = \sum_{i=1}^{n_{d}} C^{(d)}_{i} \frac{\rho^{i-1}}{(i-1)!}\left(-\frac{\text{d}}{\text{d}\rho}\right)^{i-1}.
\end{split}
\end{equation}
If we partition a manifold $M$ into the union of two non-intersecting regions $X$ and $Y$, and sprinkle into $M$ with subcauset $A$ in region $X$ and subcauset $B$ in $Y$, we can similarly turn the bi-action (\ref{eq:biaction}) into a random variable, $\boldsymbol{S}_{\rho}^{(d)}(M;X,Y)$. \\
In this case,  
the mean of the action for the entire manifold can then be calculated using:
\begin{equation} \label{bilocal_mean}
     \langle \boldsymbol{S}_{M}\rangle  = \langle \boldsymbol{S}_{X}\rangle +\langle \boldsymbol{S}_{Y}\rangle +\langle \boldsymbol{S}_{X,Y}\rangle +\langle \boldsymbol{S}_{Y,X}\rangle 
\end{equation}
where the shorthand $\boldsymbol{S}_{M} = \boldsymbol{S}_{\rho}^{(d)}(M)$ and $  \boldsymbol{S}_{X,Y} =\boldsymbol{S}_{\rho}^{(d)}(M;X,Y) $ have been used. This can be generalised if the manifold is  partitioned into more regions: the action is the sum of the actions for each element of the partition plus bi-action  cross terms between all ordered pairs of elements of the partition.  
\subsection{Regimes}

In all the cases we consider in this paper the manifold we sprinkle into, $M$, is a finite region of Minkowski spacetime $\mathbb{M}^d$. 
When we sprinkle into a manifold ${{M}}$ that is a submanifold of a larger manifold ${\Tilde{M}}$, we can consider the sprinkled causal set $\mathcal{C}$ to be a subset of $\mathcal{\Tilde{C}}$ where 
$\mathcal{\Tilde{C}}$ is the sprinkling in ${\Tilde{M}}$. This can have  consequences for the calculation of the action of 
$\mathcal{C}$ depending on whether or not 
$M$ is \textit{causally convex} in ${\Tilde{M}}$:\\
\textbf{Definition}: $M$ is \textit{causally convex} in ${\Tilde{M}}$ if, for any 2 points in $M$, the causal interval between these 2 points in ${\Tilde{M}}$ is contained in ${{M}}$. \\

In the BDG action \eqref{action}, calculating $N_{i}$ involves counting the number of causal set elements in the order interval between 2 elements. 
If $M$ is not causally convex, when evaluating the action of $\mathcal{C}$ we must decide whether or not to include elements sprinkled in the larger manifold ${\Tilde{M}} = \mathbb{M}^d$.
 
\begin{figure} [H]
    \centering
    \includegraphics[width=50mm,scale=1]{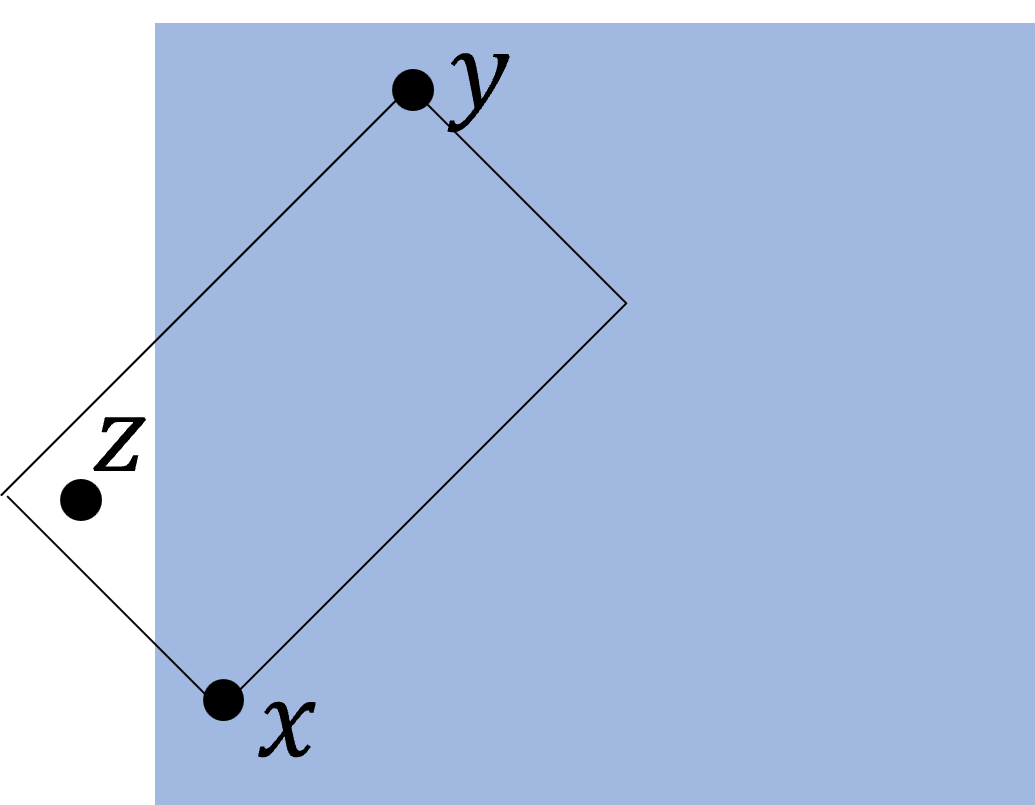}
    \caption{As an example, the blue region is $M$, a non-causally convex manifold.  $x,y$ are elements of $\mathcal{C}$ but $z$ is  an element only of ${\Tilde{C}}$. Whether the order interval $[x,y]$ counts as having cardinality 0 or 1 depends on whether we are working in the isolated or embedded regime, respectively.}
    \label{regimes}
\end{figure}
If $M$ is not causally convex we must decide between two different \textit{regimes} \cite{chevalier:2023}:
\begin{enumerate}
    \item The isolated regime: the causal set $\mathcal{C}$ and manifold $M$ are considered to be physical spacetimes in and of themselves. For example, in Fig. \ref{regimes}, $M$ is the blue rectangle and the sprinkled point $z$ does not exist so $|[x,y]|=0$ and the pair $(x,y)$ is counted in $N_1$. In the integral for the mean of the action, the area $V_{xy}$ equals the area of the diamond between $x$ and $y$ intersected with $M$.
    \item The embedded regime: the manifold $M$ is a sub-manifold embedded within the larger manifold ${\Tilde{M}}$ and causal set $\mathcal{C}$ is a subset of $\Tilde{\mathcal{C}}$. Point $z$ in Fig. \ref{regimes} lies in this larger manifold and so is included in the order interval between $x$ and $y$. As a result, $|[x,y]|=1$ and the pair $(x,y)$ is counted in $N_2$. $V_{xy}$ is the area of the full diamond.
\end{enumerate}
The isolated regime is more relevant to our purpose of investigating the actions of causal sets corresponding to manifolds with different boundary properties and understanding how they contribute to the sum-over-histories. However, the embedded regime sometimes allows analytic results to be obtained because the volume $V_{xy}$ is simpler and so we will also investigate the action in the embedded regime. In most cases, we find that the two regimes give similar results. 

\subsection{Conjectures}\label{sec:conjectures}
For globally hyperbolic manifolds, in the limit $\rho$ tends to infinity, the mean of the random action was conjectured to be local and to equal the Einstein-Hilbert action plus a boundary term \cite{Benincasa:thesis}:\\
\noindent\textbf{Conjecture 1:}\\
\begin{equation} \label{conjecture1}
\lim_{\rho\to\infty}\frac{1}{\hbar}\langle\boldsymbol{S}_{\rho}^{(d)}(M)\rangle = \frac{1}{l_{p}^{d-2}}\int_{M} \,d^{d}x \sqrt{-g}\frac{R}{2} + \frac{1}{l_{p}^{d-2}} \text{Vol}_{d-2}(J).
\end{equation}
$\text{Vol}_{d-2}(J)$ is the volume of the so-called \textit{joint}. A joint is a co-dimension 2 surface that is the intersection of the 
the future boundary and the past boundary of the spacetime, where the future (past) boundary is the set of points where future directed causal curves in $M$ leave $M$. (\ref{conjecture1})  has been proved for causal diamonds in Minkowski space in any dimension \cite{Buck:2015oaa}: for example, for $d=2$ the volume of the joint is 2, there is no Planck length and the limit of the mean of the action equals 2. Further evidence has been provided for the conjecture in the case of causal intervals and other regions in curved spacetime with low curvature \cite{10.1088/1361-6382/abc2fd, MachetWang_2021}.\par

However, Conjecture 1 had not been investigated for joints between 2 spacelike boundaries and in Section \ref{spacelikejoints} we will show that the joint term in the conjecture above is not correct in general and must be modified to: \\
\noindent\textbf{Conjecture 1':}\\
\begin{equation}\label{conjecture1prime}
    \begin{split}
          \lim_{\rho\to\infty}\frac{1}{\hbar}\langle\boldsymbol{S}_{\rho}^{(d)}(M)\rangle &= \frac{1}{l_{p}^{d-2}}\int_{M} \,d^{d}x \sqrt{-g}\frac{R}{2}+ \frac{1}{l_p^{d-2}}\int_{J} d\mu(\lambda)  \coth{(\theta(\lambda))}.
    \end{split}
\end{equation}
where $\theta(\lambda)$ is the Lorentzian angle between the tangent vectors to the past and future boundaries respectively at the point $\lambda$ on the joint and 
$d\mu(\lambda)$ is the (d-2)-dimensional  volume measure on the joint. This is consistent with Conjecture 1 in the cases in which Conjecture 1 was tested because in those cases $\theta$ is constant and equal to infinity and so 
$\coth(\theta) = 1$.\\

\noindent For manifolds with timelike boundaries (an example of non-globally hyperbolic manifolds), the conjecture is \cite{Benincasa:thesis}:\\
\noindent\textbf{Conjecture 2:}\\
\begin{equation} \label{timelike_conjecture}
\lim_{{\rho \to \infty}} \frac{1}{\hbar} \langle \boldsymbol{S}^{(d)}_{\rho}(M) \rangle \, \rho^{-1/d}\, = \, \frac{1}{l_{p}^{d-2}} a_d^{\text{emb/iso}} \text{Vol}_{d-1}(T) 
\end{equation}
where $a_d^{\text{emb/iso}}$ is a $d$-dependent dimensionless coefficient, different according to whether the action is calculated in the embedded or isolated regime, and $\text{Vol}_{d-1}(T)$ is the volume of the timelike boundary $T$.  This implies that the mean of the random action diverges like $\rho^{1/d} = l^{-1}$ as the discreteness scale $l$ goes to zero. In the sum-over-histories, then, such causal sets will have very large actions, with potential consequences that we will discuss in the final section. For $d=2$, these conjectures were first investigated numerically in \cite{Benincasa:2010as} and have been tested on a rectangle and a disc in \cite{chevalier:2023}, and the coefficients found to be:
\begin{equation}
    \begin{split}
        &a_2^\text{emb} = \frac{1}{2}\sqrt{\frac{\pi}{2}}\\
        &a_2^\text{iso} = 0.6959.\\
    \end{split}
\end{equation}
$a_2^\text{iso}$ is  a value calculated from numerical integrations and there is no error estimation \cite{chevalier:2023}. 

\subsection{Weighted Integral Method in flat space}

To compute the mean of the action (equation \eqref{mean_action}) in flat spacetime ($dV_x = d^dx$), we must evaluate the integral $X_{\rho}$ from equation \eqref{mean_action_operator}. A way of calculating this integral in the embedded regime, that we call the weighted integral method, was developed in \cite{chevalier:2023} for flat space. The original $d=2$ integration variables in \eqref{mean_action} are $\Vec{x}$ and $\Vec{y}$. These variables are transformed to $\Vec{x}$ and $\Vec{c} = \Vec{y}-\Vec{x}$. The jacobian for this transformation equals 1. One interprets $\Vec{c}$ as a timelike future-pointing ``defining vector'' for the causal interval from $x$ to $y$ in $M$. The variables $\Vec{x}$ then represent the position of the past endpoint of the interval. The point is that in the embedded regime the integrand  does not depend on $\Vec{x}$ because the volume of the interval between  $x$ and $y$ is a function of $\Vec{c}$ only and so the integral over $\Vec{x}$ can be done first, resulting in a factor we call the ``volume of realisation'' $a(\Vec{c})$. This is the volume in which the point $x$ can be such that the defining vector $\Vec{c}$ starting at $x$ fits within the manifold. \par

\begin{figure} [H]
    \centering
    \includegraphics[width=100mm,scale=1]{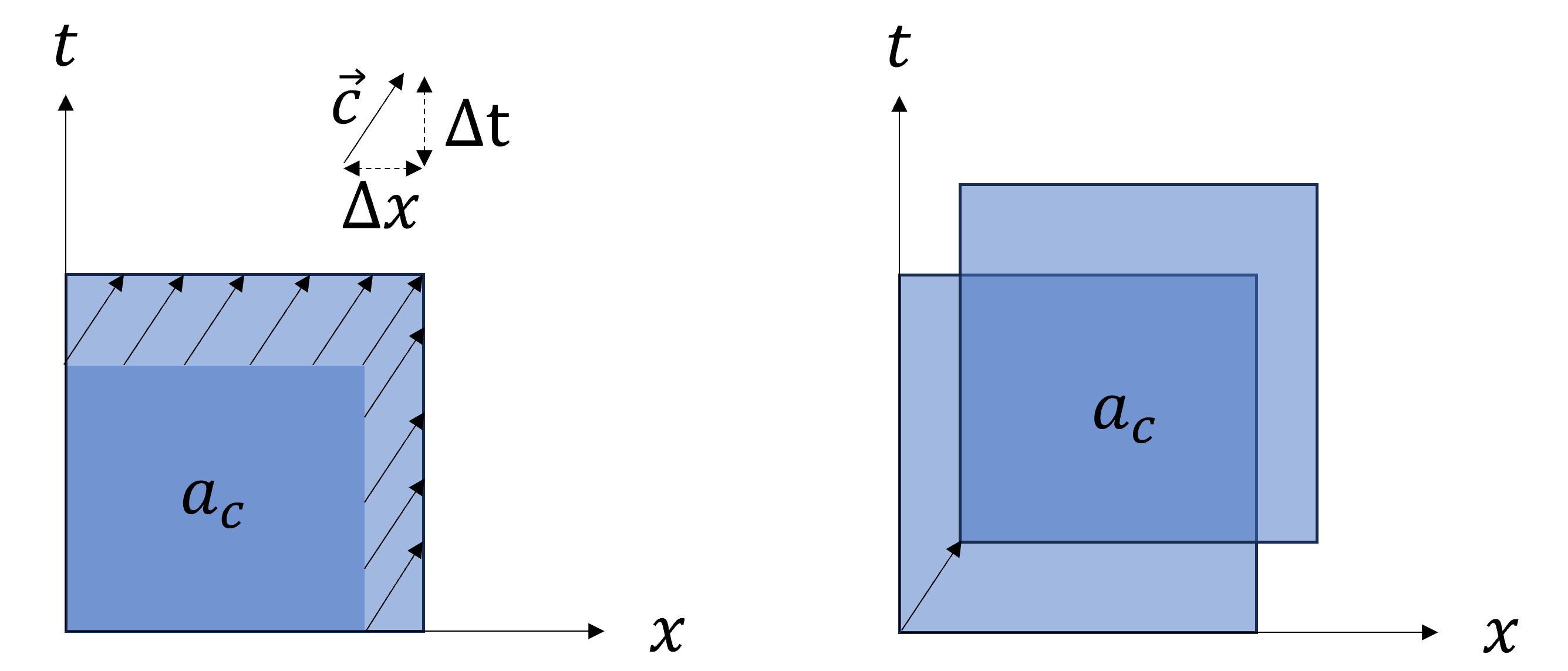}
    \caption{A d=2 example, showing how by cloning the manifold and displacing it by the defining vector $\vec{c}=(\Delta t, \Delta x)$, the volume of realisation can be found from the overlapping area. The dark blue region on the left has an identical area to the overlapping area on the right.}
    \label{cloning}
\end{figure}

To find $a(\Vec{c})$, we clone and displace the manifold by the defining vector. The volume of realisation equals the volume of the intersection of the two isometric shapes. This can be seen in Fig. \ref{cloning} for a simple example, but will be true for any manifold in any dimension. Sometimes $a(\Vec{c})$ is easy to calculate and then the remaining integration is performed over all future-pointing timelike defining vectors that can fit within the manifold.
Following this method, the integral \eqref{mean_action} becomes 
\begin{equation} \label{X_rho}
    X_\rho = \int_{c\in M}d^dc \text{ }a(\Vec{c})e^{-\rho V_c}.
\end{equation}
 $V_c$ is the volume of the causal interval defined by the vector $\Vec{c}$. Note: we will sometimes use the notation $ \vec{c} = (\Delta t, \Delta x_1, \dots, \Delta x_{d-1})$. 
\par
This method only works if the volume of the causal interval between $\vec{x}$ and $\vec{y}$ depends only on $\vec{c}= \vec{y} - \vec{x}$.  This is true in the embedded regime and for globally hyperbolic manifolds -- which are causally convex and so the two regimes coincide -- but not for non-causally convex manifolds in the isolated regime. In the latter case, the causal interval will no longer be a complete diamond for intervals too close to a timelike boundary. We also cannot use this method in curved spacetime because defining vectors do not exist. \par

We can also use the weighted integral method to calculate the bi-action contributions. If we have two regions of the manifold $X$ and $Y$, in order to find $\langle \boldsymbol{S}_{\rho}^{(d)}(M;X,Y)\rangle $, we clone region $Y$ and displace it by the defining vector $\vec{c}$. The volume of realisation for this defining vector equals the volume of the intersection of the displaced region $Y$ and original region $X$. Then, defining vectors that can start in region $Y$ and end in region $X$ are integrated over. 

\subsection{Smearing to tame the fluctuations}\label{sec:smearing}
The fluctuations of the random action around its expected value increase as $\rho$ increases which makes it extremely challenging to discover the expected value in the $\rho \rightarrow \infty$ limit by performing simulations.  In order to obtain useful results from simulations, it is useful to dampen these fluctuations by using a smeared form of the action, which introduces an intermediate length scale $l_k \geq l$. This smearing scale is parametrised by $\epsilon\le 1$ which is defined as
\begin{equation} \label{epsilon}
    \epsilon=\left(\frac{l}{l_k}\right)^d=\frac{K}{\rho},\quad \textrm{where} \quad K=\frac{1}{l_k^d}.
\end{equation}

The smeared form of the action is based on the smearing of the causal set D'Alembertian introduced by Sorkin to dampen its fluctuations  \cite{Sorkin:2007qi}. The smeared action is given by 
\cite{Benincasa:thesis,dalembertians_various_dimensions}:
\begin{equation} \label{smeared_action}
   \frac{1}{\hbar} S^{(d)}_\epsilon (\mathcal{C}) =- \alpha_d \left(\frac{l}{l_p}\right)^{d-2}\left( \epsilon^{2/d} N+\frac{\beta_d}{ \alpha_d} \epsilon^{(d+2)/d} \sum_{x\in \mathcal{C}}\sum_{y\prec x}f_d(n(x,y),\epsilon) \right)
\end{equation}
where $n(x,y)=|[y,x]|$. Now the sum is over all pairs of related elements $y \prec x$  and
\begin{equation} \label{function}
    f_d(n,\epsilon)=(1-\epsilon)^n \sum_{i=1}^{n_d}C_i^{(d)}\binom{n}{i-1}\left(\frac{\epsilon}{1-\epsilon}\right)^{i-1}.
\end{equation}
When $\epsilon =1$, the smeared action equals the BDG action. 
The explicit forms of equation \eqref{smeared_action} for $d=2,3,4$ that we used in the simulations are given in Appendix \ref{App:A}. \par
By Poisson sprinkling a manifold at density $\rho$ and and evaluating $S^{(d)}_\epsilon (\mathcal{C})$ for a given $\epsilon$, $S^{(d)}_\epsilon (\mathcal{C})$ becomes the random variable $\boldsymbol{S}^{(d)}_{\rho,K}(M)$. In the limit of infinite $\rho$, with $K$ fixed, the mean of $\boldsymbol{S}^{(d)}_{\rho,K}(M)$ equals the mean of the random BDG action with $\rho$ replaced by $K$:
\begin{equation} \label{limit_smeared_action}
\lim_{{\rho \to \infty}} \frac{1}{\hbar} \langle \boldsymbol{S}^{(d)}_{\rho,K}(M) \rangle = \frac{1}{\hbar}\langle \boldsymbol{S}^{(d)}_{K}(M) \rangle \,. 
\end{equation}
Now the fluctuations of $\boldsymbol{S}^{(d)}_{\rho,K}(M)$ around its expected value die away as $\rho \rightarrow \infty$ (as $\epsilon\rightarrow 0$) so it is more suitable for simulations.\footnote{It is an open question whether the smeared version of the action should be used in the path integral instead of the BDG action. We will not take a position on this here, but will use the smeared action to do simulations to obtain evidence for the conjectures.} This advantage is gained however at the expense of needing to know the form of finite $\rho$ corrections to the conjectures as we explain.\par

We expect that the mean of the original unsmeared random action at large, finite $\rho$ will be an expansion like 
\begin{equation} \label{rho_action}
 \frac{1}{\hbar} \langle \boldsymbol{S}^{(d)}_{\rho}(M) \rangle = \frac{1}{l_{p}^{d-2}} (c_1 \rho^{\lambda_1} + c_2 \rho^{\lambda_2}  +... + c_{n-1} \log{\rho} + c_n)
\end{equation}
up to terms that are exponentially small for large $\rho$. The $c_i$ are--bulk and boundary--geometric quantities of the manifold times dimensionless constants and the powers $\lambda_i$ of $\rho$ ensure dimensional consistency. There may be other terms such as other $\log$ terms. This is a conjecture that includes both original conjectures \eqref{timelike_conjecture} and \eqref{conjecture1} and  that includes subleading terms in the expansion in $\rho$ for large $\rho$. \par

The mean of the random smeared action will be dependent on the new scale $K$ as well as on $\rho$. Equations (\ref{limit_smeared_action}) and (\ref{rho_action}) imply that in the limit of infinite $\rho$, with $K$ fixed, the conjecture is 
\begin{equation} \label{K_action}
\lim_{{\rho \to \infty}} \frac{1}{\hbar} \langle \boldsymbol{S}^{(d)}_{\rho,K}(M) \rangle = \frac{1}{l_{p}^{d-2}} (c_1 K^{\lambda_1} + c_2 K^{\lambda_2} + ... c_{n-1} \log{K} + c_n)
\end{equation}
where the geometric quantities and exponents are the same as before and the only difference is that $K$ has replaced $\rho$. For example, in the case of a manifold with timelike boundaries, Conjecture 2 becomes 
\begin{equation}\label{K_conj}
\lim_{{\rho \to \infty}} \frac{1}{\hbar} \langle \boldsymbol{S}^{(d)}_{\rho,K}(M) \rangle = \frac{1}{\hbar} \langle \boldsymbol{S}^{(d)}_{K}(M) \rangle = \frac{1}{l_{p}^{d-2}} a_d^\text{emb/iso} \text{Vol}_{d-1}(T) K^{\frac{1}{d}} + \text{subleading}
\end{equation}
When  $\rho$ is finite, we expect there to be corrections to this expression that depend on both $K$ and $\rho$,  $E_{K,\rho}$:
\begin{equation} \label{smear_error}
\frac{1}{\hbar} \langle \boldsymbol{S}^{(d)}_{\rho,K}(M) \rangle = \frac{1}{\hbar} \langle \boldsymbol{S}^{(d)}_{K}(M) \rangle + E_{K,\rho}\,.
\end{equation}

In all of the simulations that follow, we fixed $\epsilon = 0.1$,  i.e. $K = \rho/10$,  and we varied the sprinkling density $\rho$. Increasing $\rho$ therefore increases $K$. 

\section{Timelike Boundaries in 1+1 Dimensions}\label{sectionTL2d}
All the examples in this section are in the embedded regime. 
\subsection{Triangle in 1+1}\label{trianglesec}
\begin{figure} [H]
    \centering
    \includegraphics[width=100mm,scale=0.5]{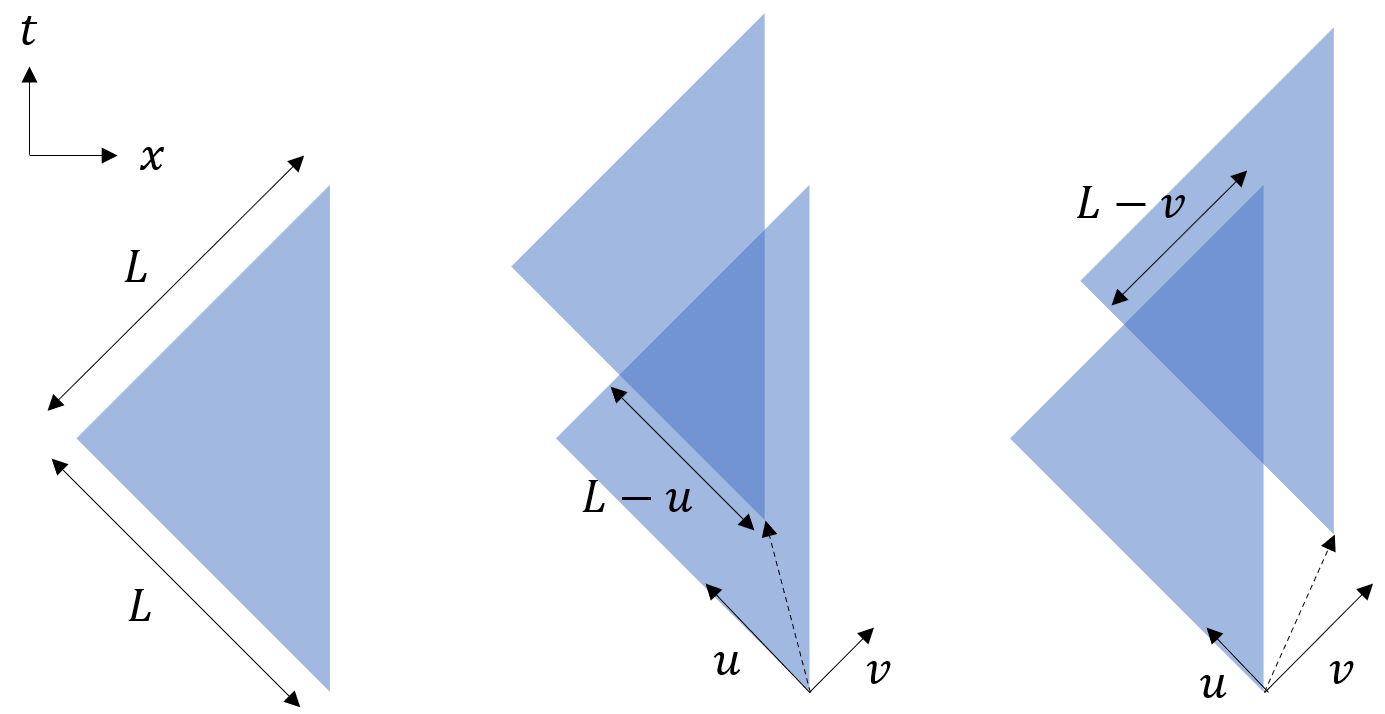}
    \caption{Triangular manifold (left) and its area of realisation (= area of overlapping region) with $u> v$ (middle) and $v> u$ (right).}
    \label{overlap}
\end{figure}

As our first example of a spacetime region with timelike boundaries, we considered the triangle shown in the far left of Fig. \ref{overlap}. This has two null and one timelike edge. Coordinates $(t,x)$ have origin at the bottom corner of the triangle and we use null coordinates 
$u = \frac{1}{\sqrt{2}}(t - x)$ and $v = \frac{1}{\sqrt{2}}(t + x)$.
Using the weighted integral method, we consider the vector $(u,v)$ (working in null coordinates) that defines an interval that can fit within this region. To find the volume of realisation, the situations $u >  v$ and $v >  u$ must be considered separately. This is shown in Fig. \ref{overlap}
We find,
\begin{equation}
    \begin{split}
        X_\rho &=  \int_{0}^L \,du \int_{0}^u \,dv \text{ }\frac{1}{2}(L-u)^2 e^{-\rho uv}+\int_{0}^L \,dv \int_{0}^v   \,du \text{ }\frac{1}{2}(L-v)^2 e^{-\rho uv}\\
        & =2\int_{0}^L \,du\int_{0}^u   \,dv \text{ }\frac{1}{2}(L-u)^2 e^{-\rho uv}\,
    \end{split}
\end{equation}
and \eqref{mean_action_operator} gives
\begin{align}
        \frac{1}{\hbar}\langle\boldsymbol{S}_{\rho}(M)\rangle &=  2\rho V -4\rho^2 \hat{\mathcal O}_2 X_\rho\label{eq:trianglemean}
        \\
        &=\frac{1}{2} \sqrt{\pi} L \sqrt{\rho} \text{Erf}(L \sqrt{\rho})\nonumber
\end{align}
where $\text{Erf}$ is the error function. 
This expression does not have the $+1$ contribution which we might have expected from the joint (left corner). This is further investigated in Section \ref{inf_t_sec} and Appendix \ref{crossterm}.
The length of the timelike boundary for this manifold is $T = \sqrt{2} L$ and, we have in the limit,
\begin{equation} 
\lim_{\rho\to\infty}\frac{1}{\hbar}\langle\boldsymbol{S}_{\rho}(M)\rangle \sim \frac{1}{2} \sqrt{\frac{\pi}{2}} T \sqrt{\rho}\,
\end{equation}
which agrees with conjecture 2 and the value of $a_2^{emb} = \frac{1}{2}\sqrt{\frac{\pi}{2}}$ found in \cite{chevalier:2023}.

\subsection{1+1 Infinite t} \label{inf_t_sec}
\begin{figure} [H]
    \centering
    \includegraphics[width=80mm,scale=0.5]{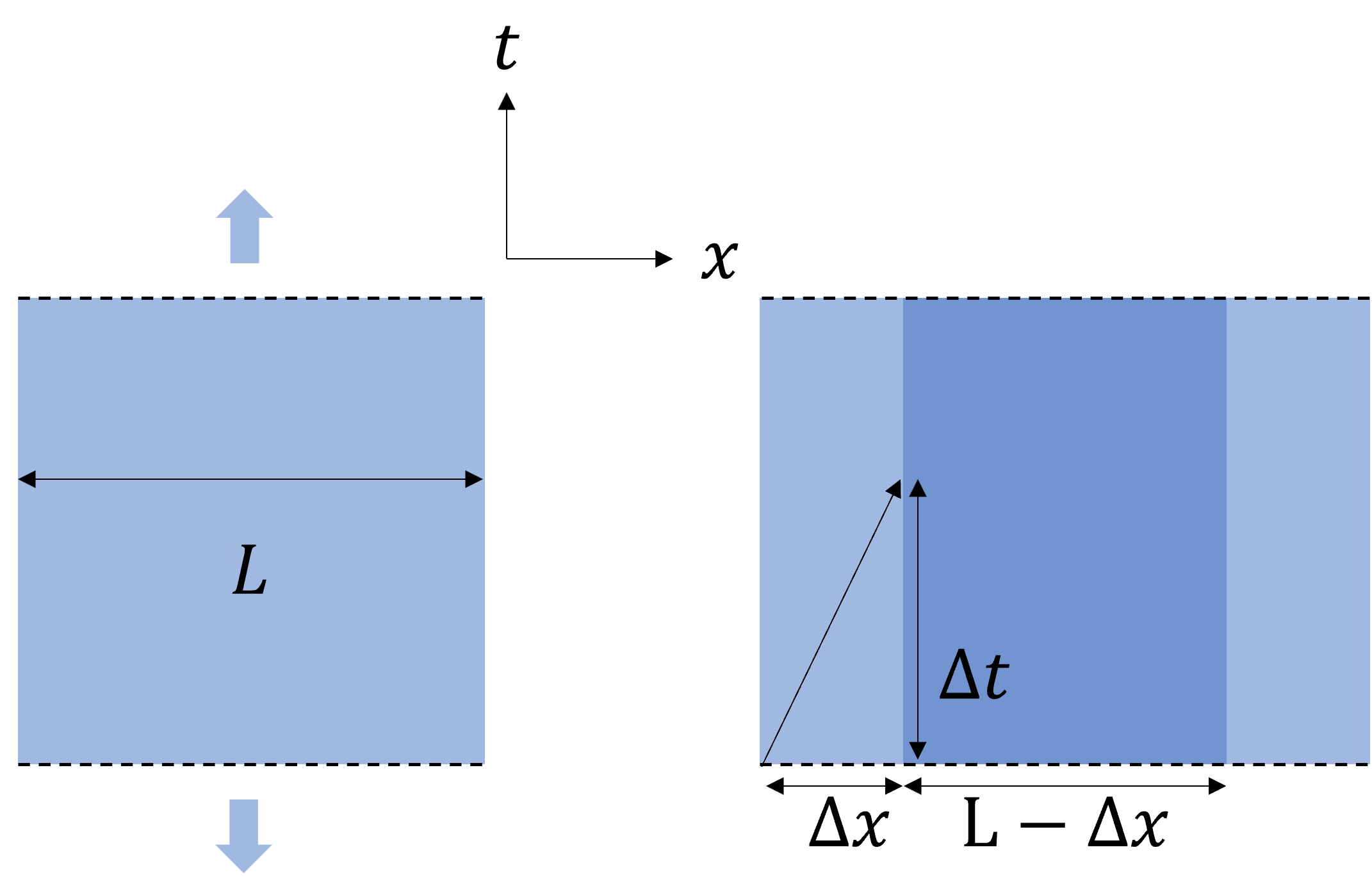}
    \caption{Left: a spacetime region with infinite timelike boundaries. Right: cloning the region to obtain the volume of realisation, shown in dark blue.}
    \label{infinite_t}
\end{figure}

The next shape for which we compute the action is a slab, which is finite in the $x$ direction and infinite in the $t$ direction. This manifold is depicted in Fig. \ref{infinite_t}.  
The volume of realisation is 
\begin{equation}
    a(\Vec{c})=(L-\Delta x)\left(\int_{-\infty}^{\infty} dt\right)
\end{equation}
as the length in the $t$ direction of this volume is infinite. For the vector to define an interval, $|\Delta x|  < \Delta t$. However, the defining vector will only fit within the manifold if $|\Delta x|  < L$. The integral is a sum of 2 integrals for $\Delta t< L$ and $\Delta t> L$:
\begin{equation}
    \begin{split}
        X_\rho &=  2 \int_{-\infty}^\infty \,dt\left(\int_{0}^L \,d\Delta t \int_{0}^{\Delta t} \,d\Delta x +\int_{L}^\infty \,d\Delta t \int_{0}^{ L} \,d\Delta x\right)(L-\Delta x) e^{-\frac{1}{2} \rho  \left(\text{$\Delta $t}^2-\text{$\Delta
   $x}^2\right)}.
    \end{split}
\end{equation}
The factor of 2 is  due to symmetry when including the contribution from $\Delta x< 0$. The action is then given by
\begin{equation}\label{moa rectangle}
    \begin{split}
    \frac{1}{\hbar}\langle\boldsymbol{S}_{\rho}(M)\rangle=2\rho V-4\rho^2 \hat{\mathcal O}_2 X_\rho =\int_{-\infty}^\infty \,dt \text{ } \sqrt{\frac{\pi }{2}} \sqrt{\rho }+\mathcal{O}\left(\frac{1}{\rho }\right).
    \end{split}
\end{equation}
For this manifold, the volume of the timelike boundary is $T = 2 \left(\int_{-\infty}^\infty dt \right)$ since there are 2 boundaries. We therefore find 
\begin{equation} 
\lim_{\rho\to\infty}\frac{1}{\hbar}\langle\boldsymbol{S}_{\rho}(M)\rangle \sim \frac{1}{2} \sqrt{\frac{\pi}{2}} T \sqrt{\rho}\,,
\end{equation}
 consistent with the conjecture and $a_2^{emb} = \frac{1}{2}\sqrt{\frac{\pi}{2}}$.  
\par
Moreover, from Eqn. \eqref{moa rectangle}, we see that the mean of the action does not contain any constant terms. This is an interesting result, suggesting the boundary itself does not give any constant terms, and only contributes the conjectured term in the infinite density limit. Therefore, we might suspect the missing +1 term from the causal triangle in Section \ref{trianglesec} was not cancelled out by the lower ordered timelike boundary terms, but by something else. A good candidate is the top and bottom corners of the triangle (the intersection between a null boundary and a timelike boundary), which each contribute $-\frac{1}{2}$ to the action. This is of minor importance however, since the action is always dominated by the term from the timelike boundary. A contribution from the intersection between a general timelike and a null/spacelike boundary is conjectured in Appendix \ref{t-s_corners_section}.

\section{$d$-dimensional infinite slab in the embedded regime}\label{sectionTLdd}
\subsection{Set-up}
Assuming universality of the constant $a_d^{\text{emb}}$, we choose a convenient manifold with a timelike boundary to calculate $a_d^{\text{emb}}$: the $d$ dimensional ``infinite slab''. This manifold is given by $t\in [0,T]$, $x_1 \in [0, L]$ with $T \le L$, and all other coordinates are unbounded $-\infty < x_i < \infty$, $ i= 2,3 \dots d-1$.
For such a slab in $d$ dimensions, the conjecture \eqref{timelike_conjecture} becomes:
\begin{equation} \label{infinite_slab_conjecture}
\lim_{{\rho \to \infty}} l_p^{d-2}\frac{1}{\hbar} \langle \boldsymbol{S}^{(d)}_{\rho}(M) \rangle \rho^{-\frac{1}{d}}= a_d^\text{emb} \int_{-\infty}^{\infty} dx_2...\int_{-\infty}^{\infty} dx_{d-1} \text{ } 2 T \,.
\end{equation}
We again use the weighted integral method. The defining vector has components $\vec{c} = (\Delta t, \Delta x_1,   \Delta x_2,... \Delta x_{d-1})$. We find
\begin{equation}\label{eq:28}
    \begin{split}
        X_\rho(d) &=  \int_{0}^{T} d \Delta t \int_{-\Delta t}^{\Delta t} d\Delta x_1 \int_{-\sqrt{\Delta t^2 - \Delta x_1^2}}^{\sqrt{\Delta t^2 - \Delta x_1^2}} d\Delta x_2...\int_{-\sqrt{\Delta t^2 - \Delta x_1^2...-\Delta x_{d-2}^2}}^{\sqrt{\Delta t^2 - \Delta x_1^2...-\Delta x_{d-2}^2}} d\Delta x_{d-1}\\
        &\int_{-\infty}^{\infty} dx_2...\int_{-\infty}^{\infty} dx_{d-1}(L-|\Delta x_1|)(T-\Delta t) e^{-\rho V_{c}(d)}
    \end{split}
\end{equation}
where the limits of the integral in the first line ensure that the defining vector is timelike and future pointing.\par

The volume of the interval (causal diamond) $V_{c}(d)$ is a function of the proper height of the interval i.e. the proper length of the defining vector $c = |\vec{c}|$:
\begin{equation}
    V_{c}(d) = \frac{\tau}{d} V_{ball}(d-1) = \frac{\tau}{d}\frac{\pi^{(d-1)/2}(\frac{c}{2})^{d-1}}{\Gamma((d+1)/2)}
\end{equation}
where $V_{ball}(d-1)$ is the volume of the $(d-1)$ dimensional ball with radius $c/2$. \par

For the rest of this section, we will suppress mention of the integrals $\int_{-\infty}^{\infty} dx_2...\int_{-\infty}^{\infty} dx_{d-1}$ which are implied. 
\par
We switch to radial spherical polar variables for the spatial components of $\vec{c}$ including 
\begin{equation}
    \Delta r = \sqrt{\Delta x_1^2 + \Delta x_2^2... + \Delta x_{d-1}^2}\,, 
\end{equation}
and define null radial variables 
\begin{equation}\label{eq:nullcoortr}
    v = \frac{1}{\sqrt{2}}(\Delta t+ \Delta r),\text{ } u = \frac{1}{\sqrt{2}}(\Delta t- \Delta r)\,.
\end{equation}

\begin{figure}[H]
    \centering
    \includegraphics[width=120mm,scale=0.5]{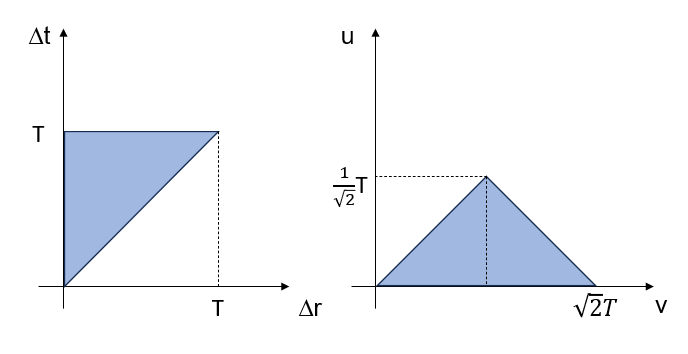}
    \caption{Left: the area of integration in $\Delta t,\Delta r$ space. Right: the area of integration in $u,v$ space.}
    \label{6.1}
\end{figure}

We find 
\begin{equation}\label{eq:6.8}
\begin{split}
  &X_\rho(d) =\\
    & - \int_{0}^{\frac{T}{\sqrt{2}}} du \int_{u}^{\sqrt{2}T-u} dv\textbf{ }e^{-\rho V_{c}(d)}\\
    & \frac{(2 \pi )^{\frac{d}{2}-2} (v-u)^{d-2} \left(\sqrt{2} (u+v)-2 T\right) \left(2
   \pi  L \Gamma \left(\frac{d}{2}\right)+\sqrt{2 \pi } \Gamma
   \left(\frac{d-1}{2}\right) (u-v)\right)}{\Gamma (d-1)} 
\end{split}
\end{equation}
where 
\begin{equation} \label{V_c}
     V_{c}(d) = \frac{2^{1-\frac{d}{2}} \pi ^{\frac{d-1}{2}} (u v)^{d/2}}{d \Gamma \left(\frac{d+1}{2}\right)}.
\end{equation}
The range of the $u,v$ integration is within the area $\int_{0}^{T} d \Delta t \int_{0}^{\Delta t} dr$, shown in Fig.  \ref{6.1} both in $\Delta t,r$ and $u,v$ coordinates. We rename $X_\rho =  X_\rho^{(1)}$ for reasons that will be clear in the next section. We are not able to evaluate this integral analytically.
\subsection{Extending the triangle}
\begin{figure}[H]
    \centering
    \includegraphics[width=60mm,scale=1]{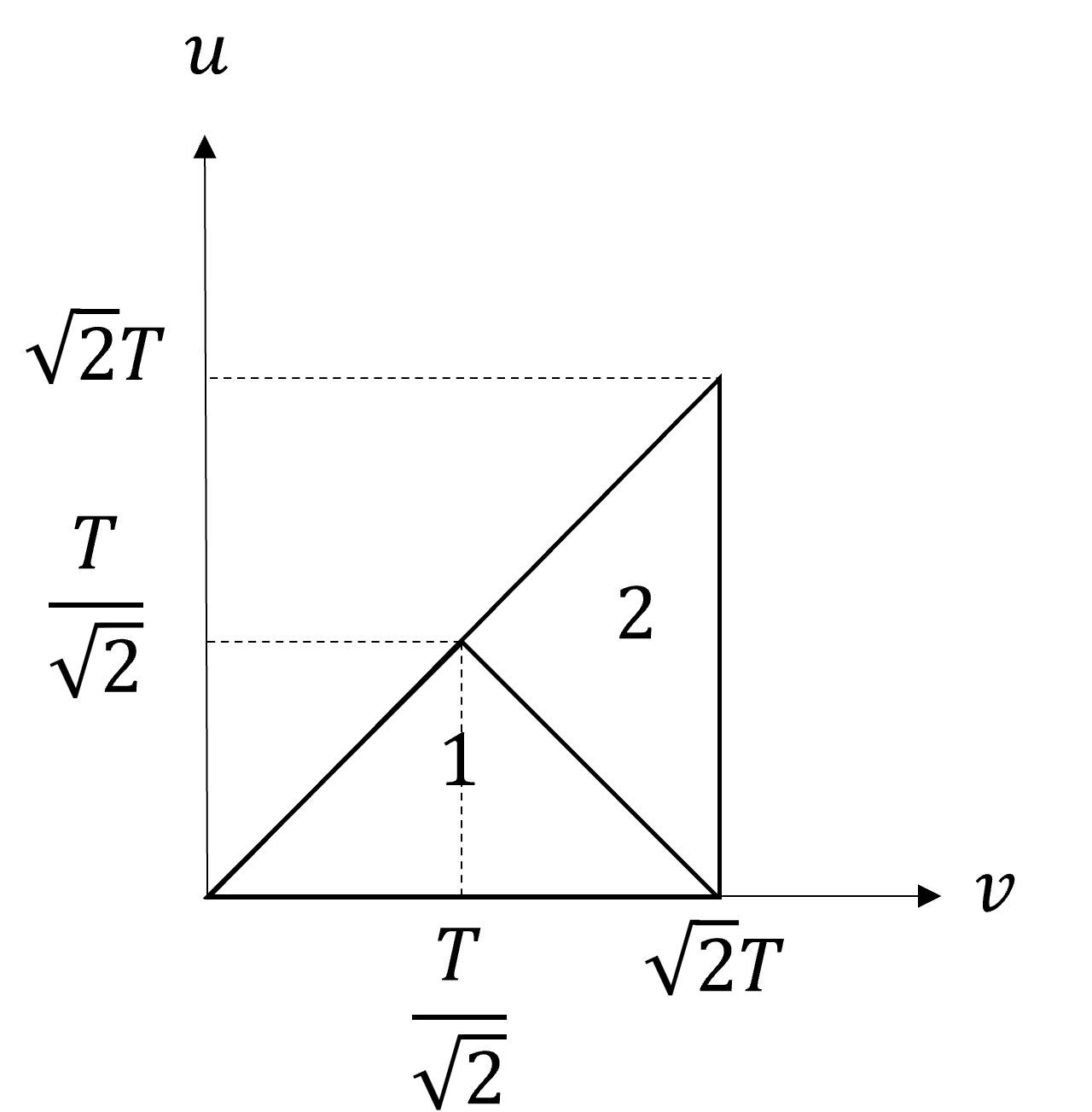}
    \caption{Extending the shape in Fig. \ref{6.1} (right).}
    \label{extended_triangle_1}
\end{figure}

For the conjecture, we are interested in the limit of large $\rho$. Observing the exponent of equation \eqref{eq:6.8} (i.e. $V_c(d)$ from equation \eqref{V_c}), we note the integrand is exponentially suppressed when both $u$ and $v$ are bounded away from zero. Therefore, the non-exponentially suppressed contribution comes from close to $u=0$ or $v=0$. This idea inspires an approximation of $X_\rho =  X_\rho^{(1)}$ in \eqref{eq:6.8} by extending the range of integration to the full triangle shown in Fig. \ref{extended_triangle_1}. This extended range is the union of region 1, the original range, and an extra region 2.  In Appendix \ref{proof}, we show that the integral over region 2 alone tends to zero as $\rho\rightarrow\infty$, using both numerical and analytical arguments. Thus, extending the range of integration does not change the leading order behaviour of $X_\rho =  X_\rho^{(1)}$ as $\rho\rightarrow \infty$. 

We have 
\begin{equation} \label{int_in_ext}
\begin{split}
    X_\rho^\text{extended}(d) &= X_\rho^{(1)}(d)+X_\rho^{(2)}(d)\\&=  -\int_{0}^{\sqrt{2}T} du \int_{u}^{\sqrt{2}T} dv\ e^{-\rho V_c(d)}\\
    & \frac{(2 \pi )^{\frac{d}{2}-2} (v-u)^{d-2} \left(\sqrt{2} (u+v)-2 T\right) \left(2
   \pi  L \Gamma \left(\frac{d}{2}\right)+\sqrt{2 \pi } \Gamma
   \left(\frac{d-1}{2}\right) (u-v)\right)}{\Gamma (d-1)}, 
\end{split}
\end{equation}
where the superscript refers to the region in which the integral is performed i.e. $X_\rho^{(1)}$ is the integral evaluated in region 1, and $X_\rho^{(2)}$ in region 2.  The integral with the extended range can be performed analytically in any chosen dimension and then using operator $\hat{\mathcal O}_d$, the leading order of the mean action can be found.  
\subsection{Results}\label{sec:slabresults}
The mean of the action calculated as described above is consistent with conjecture \eqref{infinite_slab_conjecture} in every dimension up to and including $d=17$ with coefficients $a_d^{\text{emb}}$ given by 
\begin{equation}\label{eq:slabresults}
        a_d^{\text{emb}}=
        \begin{cases}
            \frac{2^{-\frac{1}{d}} \pi ^{{-\frac{1}{2d}}} \left(\Gamma
   \left(\frac{d+1}{2}\right) d\right)^{-\frac{1}{d}}}{\Gamma
   \left(\frac{1}{2}+\frac{1}{d}\right)},& \text{odd d}\\
            \frac{2^{-\frac{1}{d}-1} \pi ^{{-\frac{1}{2d}+1}} \left(\Gamma
   \left(\frac{d+1}{2}\right) d\right)^{-\frac{1}{d}}}{\Gamma
   \left(\frac{1}{2}+\frac{1}{d}\right)} ,              & \text{even d}.
        \end{cases}
\end{equation}
where $\Gamma(x)$ is the Gamma function. We conjecture that these formulae hold for all $d$.
\par
For $d=3$ specifically, if we integrate over the extended triangle, and then subtract the leading order contribution from region 2, then we find that the mean of the action has the form of 
\begin{equation}\label{eq:finiterho}
    \frac{1}{\hbar} \langle \boldsymbol{S}^{(d=3)}_{\rho}(M) \rangle = \frac{1}{l_{p}} (c_1 \rho^{1/3} + c_2 \rho^{-1/3} + ...),
\end{equation}
for finite $\rho$ in the large $\rho$ expansion, where $c_1$ and $c_2$ are constants. In what follows we will assume this form of finite $\rho$ correction when we assess and fit computational results for other $d=3$ manifolds, as the corrections to the leading term will not be negligible. 
\section{Simulations of Timelike Boundaries}\label{sectionTLsimu}
\subsection{Simulation of $d=3$ Ball in Embedded Regime}
\begin{figure}[H]
    \centering
    \includegraphics[width=50mm,scale=1]{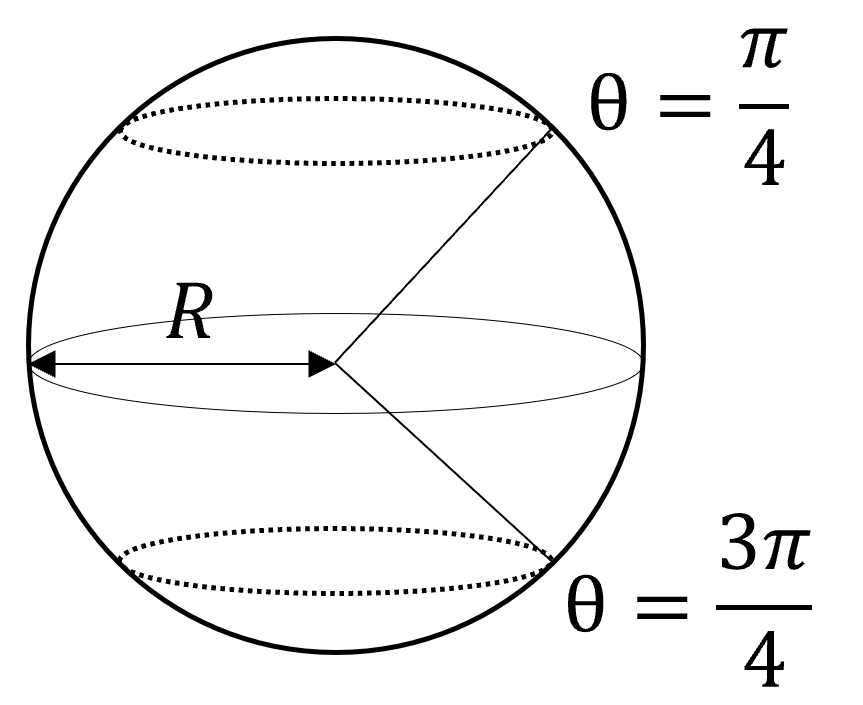}
    \caption{The timelike boundary is the surface  between the 2 dotted lines.}
    \label{timelike_sphere_volume}
\end{figure}

In order to test the conjecture with a manifold with curved timelike boundaries, using (\ref{eq:slabresults}), 
we sprinkled a region of  $d=3$ 
Minkowski space that is a ball of Euclidean coordinate radius $R$ in the embedded regime, see Fig. \ref{timelike_sphere_volume}.  

The area of the timelike boundary is 
$ V = {\pi^2}R^2/{\sqrt{2}}$
and the conjecture for the action for this manifold is:
\begin{equation}
\lim_{{\rho \to \infty}} l_p\frac{1}{\hbar} \langle \boldsymbol{S}^{(3)}_{\rho}(M) \rangle \rho^{-\frac{1}{3}} = \frac{\pi^2}{\sqrt{2}}R^2 a_3^{\text{emb}} \,.
\end{equation}
\subsubsection{Results}
The random smeared action was sampled by sprinkling, for $R=0.5$. As described in  Section 
\ref{sec:smearing}, the conjecture implies for fixed $K$ that (\ref{K_conj})
\begin{equation}\label{eq:form}
\lim_{{\rho \to \infty}} l_p\frac{1}{\hbar} \langle \boldsymbol{S}^{(3)}_{\rho, K}(M) \rangle = \frac{\pi^2}{4\sqrt{2}} a_3^{\text{emb}} K^{\frac{1}{3}}  + c_2 K^{-\frac{1}{3}} + o(K^{-\frac{1}{3}})
\end{equation}
where we have assumed the first correction has the same form as in the calculation for the $d=3$ infinite slab (\ref{eq:finiterho}). \par
\begin{figure}[H]
    \centering
    \includegraphics[width=150mm,scale=1]{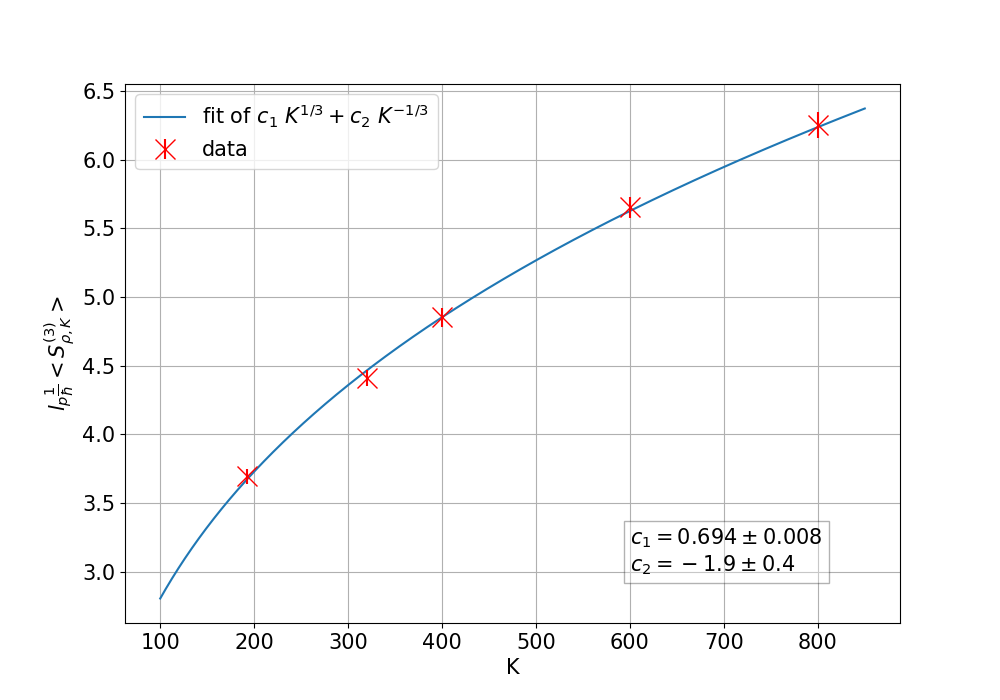}
    \caption{The sample mean of the action of the sphere against $K$, with a curve fitted to the data (blue). The sphere has radius 0.5.}
    \label{sphere_graph}
\end{figure}
In the simulations $\epsilon = 0.1$ is fixed so that $K = \rho/10$ and the sprinkling density $\rho$ was varied.  
For each $\rho$, the sample mean of the action over 100 sprinklings was calculated, the results are shown in Fig. \ref{sphere_graph} and the data are fitted by a function with the expected form $c_1 K^{1/3} + c_2 K^{-1/3}$. This gives best fit  values $c_1=0.694\pm0.008$ and $c_2=-1.9\pm0.4$. This gives $a_{3}^{\text{emb}}=0.398\pm0.005$ which is consistent with the analytic value from the $d=3$ infinite slab $a_3^{\text{emb}} = \frac{\sqrt[3]{\frac{2}{3}} \Gamma \left(\frac{1}{3}\right)}{3 \pi^{2/3} \Gamma \left(\frac{5}{3}\right)} \simeq 0.403$. 
\par
The data are consistent with the conjecture, however it is worth keeping in mind, here and for the other simulations, that the behaviour explored is for fixed $\epsilon = K/\rho$ and there should be unknown $\epsilon$ dependent corrections. 

\subsection{Simulation of $d=3$ Cube, Isolated}
The isolated regime is the physically relevant one when considering the sprinkled causal sets as contributing to the causal set Sum Over Histories  as individual discrete spacetimes in their own right. We chose to sample the random smeared action in the isolated regime for a unit cube in $d=3$. The volume of the timelike boundary equals 4. 
\par
\begin{figure}[H]
    \centering
    \includegraphics[width=150mm,scale=1]{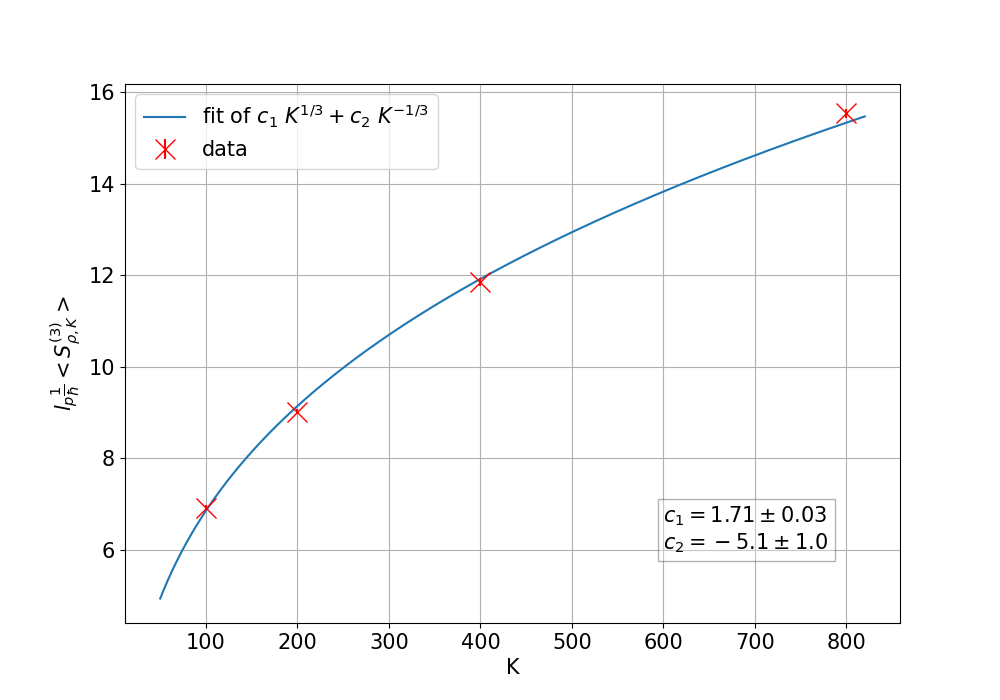}
    \caption{The sample mean of the action of the cube against $K$, with a curve fitted to the data (blue). The cube has side length 1.}
    \label{cube_graph}
\end{figure}

For each $\rho$, the sample mean of the action over 200 sprinklings is shown in Fig. \ref{cube_graph} and the data are fitted by a function with the same form as before $c_1 K^{1/3} + c_2 K^{-1/3}$. This gives best fit  values $c_1=1.71\pm0.03$ and $c_2=-5.1\pm1.0$ which gives $a_{3}^{\text{iso}}=0.428\pm0.008$. In this case we do not have an analytic calculation of $a_{3}^{\text{iso}}$ to compare to. \par
 We conclude that the $d=3$ cube in the isolated regime is consistent with the same conjecture as the embedded regime but with a different coefficient. Note that $a_{3}^{\text{iso}}>a_{3}^{\text{emb}}$, as was also the case for $d=2$ case. This make sense because $V_{xy}$ is smaller in the isolated regime, so the exponent in the integral is less negative, resulting in a larger action.

\section{Holes}\label{sectionHoles}
We turn to a different class of non-globally hyperbolic spacetimes: spacetimes with holes. 
\subsection{Isolated Annulus} \label{isolated_annulus}
\begin{figure} [H]
    \centering
    \includegraphics[width=70mm,scale=1]{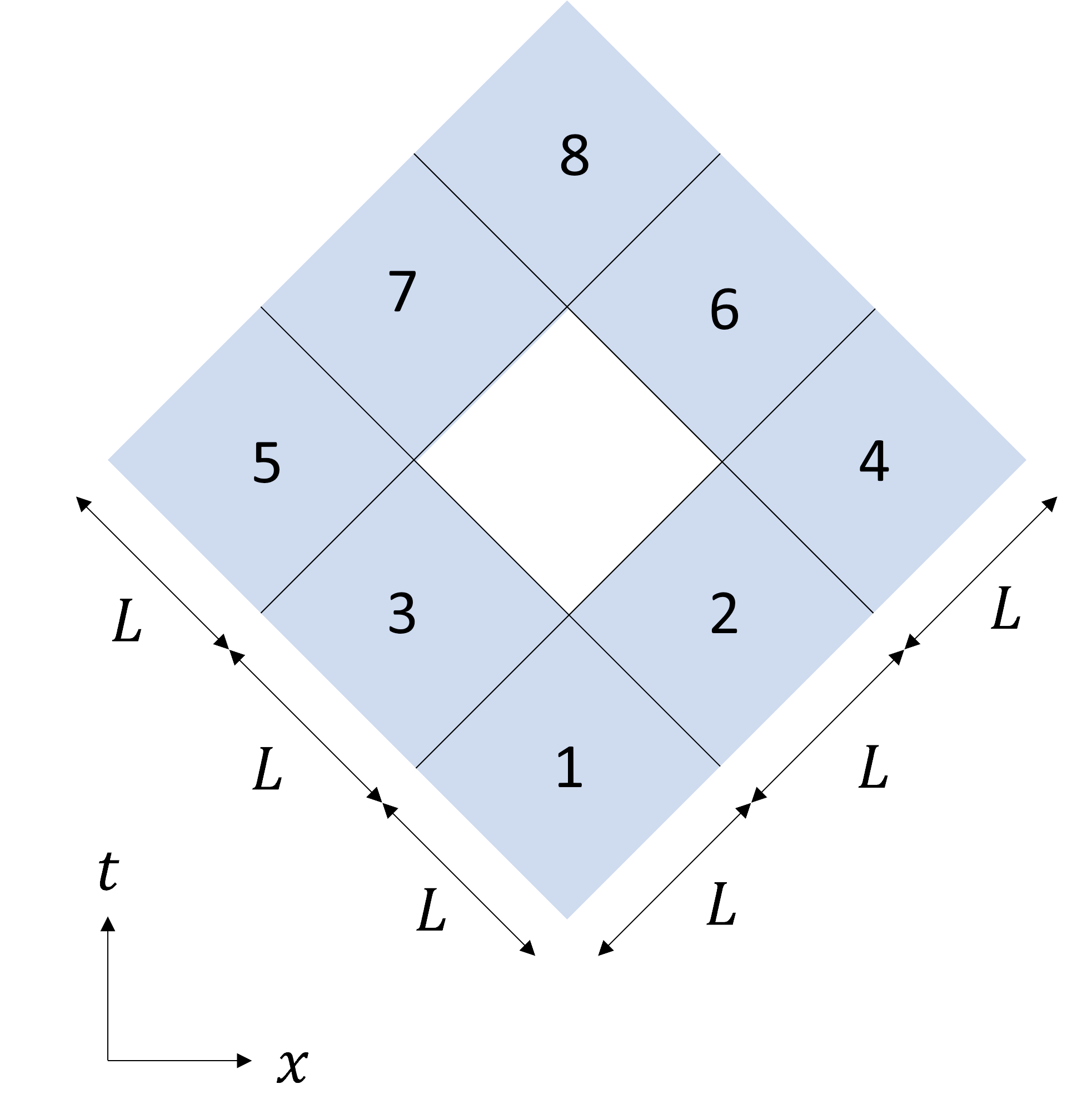}
    \caption{The annulus in Cartesian coordinates. The hole is centred and is a ninth of the total area of the diamond.}
    \label{nulldou}
\end{figure}

The null annulus--a causal diamond with a diamond removed from the centre--is the simplest example of a hole for $d=2$. An example is shown in Fig.\ref{nulldou}.  
In the embedded regime the continuum limit of the mean random action of a null annulus can be calculated using the ``topological calculus" of \cite{Benincasa:2010as} and equals 4 no matter the size of the diamond, the size of the hole or where the hole is \cite{chevalier:2023}.\par

In the isolated regime, the calculation can still be done analytically by partitioning the manifold into the regions shown in Fig. \ref{nulldou} and using the bilocal action formula:
\begin{equation} \label{annulus_action}
\begin{split}
\langle \boldsymbol{S}_{\text{annulus}}\rangle  &= \sum_{i=1}^{8} \langle \boldsymbol{S}_{i}\rangle  + \sum_{\substack{i,j=1\\ j< i}}^{8} \langle \boldsymbol{S}_{i,j}\rangle .\\
\end{split}
\end{equation}
$ \langle \boldsymbol{S}_{i}\rangle $ is the action of a causal diamond and is the same for all 8. From symmetry, many of the $\langle \boldsymbol{S}_{i,j}\rangle $ are identical. We can separate these into 5 different categories and within each category, the terms are equal:  
\begin{itemize}
    \item $(i,j)=$(2,1),(4,2),(6,4),(8,6),(3,1),(5,3),(7,5) and (8,7). 
    \item $(i,j)=$(4,1),(8,4),(5,1) and (8,5). 
    \item $(i,j)=$(6,1)(8,2),(7,1) and (8,3)
    \item  $(i,j)=$(6,2),(7,3)
    \item $(i,j)$=(8,1).
\end{itemize}
Therefore \eqref{annulus_action} can be simplified to:
\begin{equation}
\begin{split}
     \lim_{{\rho \to \infty}}\frac{1}{\hbar}\langle \boldsymbol{S}_{\text{annulus}}\rangle  &= \lim_{{\rho \to \infty}}\frac{1}{\hbar}\Big(8 \langle \boldsymbol{S}_{\text{diamond}}\rangle  +8\langle \boldsymbol{S}_{2,1}\rangle +4\langle \boldsymbol{S}_{4,1}\rangle \\
     &+4\langle \boldsymbol{S}_{6,1}\rangle +2\langle \boldsymbol{S}_{6,2}\rangle + \langle \boldsymbol{S}_{8,1}\rangle \Big). \\
\end{split}
\end{equation}
 From \cite{Benincasa:2010as} we have\footnote{Note the action we use here has an additional factor of 2 compared to the action in \cite{Benincasa:2010as}} 
\begin{equation}
\begin{split}
& \lim_{{\rho \to \infty}}\frac{1}{\hbar} \langle \boldsymbol{S}_{\text{diamond}}\rangle =2\,,\quad
    \lim_{{\rho \to \infty}}\frac{1}{\hbar} \langle \boldsymbol{S}_{2,1}\rangle =-2\,,\quad
    \lim_{{\rho \to \infty}}\frac{1}{\hbar} \langle \boldsymbol{S}_{4,1}\rangle =0\,,\\
    &\lim_{{\rho \to \infty}}\frac{1}{\hbar} \langle \boldsymbol{S}_{6,2}\rangle =\lim_{{\rho \to \infty}} 2\left(\gamma -1 +\ln{L^2\rho} + \mathcal{O}\left(\frac{1}{\rho}\right)\right)\sim2\ln{\rho}\,.\\
\end{split}
\end{equation}
In the last expression, the $\ln{\rho}$ term dominates over the constant terms and the limit is divergent. It remains to compute $\langle \boldsymbol{S}_{8,1}\rangle $ and $\langle \boldsymbol{S}_{6,1}\rangle $.
\subsubsection{Region 1 to 8}
\begin{figure} [H]
    \centering
    \includegraphics[width=80mm,scale=1]{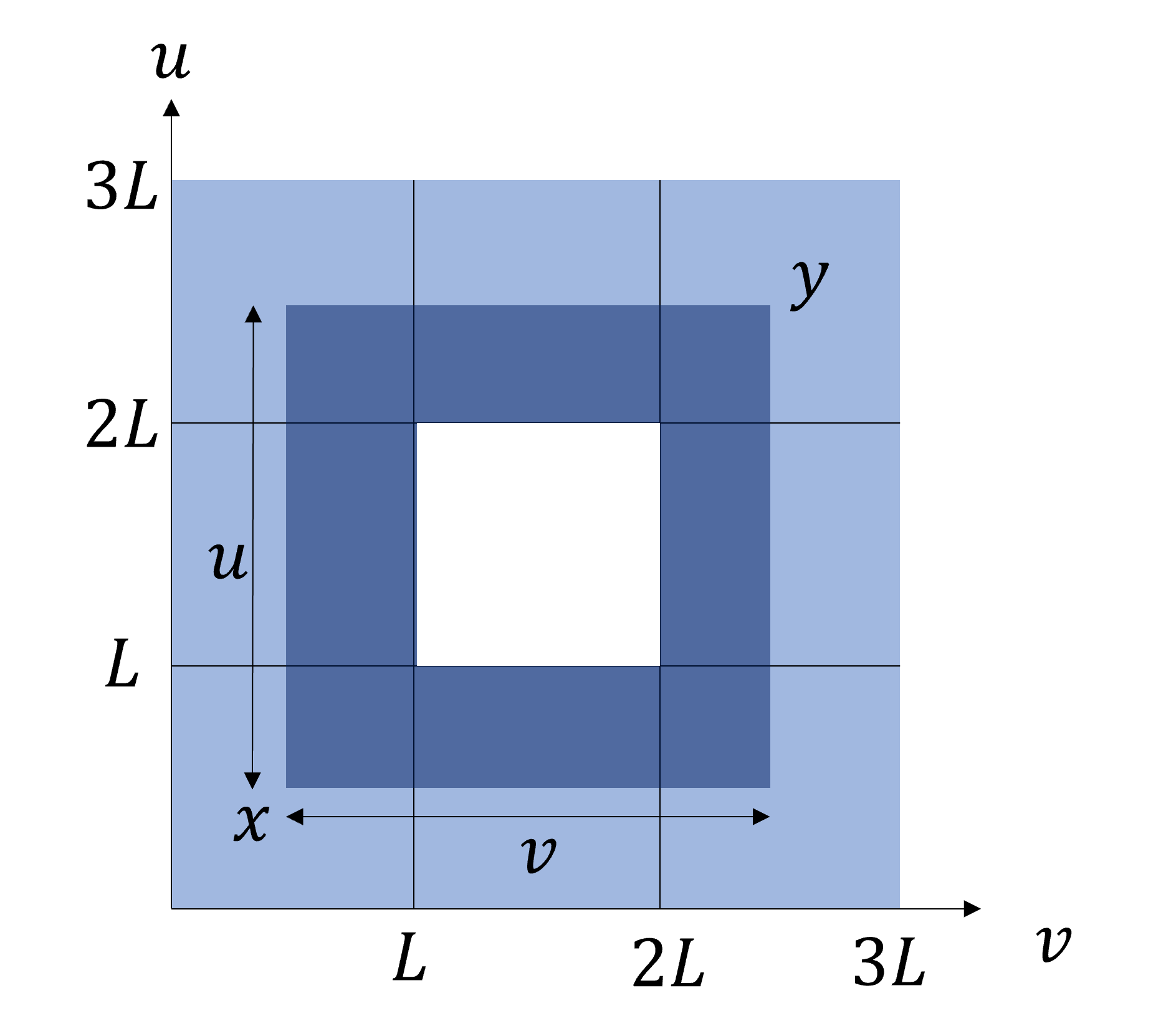}
    \caption{For an interval starting at any point $x$ in region 1 and ending  at any point $y$ in region 8, the volume of the interval (shown in dark blue) will always be $uv$ minus the area of the hole (white).}
    \label{nh15}
\end{figure}

\begin{figure} [H]
    \centering
    \includegraphics[width=150mm,scale=1]{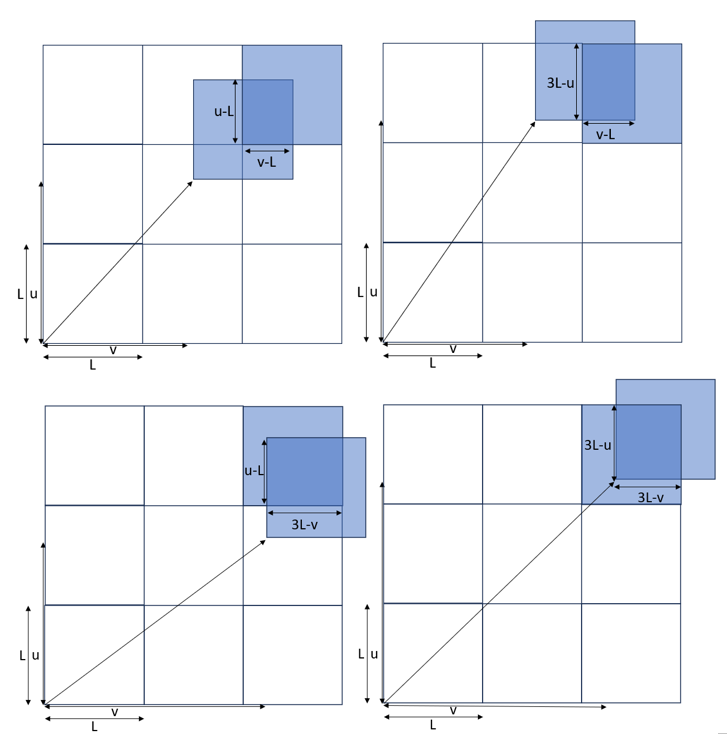}
    \caption{The 4 possible volumes of realisation for different ranges of $u$ and $v$.}
    \label{nh151}
\end{figure}

We use the weighted integral method. From Fig. \ref{nh15} we see the interval defined by the vector $(u,v)$ will always have an area of $uv-L^2$, where $u = u_y-u_x$ and $v = v_y-v_x$. 
There are 4 separate cases for the volume of realisation shown in Fig. \ref{nh151} and $X_{\rho}$ equals
\begin{equation}
\begin{split}
    X_{\rho} &= \int_{L}^{2L} du \left[\int_{L}^{2L} dv \text{ } (u-L)(v-L)e^{-\rho(uv-L^2)}
+  \int_{2L}^{3L}dv  \text{ } (u-L)(3L-v)e^{-\rho(uv-L^2)}\right]\\
& + \int_{2L}^{3L} du\left[ \int_{L}^{2L} dv  \text{ } (3L-u)(v-L)e^{-\rho(uv-L^2)} 
+ \int_{2L}^{3L} dv  \text{ } (3L-u)(3L-v)e^{-\rho(uv-L^2)}\right]. \\
\end{split}
\end{equation}

All the integrals are exponentially small as $\rho\rightarrow \infty$ and 
\begin{equation}
    \lim_{{\rho \to \infty}}\frac{1}{\hbar}\langle \boldsymbol{S}_{8,1}\rangle  =\lim_{{\rho \to \infty}}\ (-4\rho^2 \hat{\mathcal O}_2 X_\rho) = 0\,.
\end{equation}

\subsubsection{Region 1 to 6}
\begin{figure} [H]
    \centering
    \includegraphics[width=80mm,scale=1]{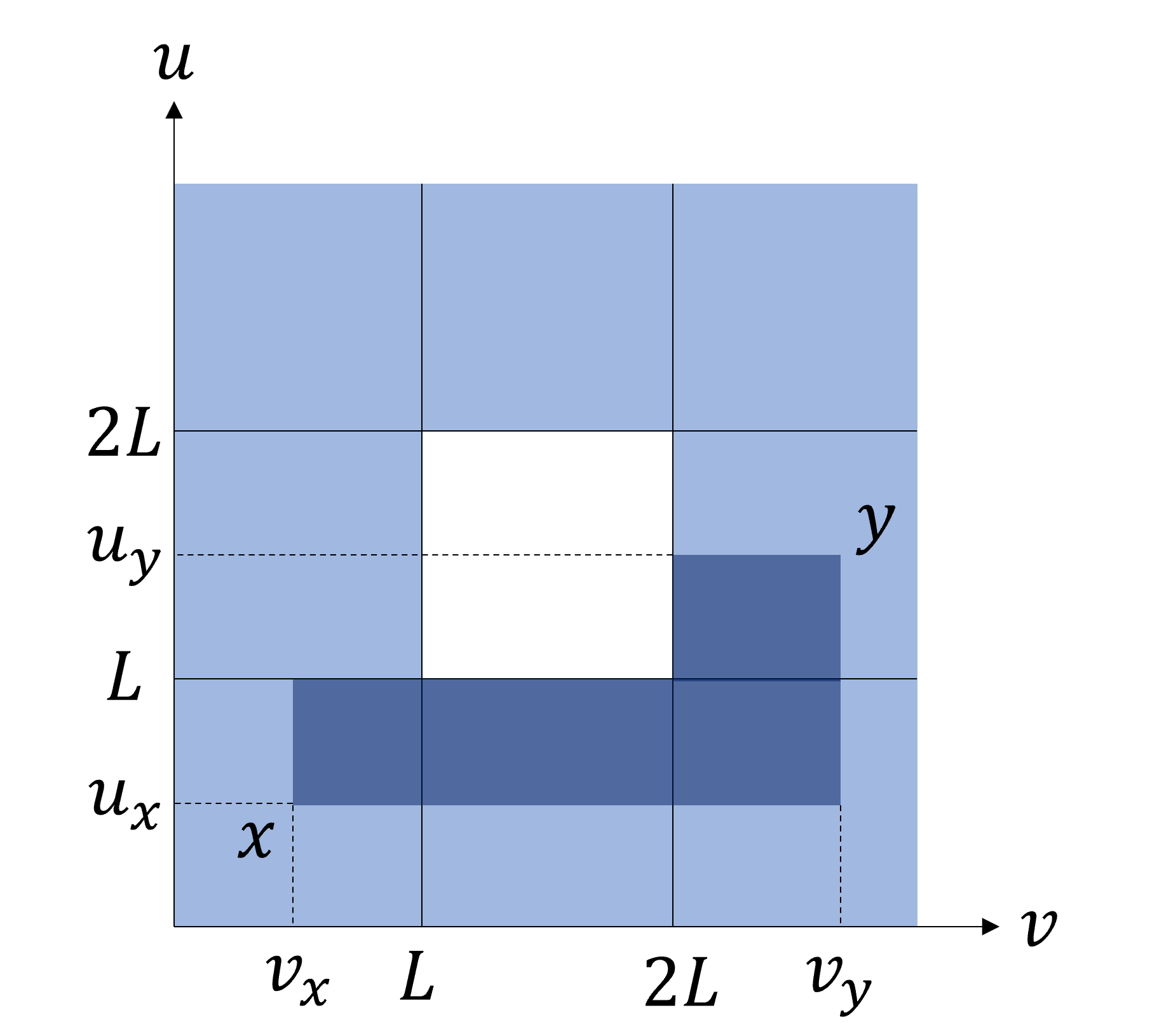}
    \caption{For an interval starting at any point $x$ in region 1 and ending  at any point $y$ in region 6, the volume of the interval is shown in dark blue.}
    \label{nh16}
\end{figure}
We have 
\begin{align}
     X_{\rho}=\int_0^{L} dv_x \int_0^{L} du_x \int_{L}^{2L} du_y\int_{2L}^{3L} dv_y \text{ }e^{-\rho ((u_x-u_y)(v_x-v_y)-(2L-v_x)(u_y-L))}\,,
\end{align}
which gives
\begin{equation}
  \lim_{{\rho \to \infty}}\frac{1}{\hbar}\langle \boldsymbol{S}_{6,1}\rangle  =\lim_{{\rho \to \infty}}\ (-4\rho^2 \hat{\mathcal O}_2 X_\rho) = \ln{4}.
\end{equation}
\subsubsection{Full Result}
Combining these results, we obtain
\begin{equation} \label{null_hole_eq}
\begin{split}
     \lim_{{\rho \to \infty}}\frac{1}{\hbar}\langle \boldsymbol{S}_{\text{annulus}}\rangle 
     = \lim_{{\rho \to \infty}}4(\ln{L^2\rho} + \ln{4} + \gamma -1)
\end{split}
\end{equation}
which diverges logarithmically.
\subsubsection{Simulations}
\begin{figure} [H]
    \centering
    \includegraphics[width=150mm,scale=1]{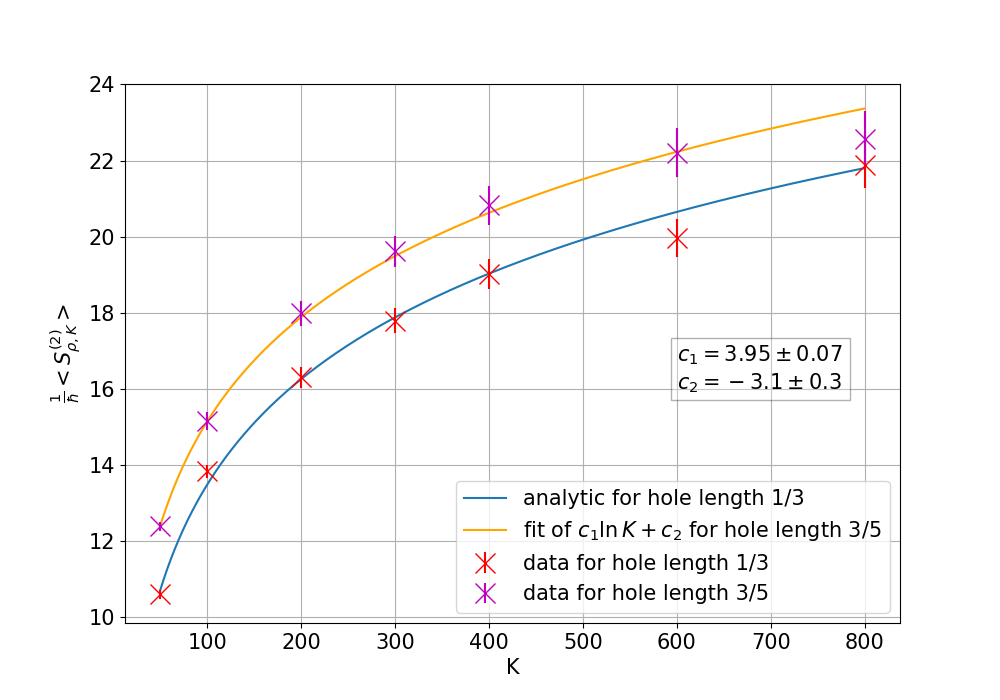}
    \caption{Sample mean action vs $K$ for the null annulus with different sized holes.}
    \label{null_hole_graph}
\end{figure}

The results of our simulations for this manifold and one variant are shown in Fig. \ref{null_hole_graph}. The causal diamond has side length 1 and the size of the hole was varied. The case of the previous section corresponds to $L=\frac{1}{3}$. 
The equation of the blue line in Fig. \ref{null_hole_graph} is given by:
\begin{equation}
     4\left(\ln{\frac{K}{9}} + \ln{4} + \gamma -1\right)
\end{equation}
and the data is consistent with it. \\
We fit $c_1 \ln K + c_2$ to the data for the hole of length $\frac{3}{5}$ to find $c_1=3.95 \pm 0.07$ (and $c_2= -3.1\pm0.3$). This suggests that the leading order term for any sized hole will be $4 \ln{\rho}$ which is consistent with the log term originating from $\langle \boldsymbol{S}_{6,2}\rangle $. This is therefore likely to be the dominant term for any sized hole. We conclude that for a null annulus, the mean action diverges logarithmically with the density of sprinkling, regardless of the size of the hole.  
\subsection{Spacelike Holes} \label{spacelike_holes}
\begin{figure}[H]
    \centering
    \includegraphics[width=140mm,scale=0.5]{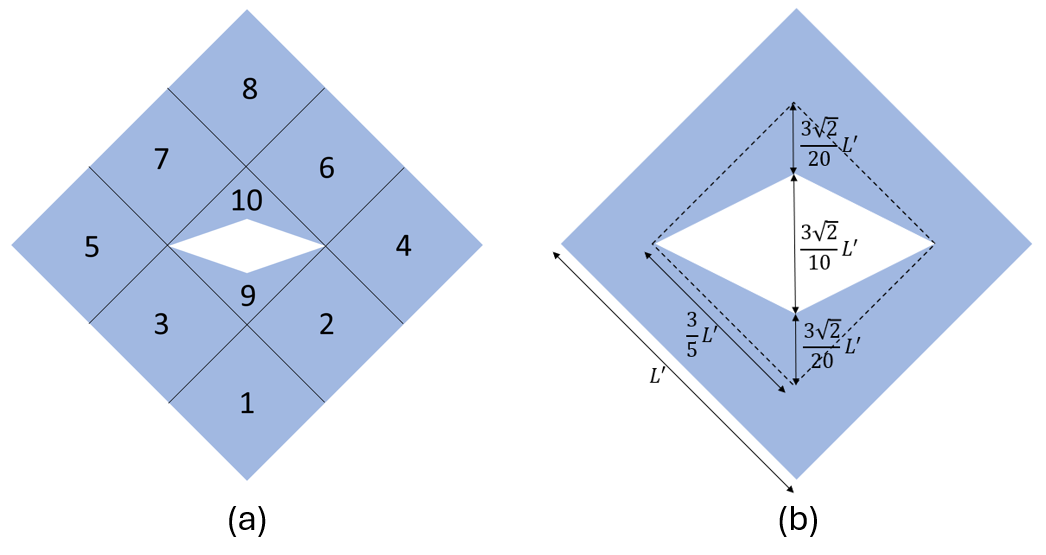}
    \caption{(a) A $d=2$ causal diamond with a spacelike hole. Note the new regions 9 and 10. (b) A $d=2$ causal diamond with side $L'$ with a spacelike hole. The dimensions of the hole are labelled.}
    \label{7.10}
\end{figure}

We investigate another type of hole, one which has only spacelike boundaries such as Fig. \ref{7.10} (a). We argue that the  contribution  of such a hole to the action will be logarithmically divergent, as for a null hole. 
\par
The argument is as follows. A causal curve involving points in region 9 can only begin in regions 1, 2, 3, or 9, and must end in region 9. Let us denote the union of these four regions $1\cup 2 \cup 3 \cup 9 $. The contribution of $1\cup 2 \cup 3 \cup 9 $ to the limiting mean action equals the contribution of $1\cup 2 \cup 3 $  to the limiting mean action of the null annulus (without 9) because 
 (i) $1\cup 2 \cup 3 \cup 9 $ is globally hyperbolic in itself, so its mean action equals the mean action of $1\cup 2 \cup 3$ by conjecture \eqref{conjecture1}, and (ii) the bilocal contribution to the action from $1\cup 2 \cup 3 \cup 9 $ is unchanged by deleting $9$, since there are no causal curves between 9 and the complement of $1\cup 2 \cup 3 \cup 9 $. There's a similar argument for region 10. Therefore, the limiting mean action for the diamond with the spacelike hole equals that for the null annulus.
We performed a simulation for the diamond with $L'=1$ and a spacelike hole as shown in Fig. \ref{7.10} (b). The joints of the hole have a small Lorentzian angle (to be defined in the next section), so this manifold and the null annulus are distinguishable at finite
$\rho$. 
\begin{figure}[H]
    \centering
    \includegraphics[width=150mm,scale=1]{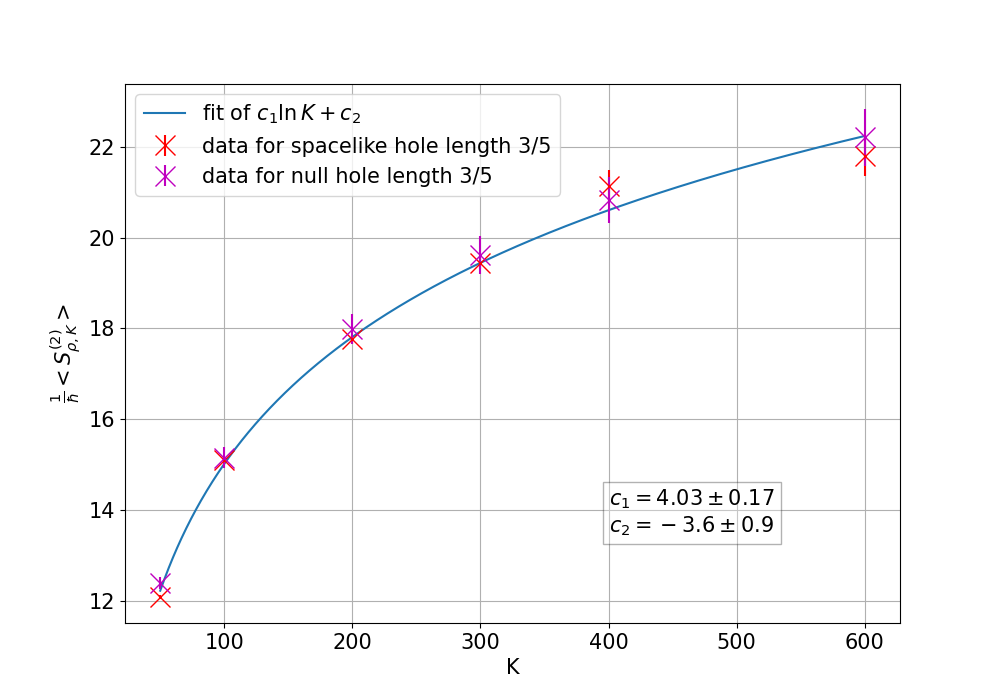}
    \caption{Sample mean and standard error of the action for the $L'=1$ diamond with a spacelike and also with a null hole. The blue curve is fitted to the spacelike data (red).}
    \label{7.11}
\end{figure}
The data is shown in Fig. \ref{7.11}, and is fitted to the function $c_1\ln(L'^2 K)+c_2$. The best fit is for $c_1=4.03\pm0.17$, which is consistent with the argument given above, and $c_2=-3.6\pm0.9$. 
The data from the null hole and spacelike hole have overlapping error bars, consistent with the argument above that adding/deleting regions 9 and 10 does not change the mean action. One could flatten the hole to just a deleted line segment and the mean action would be unchanged. \par 
In the statement of Conjecture 1 for globally hyperbolic spacetimes, the joint was defined as the co-dimension 2 intersection of 
the future boundary and the past boundary of the spacetime. 
In the non-globally hyperbolic spacetimes with spacelike holes that we have just considered, we can extend the definition of joint. For, a spacetime with a spacelike hole has a past boundary and a future boundary at the hole, where causal curves enter and leave the spacetime respectively, and these boundaries also intersect.  We can distinguish between the two cases as follows. We define a joint to be a \textit{convex joint} when there exists a future directed causal curve from the past boundary to the future boundary---the globally hyberbolic case---and otherwise the joint is a \textit{concave joint}.\par
We propose a new conjecture for  a concave joint, that its contribution to the mean action obeys: 
\begin{equation} \label{hole_conjecture}
    \lim_{\rho\rightarrow\infty}\frac{1}{\hbar}\langle \mathbf{S}^{(d)}[\mathcal{M}]\rangle\sim \frac{1}{l_p^{d-2}}b_d\text{Vol}_{d-2}(J')\ln(\rho V)
\end{equation}
where $b_d$ is a dimensional coefficient, and $J'$ is the concave joint.  The contribution \eqref{hole_conjecture} is dimensionally correct and is proportional to the volume of the concave joint as one might expect. The results of sections \ref{isolated_annulus} and \ref{spacelike_holes} are evidence for the conjecture for $d=2$ with $b_{d=2}=2$.  More evidence is needed, especially for more than 2 dimensions.
\subsection{Timelike Holes}
\begin{figure} [H]
    \centering
    \includegraphics[width=100mm,scale=1]{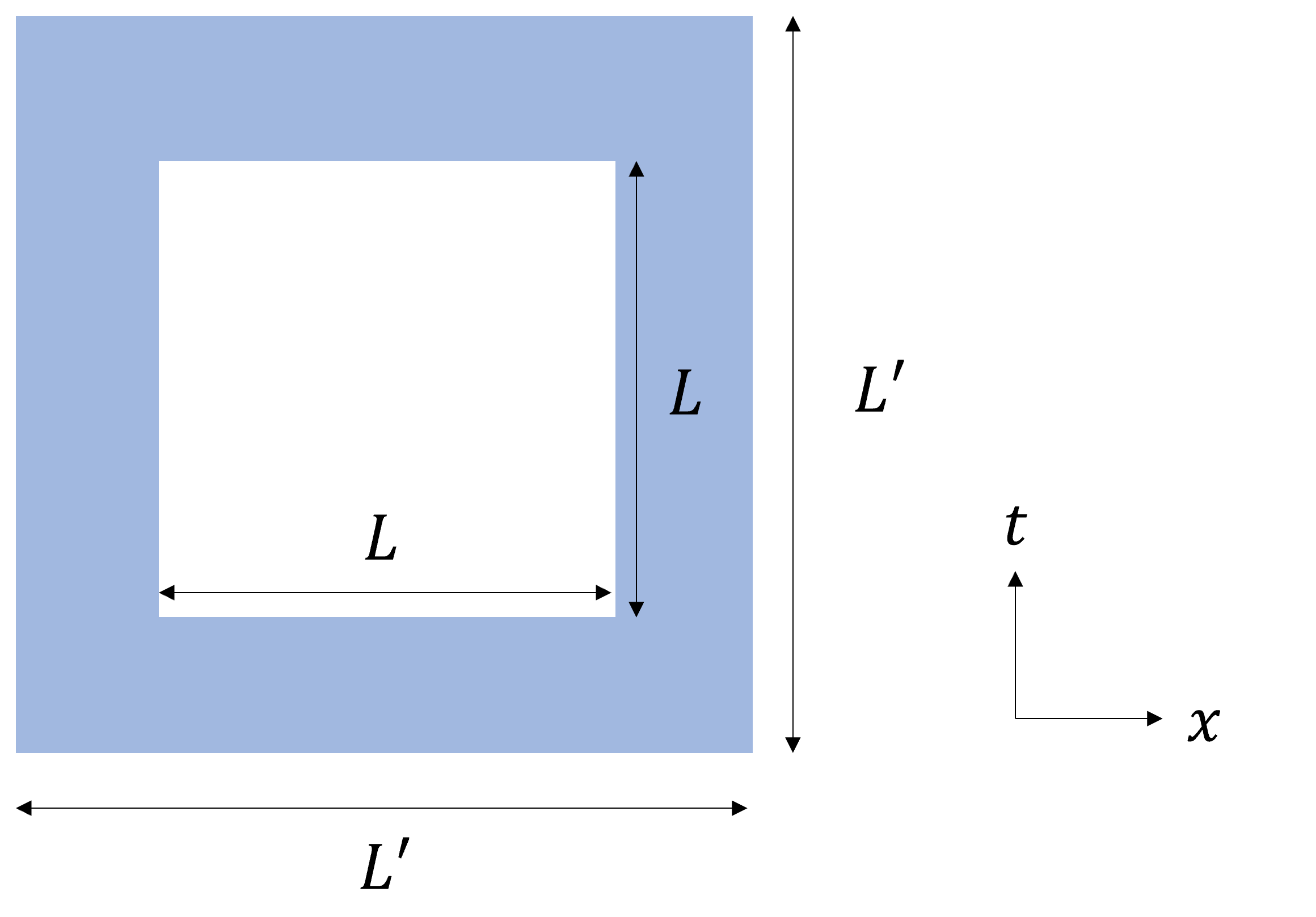}
    \caption{The square hole}
    \label{sq_hole_new}
\end{figure}
A square with a square hole in its centre is an example of a manifold with a timelike hole (shown in Fig. \ref{sq_hole_new}). In \cite{chevalier:2023},  the mean action in the embedded regime was found to be consistent with the timelike boundary conjecture, i.e. the leading order term was of the form $2(L'+L)a_2^\text{emb}\rho^{\frac{1}{2}}$. $L'$ is the side length of the square and $L$ the side length of the hole, so $2(L'+L)$ is the length of the full timelike boundary. The same leading behaviour--with a different coefficient--is expected in the isolated regime, and we investigated this using simulations.\par

\begin{figure} [H]
    \centering
    \includegraphics[width=150mm,scale=1]{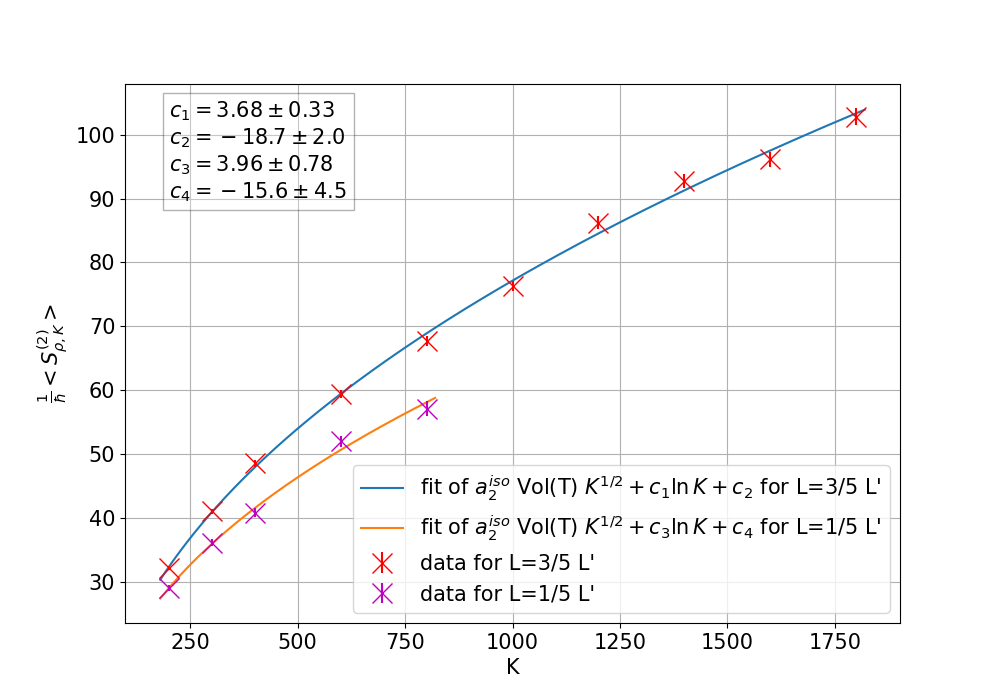}
    \caption{Sample mean of the action against $K$ for square holes with lengths $L=\frac{3}{5}L'$ and $L=\frac{1}{5}L'$. $\text{Vol(T)}=2(L+L')$ as before.}
    \label{sq_hole_graph_combined}
\end{figure}

For square holes with lengths $L=\frac{3}{5}L'$ and $L=\frac{1}{5}L'$ (with $L'=1$) the results are shown in Fig. \ref{sq_hole_graph_combined}.
We do not fit for $a_2^{iso}$ but assume its value is $a_2^{iso} = 0.6959$ from \cite{chevalier:2023}.
The subleading $\ln{\rho}$ term in the fitting function is taken from the results for the null hole and we conjecture that it originates from the corners of the hole. Including this term is not enough, however, to give a good fit to the data and we also included a constant term. 
\par
The fitted coefficients of the log terms were $c_1=3.68\pm0.33$ for $L=\frac{3}{5}L'$ and $c_3=3.96\pm0.78$ for $L=\frac{1}{5}L'$. These two values are consistent, implying that the $\log$ contribution is the same regardless of the hole size, as was the case for the null hole. The fitted values of the constant term were $c_2=-18.7\pm2.0$ for $L=\frac{3}{5}L'$ and $c_4=-15.6\pm4.5$ for $L=\frac{1}{5}L'$.  
\par

The main conjecture is the behaviour of the leading order term, $a_2^\text{iso}\text{Vol(T)}K^{\frac{1}{2}}$, consistent with the timelike boundary conjecture. We can say that the data are consistent with this but for stronger evidence we will need larger simulations and more knowledge of the subleading corrections.

\section{Spacelike Joints}\label{spacelikejoints}
\subsection{The Lozenge} \label{lozenge subsection}
\begin{figure} [H]
    \centering
    \includegraphics[width=100mm,scale=1]{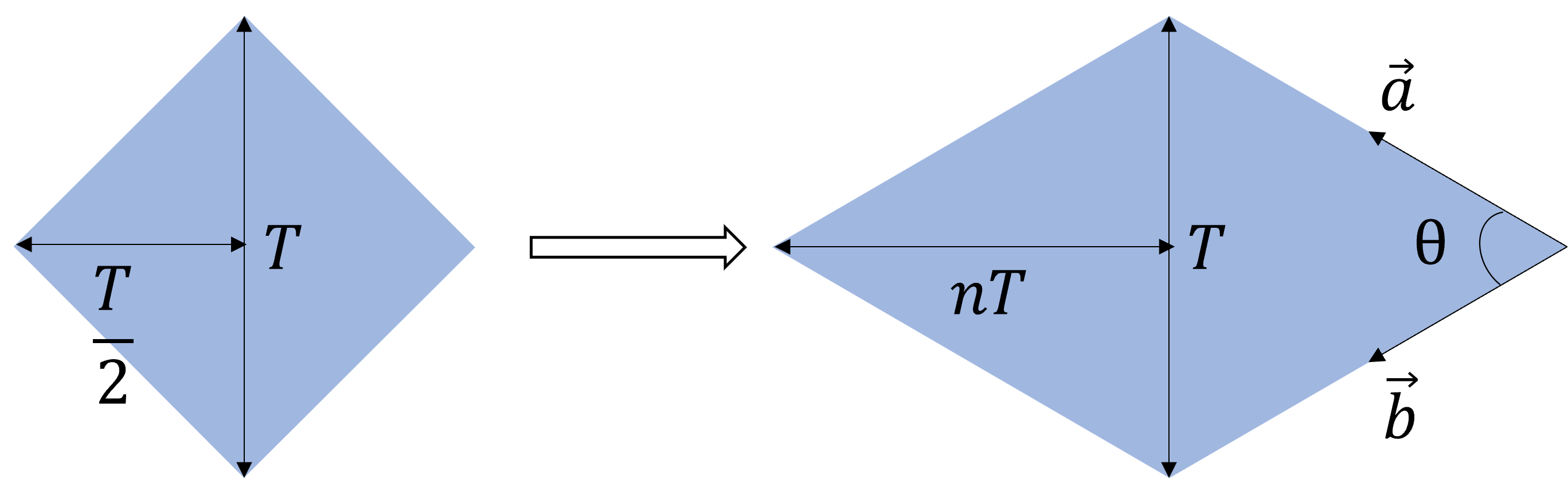}
    \caption{The $d=2$ causal diamond (left) stretched to give a lozenge (right), where $n>\frac{1}{2}$. The vectors $\vec{a}$ and $\vec{b}$ are used to define the Lorentzian angle $\theta$.}
    \label{lozenge}
\end{figure}

\begin{figure} [H]
    \centering
    \includegraphics[width=100mm,scale=1]{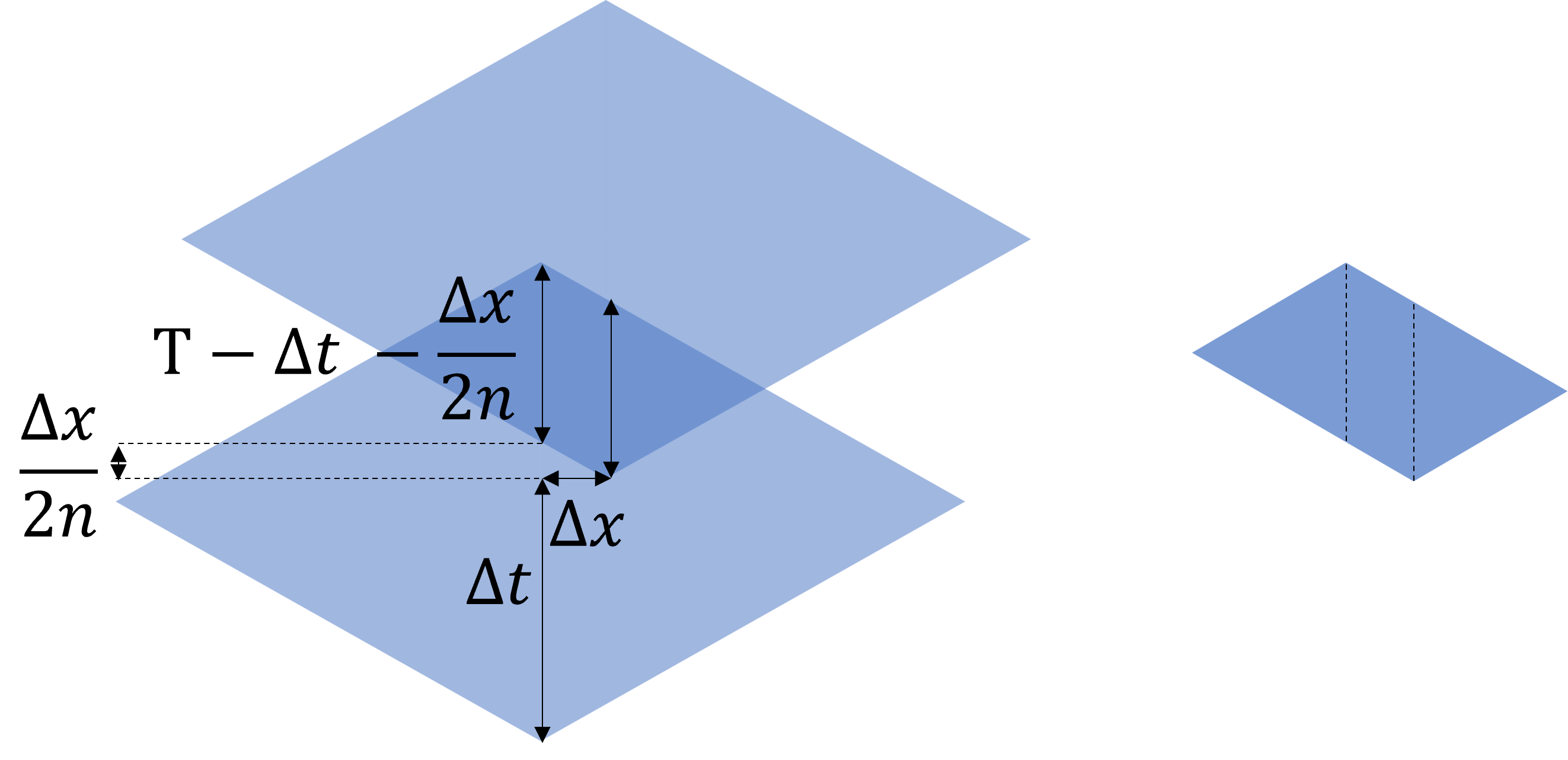}
    \caption{Left: showing how the volume of realisation (the shaded region) can be found from cloning. The length $\frac{\Delta x}{2n}$ is found using similar triangles. Right: showing how the volume of realisation can be broken into 2 identical triangles either side of a parallelogram.}
    \label{lozenge_vor}
\end{figure}

\begin{figure} [H]
    \centering
    \includegraphics[width=130mm,scale=1]{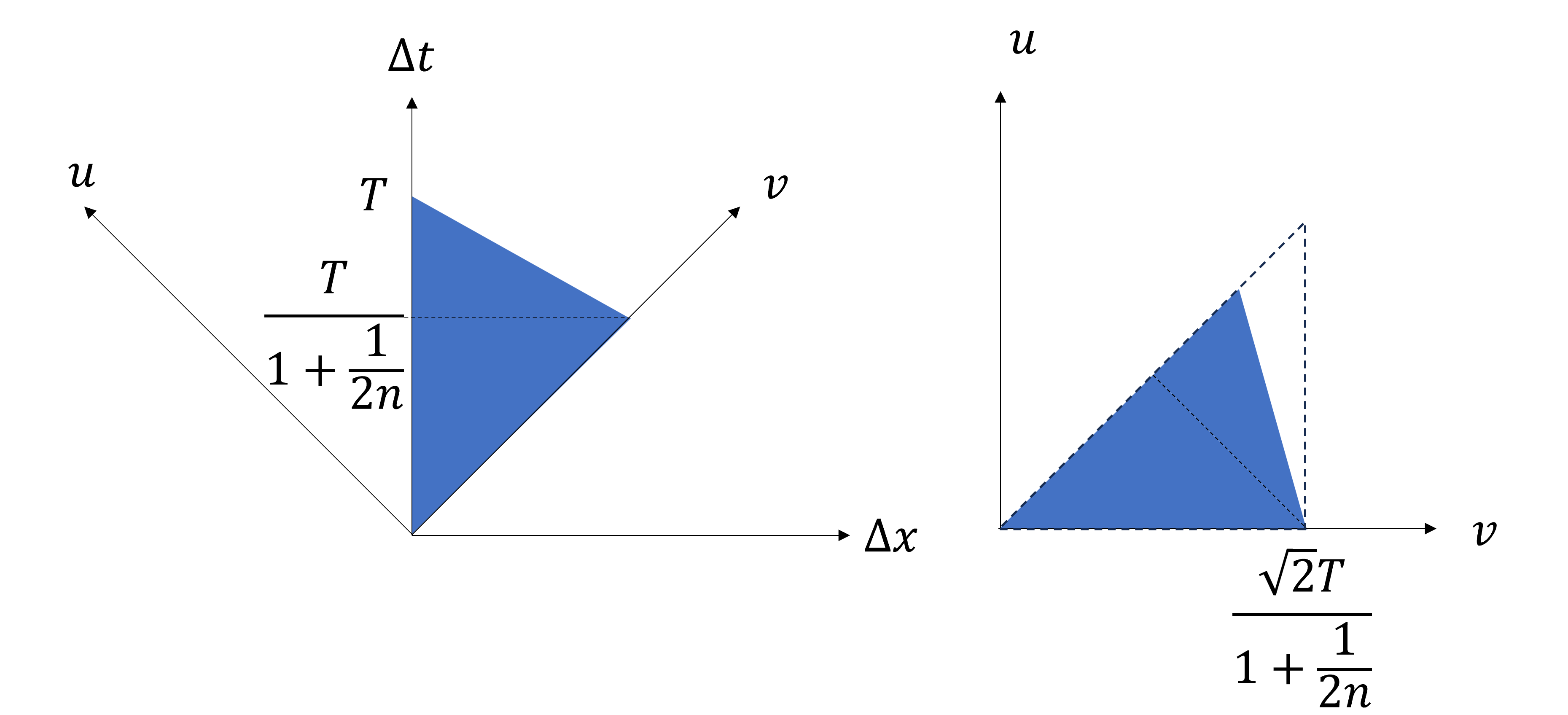}
    \caption{Converting the integration region from Cartesian (left) into null coordinates (right) and approximating the integral using the larger, dashed triangle.}
    \label{lozenge_rotate}
\end{figure}

If we stretch the causal diamond so that its null boundaries become spacelike (shown in Fig. \ref{lozenge}), we obtain a globally hyperbolic lozenge  and according to Conjecture 1, it should have the same mean action in the infinite limit as the diamond i.e. 2 for $d=2$.  
We used the weighted integral method to compute the mean action. The cloning is shown in Fig. \ref{lozenge_vor}, from which the volume of realisation is 
\begin{equation}
    a(\Vec{c})=2\times \frac{n}{2}\left(T-\Delta t - \frac{\Delta x}{2n}\right)^2 + \left(T-\Delta t - \frac{\Delta x}{2n}\right)\Delta x
\end{equation}
where the first term is the area of the 2 triangles and the second term is the area of the parallelogram between them.
 For the defining vector $ (\Delta t , \Delta x)$ to define an interval, it must satisfy $|\Delta x| <\Delta t$. 
\begin{figure} [H]
    \centering
    \includegraphics[width=100mm,scale=1]{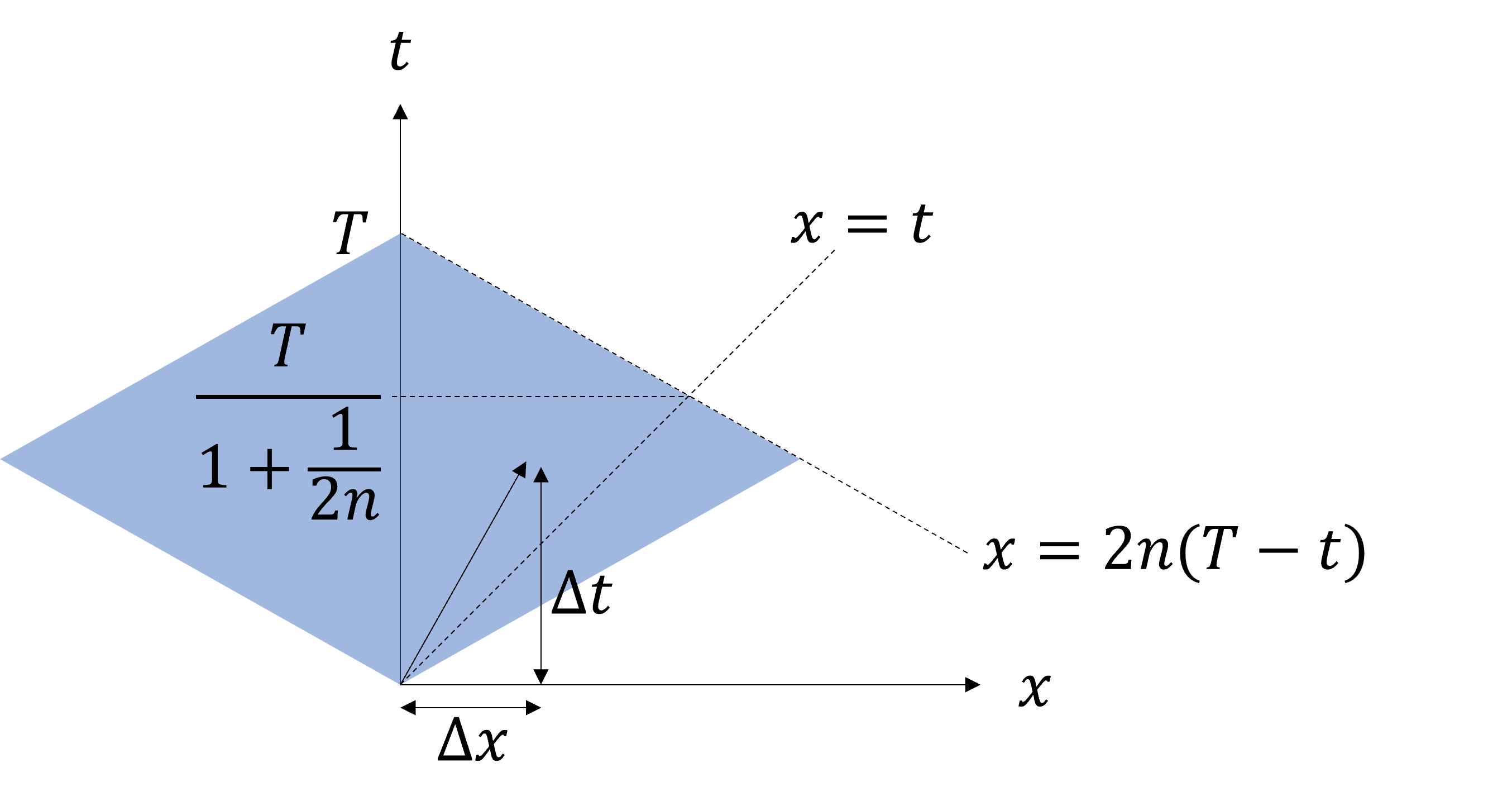}
    \caption{Showing the constraints on the defining vector $ (\Delta t , \Delta x) $ for $\Delta x>0$.}
    \label{lozenge_limits}
\end{figure}
Using Fig. \ref{lozenge_limits} and the volume of realisation, the integral for $X_\rho$ is
\begin{equation} \label{cartesian_lozenge}
\begin{split}
    X_\rho &= 2 \int _{0}^{\frac{T}{1+\frac{1}{2n}}} d\Delta t \int_{0}^{\Delta t} d\Delta x \text{ }\Bigg[n\left(T-\Delta t - \frac{\Delta x}{2n}\right)^2\\
    &+ \Delta x \left(T-\Delta t - \frac{\Delta x}{2n}\right) \Bigg] e^{-\frac{\rho}{2}(\Delta t^2-\Delta x^2)}\\
    &+2 \int _{\frac{T}{1+\frac{1}{2n}}}^{T}d\Delta t \int_{0}^{2n(T-\Delta t)} d\Delta x \text{ }\Bigg[n\left(T-\Delta t - \frac{\Delta x}{2n}\right)^2\\
    &+ \Delta x\left(T-\Delta t - \frac{\Delta x}{2n}\right) \Bigg] e^{-\frac{\rho}{2}(\Delta t^2-\Delta x^2)}\,.\\
\end{split}
\end{equation}
 The factor of 2 accounts for the contribution from $\Delta x<0$. To compute this integral, we change to $u$ and $v$ coordinates and make use of a similar approximation as in the infinite slab calculation. In this approximation, we again extend the region of integration to a larger triangle, shown in Fig. \ref{lozenge_rotate}.
The expression for $X_{\rho}$ becomes:
\begin{equation}
\begin{split}
    X_\rho &= 2 \int_0^{\frac{\sqrt{2}T}{1+\frac{1}{2n}}} dv\int_0^v du \text{ } \Bigg( \frac{1}{2} n \left(2 T^2-2 \sqrt{2} T (u+v)+(u+v)^2\right)\\
    &-\frac{(u-v)^2}{8 n}\Bigg) e^{-\rho u v}\,.\\
\end{split}
\end{equation}
Using equation (\ref{eq:trianglemean}) for the mean action in terms of $X_\rho$ and the operator $\hat{\mathcal O}_2$ we find the limit is
\begin{equation}
    \lim_{{\rho \to \infty}}\frac{1}{\hbar}\langle\boldsymbol{S}^{(2)}_{\rho}(M)\rangle=2n+\frac{1}{2 n}.
\end{equation}
For $n=\frac{1}{2}$ we recover the correct value of 2 for the limiting mean action of the causal interval. For $n>\frac{1}{2}$, however, this result violates the original Conjecture 1 for globally hyperbolic manifolds.

\subsection{Lorentzian Angle} \label{lorentzian_angle_1}

We can generalise this result for different manifolds. If we assume that each joint of the lozenge contributes half of the action, we can state that:
\begin{equation}
    \lim_{{\rho \to \infty}}\frac{1}{\hbar}\langle\boldsymbol{S}^{(2)}_{\rho}(\text{joint})\rangle=n+\frac{1}{4 n}.
\end{equation}
We can write this in terms of a Lorentzian angle $\theta$, defined as \cite{Sorkin2019LorentzianAA} 
\begin{equation}\label{eq:lorangle}
    \theta=\cosh^{-1}\left(\frac{\vec{a}\cdot\vec{b}}{|\vec{a}||\vec{b}|}\right)
\end{equation}
where $\vec{a}$ ($\vec{b}$) is the tangent vector to the future (past) boundary at the joint as shown as in Fig. \ref{lozenge}, $|\vec{a}|=|\sqrt{\vec{a}\cdot\vec{a}}|$ and $\vec{a}\cdot\vec{b}=-a_t b_t + a_x b_x$. 
\par
$\theta$ is Lorentz invariant and we conjecture that any joint with the same Lorentzian angle contributes the same amount to the mean action. From Fig. \ref{lozenge}, we can see that $\vec{a}=\begin{pmatrix} \ \frac{T}{2} \\ -nT \end{pmatrix}\ $ and $\vec{b}=\begin{pmatrix} \ -\frac{T}{2} \\ -nT \end{pmatrix}\ $, giving
\begin{equation}
    \theta=\cosh^{-1}\left(\frac{n^2+\frac{1}{4}}{n^2-\frac{1}{4}}\right).
\end{equation}
This can be rearranged to give 
\begin{equation}
    n=\frac{1}{2}\sqrt{\frac{\cosh{\theta}+1}{\cosh{\theta}-1}}
\end{equation}
and so we find
\begin{equation}
\begin{split}
     \lim_{{\rho \to \infty}}\frac{1}{\hbar}\langle\boldsymbol{S}^{(2)}_{\rho}(\text{joint})\rangle&=\frac{1}{2}\left(\sqrt{\frac{\cosh{\theta}+1}{\cosh{\theta}-1}}+\sqrt{\frac{\cosh{\theta}-1}{\cosh{\theta}+1}}\right)\\
     &= \coth{\theta}.\\
\end{split}
\end{equation}
\subsection{Spacelike Triangle} \label{spacelike_triangle_sec}
\begin{figure} [H]
    \centering
    \includegraphics[width=60mm,scale=1]{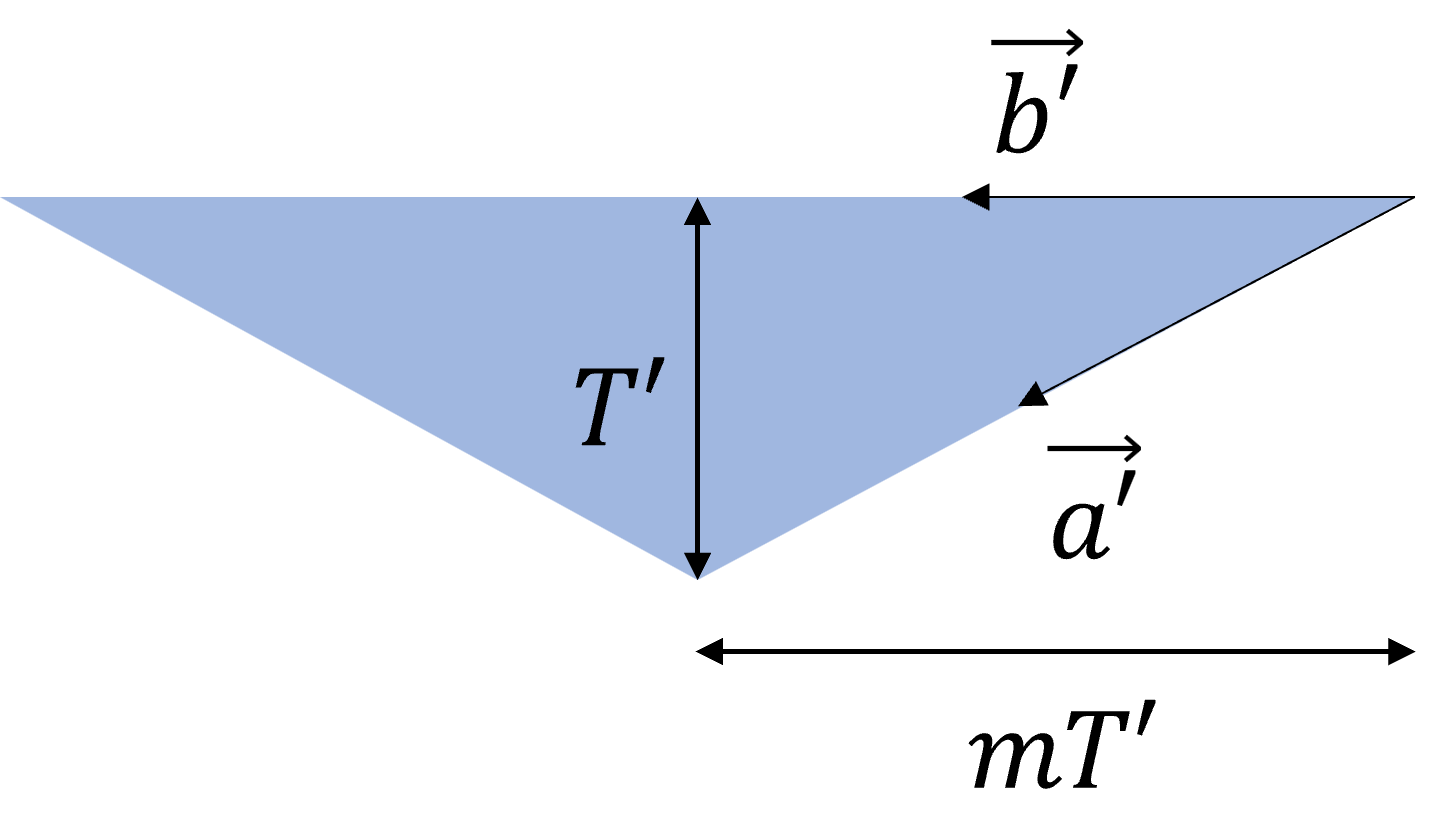}
    \caption{The spacelike triangle.}
    \label{spacelike_triangle}
\end{figure}

\begin{figure} [H]
    \centering
    \includegraphics[width=100mm,scale=1]{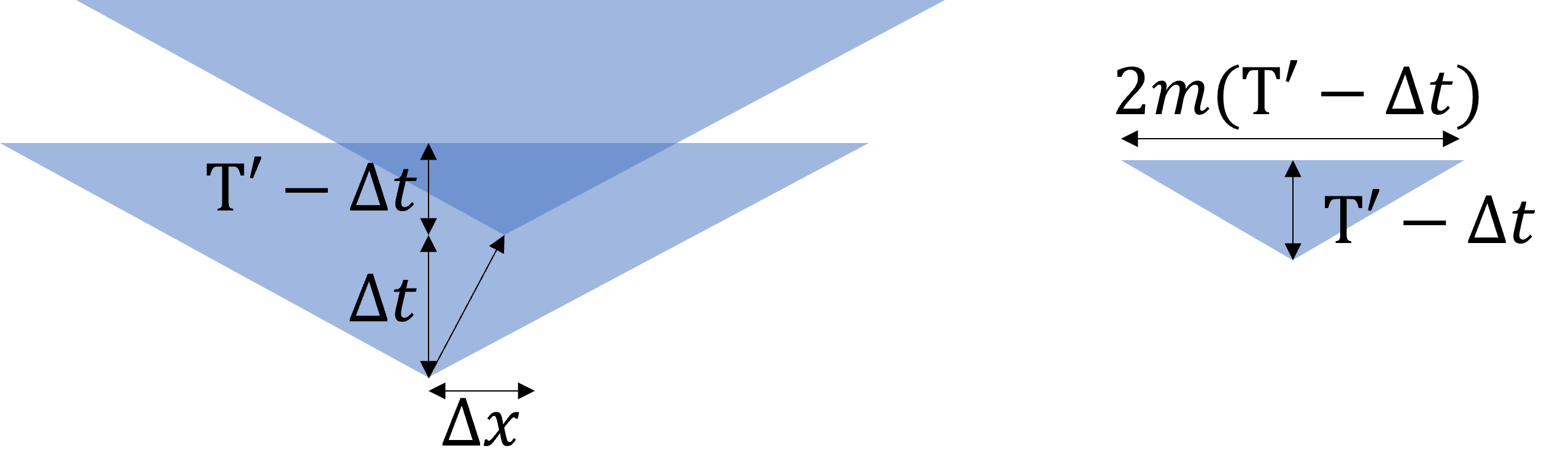}
    \caption{Left: cloning the triangle to find the volume of realisation. Right: the volume of realisation, where the base length is found using similar triangles.}
    \label{spacelike_triangle_vol}
\end{figure}

\begin{figure} [H]
    \centering
    \includegraphics[width=80mm,scale=1]{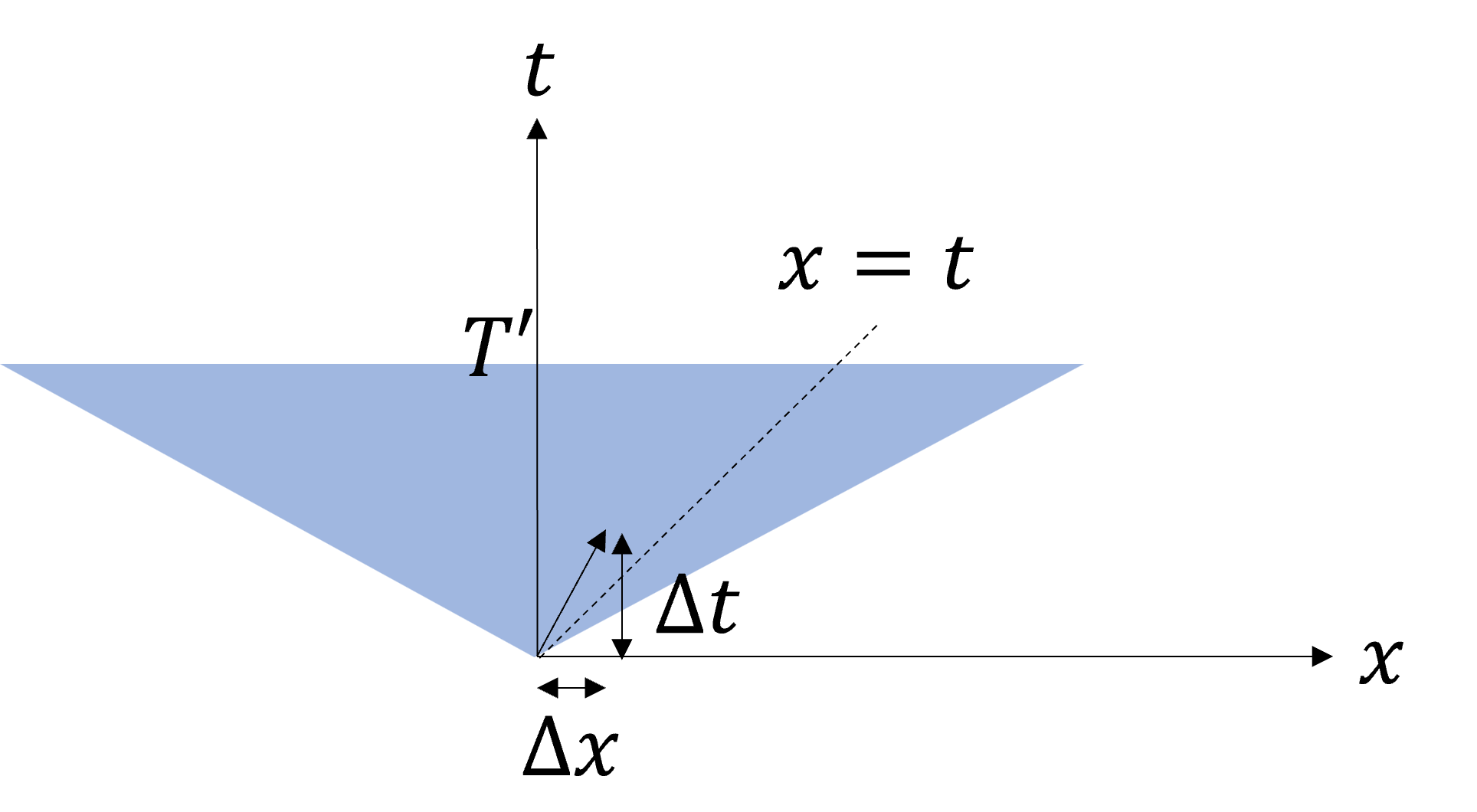}
    \caption{Showing the constraints on the defining vector, only considering $\Delta x>0$.}
    \label{spacelike_triangle_limits}
\end{figure}

To test the conjecture that the mean action depends only on 
Lorentzian angle as above, we now compute the mean action of a triangle with 3 spacelike sides, as shown in Fig. \ref{spacelike_triangle}.
Again, we find the volume of realisation by cloning, shown in Fig. \ref{spacelike_triangle_vol}. This gives the volume as:
\begin{equation}
    a(\Vec{c})=m(T'-\Delta t)^2.
\end{equation}
The limits can be found by considering the defining vector that will give a non-zero volume of realisation, shown in Fig. \ref{spacelike_triangle_limits} 
Therefore 
\begin{equation} 
\begin{split}
    X_\rho &= 2 \int _{0}^{T'} d\Delta t \int_{0}^{\Delta t} d\Delta x \text{ } m(T'-\Delta t)^2 e^{-\frac{\rho}{2}(\Delta t^2-\Delta x^2)}.\\
\end{split}
\end{equation}
\begin{figure} [H]
    \centering
    \includegraphics[width=130mm,scale=1]{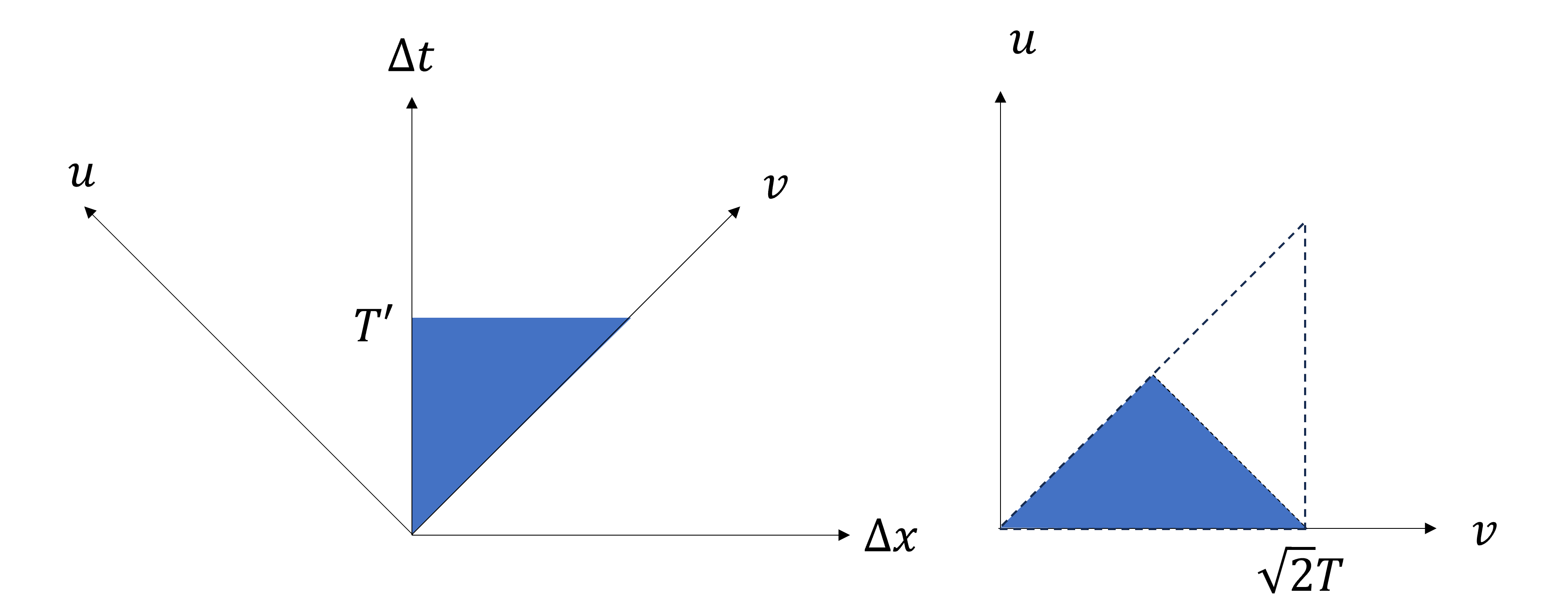}
    \caption{Converting the integration region from Cartesian (left) into null coordinates (right) and approximating the integral using the larger, dashed triangle.}
    \label{spacelike_triangle_rotate}
\end{figure}
We again use null coordinates and the same approximation as before as shown in Fig \ref{spacelike_triangle_rotate}, giving:
\begin{equation}
    X_\rho = 2 \int_0^{\sqrt{2}T'} dv\int_0^v du \text{ } m \left(T'-\frac{u+v}{\sqrt{2}}\right)^2 e^{-\rho u v}
\end{equation}
and hence find 
\begin{equation}
    \lim_{{\rho \to \infty}}\frac{1}{\hbar}\langle\boldsymbol{S}^{(2)}_{\rho}(\text{joint})\rangle=m.
\end{equation}
We now need to compute the Lorentzian angle for this joint. Taking $\vec{a'}=\begin{pmatrix} \ -T' \\ -mT' \end{pmatrix}\ $ and $\vec{b'}=\begin{pmatrix} \ 0 \\ -mT \end{pmatrix}\ $ (from Fig. \ref{spacelike_triangle}), we find that 
\begin{equation}
    \theta=\cosh^{-1}\left(\frac{m}{\sqrt{m^2-1}}\right).
\end{equation}
By inverting this, the action is found to be:
\begin{equation}
\begin{split}
     \lim_{{\rho \to \infty}}\frac{1}{\hbar}\langle\boldsymbol{S}^{(2)}_{\rho}(\text{joint})\rangle&=\frac{\cosh{\theta}}{\sqrt{\cosh^2{\theta}-1}}= \coth{\theta}.
\end{split}
\end{equation}
As expected, this is the same result as found in the previous section. Therefore, we expect that this expression will hold for the joint of any globally hyperbolic shape for $d=2$. We can rewrite the conjecture for globally hyperbolic manifolds in 2 dimensions as:
\begin{equation} 
\lim_{\rho\to\infty}\frac{1}{\hbar}\langle\boldsymbol{S}_{\rho}^{(2)}(M)\rangle = \int_{M} \,d^{2}x \sqrt{-g}\frac{R}{2} +  \sum_i \coth{\theta_i}
\end{equation}
where $\theta_i$ is the Lorentzian angle of joint $i$, and the sum is over all joints in the manifold. For a null joint, $\theta$ is infinite and so $\coth{\theta}$ will tend to 1. The sum will therefore count the number of points that make up the joint, i.e. the volume of the joint.   
\subsection{Higher Dimensions}

For $d=3$ and higher, the joint is a co-dimension 2 spacelike submanifold. At any point on the joint the vector $\vec{a}$ ($\vec{b}$) is tangent to the future (past) boundary and orthogonal to the joint. The Lorentzian angle $\theta$ of the joint at that point is defined by equation (\ref{eq:lorangle}) and it can vary over the joint. 

\begin{figure} [H]
    \centering
    \includegraphics[width=110mm,scale=1]{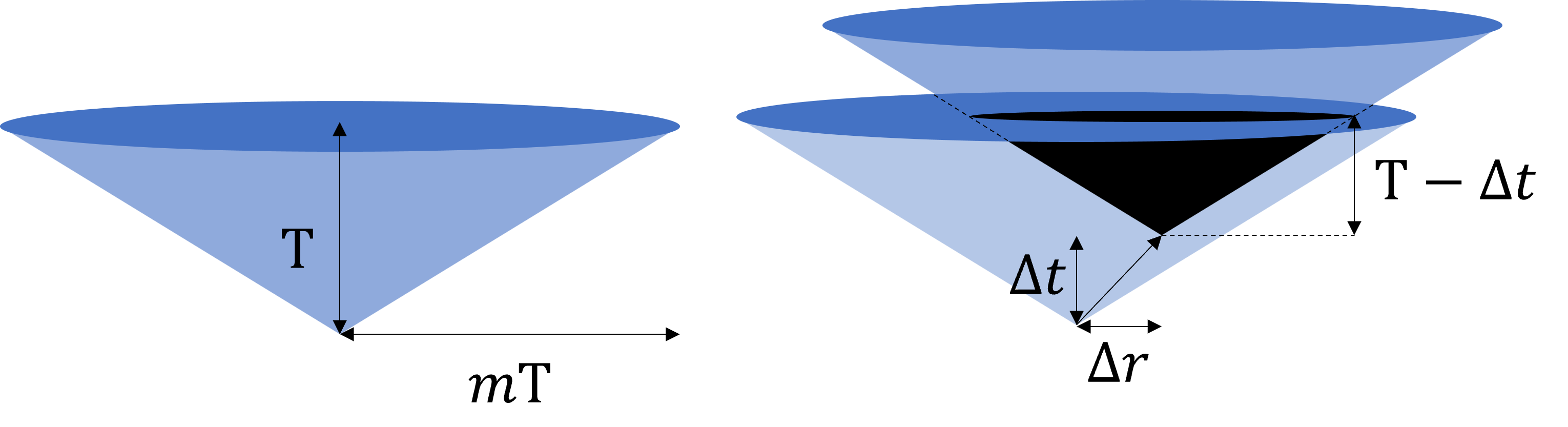}
    \caption{Left: the $d=3$ cone. Right: showing that the volume of realisation (the volume of the black cone) is equal to the volume of original cone scaled by $\left(\frac{T-\Delta t}{T}\right)^3$. This will be true in arbitrary dimensions (replacing 3 with $d$).}
    \label{hypercone_vol}
\end{figure}

If ${\lambda}$ denotes the point on the joint and 
$d\mu(\lambda)$ the $(d-2)$-volume measure of the joint at that point, then the new conjecture is:
\begin{equation}\label{new_conj}
    \begin{split}
          \lim_{\rho\to\infty}\frac{1}{\hbar}\langle\boldsymbol{S}_{\rho}^{(d)}(M)\rangle &= \frac{1}{l_{p}^{d-2}}\int_{M} \,d^{d}x \sqrt{-g}\frac{R}{2}+ \frac{1}{l_p^{d-2}}\int_{J} d\mu(\lambda)  \coth{(\theta(\lambda))}.
    \end{split}
\end{equation}
where 
\begin{equation}
    \int_{J} d\mu(\lambda) = \text{Vol}_{d-2}(J).
\end{equation}
We can test this conjecture on a $d$ dimensional cone where all the boundaries are spacelike. An example is the $d=3$ cone shown in Fig. \ref{hypercone_vol}. We use spherical coordinates. The volume of realisation is
\begin{equation}
    \begin{split}
          a(\Vec{c})=\frac{T}{d} \frac{\pi^{\frac{d-1}{2}}(mT)^{d-1}}{\Gamma(\frac{d+1}{2})} \left(\frac{T-\Delta t}{T}\right)^d,
    \end{split}
\end{equation}
which is the volume of the manifold (the volume of a hypercone with height $T$ and base radius $mT$), but scaled by a factor of $\left(\frac{T-\Delta t}{T}\right)^d$ according to Fig. \ref{hypercone_vol} . 
Then, we obtain
\begin{equation} 
\begin{split}
    X_\rho &= 2^{d-1}\int _{0}^{T} d\Delta t \int_{0}^{\Delta t} d\Delta r \prod_{i=0}^{d-3}\int_0^{\pi /2} d\Delta\phi_i \sin(\Delta\phi_i)^i \Delta r^{d-2}\\ 
    & \frac{T}{d} \frac{\pi^{\frac{d-1}{2}}(mT)^{d-1}}{\Gamma(\frac{d+1}{2})}\left(\frac{T-\Delta t}{T}\right)^d    e^{-\rho V_{c}}\\
\end{split}
\end{equation}
where $V_c$ is given by equation \eqref{V_c}. We use the null coordinates in equation \eqref{eq:nullcoortr} to obtain: 
\begin{equation} 
\begin{split}
    X_\rho &= 2^{d-1}\int _{0}^{\sqrt{2}T} dv  \int_{0}^{v} du \prod_{i=0}^{d-3}\int_0^{\pi /2} d\Delta\phi_i \sin(\Delta\phi_i)^i \\ 
    &(\frac{1}{\sqrt{2}}(v-u))^{d-2} \frac{T}{d} \frac{\pi^{\frac{d-1}{2}}(mT)^{d-1}}{\Gamma(\frac{d+1}{2})} \left(\frac{T-\frac{1}{\sqrt{2}}(v+u)}{T}\right)^d e^{-\rho V_{c}}\\
\end{split}
\end{equation}
where we have again used our large $\rho$ approximation, and integrated over the extended triangle region. We have checked for $d$ up to 11 that the mean action gives 
\begin{equation}
\begin{split}
 \lim_{\rho\to\infty}\frac{1}{\hbar}\langle\boldsymbol{S}_{\rho}^{(d)}(M)\rangle &= \frac{1}{l_{p}^{d-2}}\frac{2 \pi ^{\frac{d-1}{2}}}{\Gamma \left(\frac{d-1}{2}\right)}(mT)^{d-2}\text{ } m   \\
 &= \frac{1}{l_{p}^{d-2}}\text{Vol}_{d-2}(J)\text{ } m \\
  &= \frac{1}{l_{p}^{d-2}}\text{Vol}_{d-2}(J)\text{ } \coth{\theta}.
 \end{split}
\end{equation}
Since $\theta$ is constant around the joint in this case, this is consistent with the conjecture \eqref{new_conj}.
\section{Discussion}

Our analytical and numerical results in flat space support the conjectures in section \ref{sec:conjectures}. We also conjectured and provided evidence for the contribution  of a spacelike hole to the continuum limit of the mean action \eqref{hole_conjecture}. Testing the conjectures further is work for the future. We can, however, look forward to the implications that the conjectures might have for the sum-over-causal-sets qua path integral for quantum gravity. The starting point of the discussion is Feynman's heuristic, quoted in the introduction, that the classical action principle emerges from the quantum mechanical path integral. 
The heuristic is an interplay between the magnitude of the action (in units of $\hbar$) and the number of histories that contribute essentially the same amount to the path integral, an interplay between action and measure. The heuristic assumes that we are working in a regime in which the magnitude of the action of all the histories in the path integral under consideration is large. In the context of causal set quantum gravity, that is the regime in which the cardinality $N$ of the causal sets in the sum over histories (\ref{eq:causetsoh}) is large. \par

Our observable universe is, as far as we can currently tell, without a timelike boundary or ``edge of space''. We can ask why that is.  We argue below that the action of causal sets that are typical sprinklings into spacetimes with a timelike boundary could lead to the suppression of
such spacetimes in the gravitational path integral for cosmology, relative to spacetimes without timelike boundaries.\par

Suppose that the dimension of the causal set action in (\ref{eq:causetsoh}) $d=4$ for definiteness. The action for $d=4$ is given in equation (\ref{eq:4daction}) in Appendix A and we state it here for easy reference:
\begin{equation} \label{eq:4dactionwithoutlp}
\frac{1}{\hbar}S^{(4)}(\mathcal{C}) = \zeta (N-N_1+9N_2-16N_3+8N_4).
\end{equation}
where $\zeta$ is a fundamental constant of order 1. 
 Let us consider a cosmological scenario in which the cardinality of the causal sets in the sum over histories is $N = 10^{240}$, the volume of the observable universe in Planck units.  Suppose, for the sake of the argument, that not only the Kleitman-Rothschild orders but all non-manifoldlike causal sets are suppressed in the sum-over-histories \cite{carlip2024causalsetsemergingcontinuum}, so that all we have left to consider in (\ref{eq:causetsoh}) are the manifoldlike causal sets, i.e. the typical sprinklings into Lorentzian spacetimes that vary only slowly on the fundamental scale. In other words we assume that there exists a \textit{continuum regime} in causal set quantum gravity in which the continuum gravitational path integral (\ref{eq:continuumsoh}) for spacetime of volume $V = 10^{240}$ in fundamental units makes sense at an effective level and is to be understood as being given fundamentally by a sum (\ref{eq:causetsoh}) over manifoldlike causal sets of cardinality $N = 10^{240}$. Essentially what we are assuming here is that the (causal set version of the) cosmological constant problem has been solved.   
\par

In the Feynmannian heuristic the contribution to the gravitational path integral (\ref{eq:continuumsoh}) of a manifold $(M,g)$ is determined by the measure of the neighbourhood of $(M,g)$ over which the action, in units of $\hbar$, changes by an amount much smaller than one. A neighbourhood of $(M,g)$ in (\ref{eq:continuumsoh}) is an approximate continuum concept and corresponds in the fundamental theory to some neighbourhood of a typical sprinkling $C$ into $(M,g)$ in the set of manifoldlike causal sets of cardinality $N$. Saying that the action should vary only a little over the continuum neighbourhood of $(M,g)$ means that the action should vary only a little over the causal sets in the corresponding neighbourhood of $C$.  \par

If the dimension $d'$ of $(M,g)$ 
does not equal the dimension-parameter $d=4$ in the causal set action in (\ref{eq:causetsoh}), then the action (\ref{eq:4dactionwithoutlp}) of the sprinkled $C$ will not enjoy the cancellation between its summands that results in the conjectured values of the action in section \ref{sec:conjectures} and the order of magnitude of the action of $C$ will be the order of magnitude of the dominant summand. For an idea of what this order of magnitude is, for a causal interval in dimension $d'>2$ the number of links $N_1$ in a typical sprinkling is of order $N^{2 -2/d'}$  and for $d' =2$ the number of links $N_1$ in a typical sprinkling is of order $N\log N$ \cite{Glaser:2013pca}. A small change in $C$ that changes each summand by a few will change the action by an amount of order 1. We expect therefore that there is essentially no neighbourhood of $C$ in the set of manifoldlike causal sets over which the action varies by an amount $\ll 1$. In the effective continuum path integral this translates into the suppression of spacetimes with dimension not equal to 4 relative to spacetimes with dimension equal to 4.\par

We are then left, in the sum over causal sets, with the typical sprinklings into $4$-dimensional spacetimes. This is the scenario to which our conjectures apply.  We will ignore the fluctuations and assume that each causal set has an action given by the mean over sprinklings and further assume that the density of sprinkling $\rho = l^{-4}$ is high enough that the action is close to the limiting mean value. In this effective 4-dimensional continuum regime the constant $\zeta$ in the fundamental action is revealed to equal  
\begin{equation} \zeta =  \frac{4}{\sqrt{6}}\left(\frac{l}{l_p}\right)^2 \,,
\end{equation}
where $l_p$ is the 4-dimensional Planck length $l_p = \sqrt{8 \pi G \hbar}$. Since $\zeta$ is a constant of order 1,  our assumptions about the existence of the 4-dimensional continuum regime go hand in hand with the expectation that the fundamental discreteness length $l$ is of order the Planck length $l_p$. \par

We can compare the actions $S(C_{no\,\,edge})$ and $S(C_{edge})$ of causal sets sprinkled into realistic cosmological spacetimes $M_{no\,\, edge}$ without a timelike or null boundary and $M_{edge}$ with a timelike boundary, respectively. Our conjecture implies
\begin{align*} 
\frac{1}{\hbar}S(C_{no\,\, edge})& \approx \frac{1}{l_p^2} \int \sqrt{-g} d^4x \frac{R}{2} 
\sim \frac{V_4}{l_p^2}\Lambda \\ 
& \sim 10^{240} 10^{-120} 
\sim 10^{120} 
\end{align*}
where $V_4$ is the spacetime volume of the universe. If a manifold $(M,g)$ has a timelike boundary our conjecture implies that the action of $C$ tends, in the limit $l \rightarrow 0$, to the timelike boundary term. Physically, however, in causal set theory $l$ the fundamental length is not zero but finite and in
$S(C_{edge})$ for finite $l$
 there are competing terms, one of which is the Einstein-Hilbert (EH) term of the same order of magnitude as the above. The other is the timelike boundary term 
\begin{align*} 
\textrm{timelike boundary term of}\ \ \frac{1}{\hbar}S(C_{edge})& \sim \frac{1}{l_p^2} \frac{1}{l}  V_3 \sim \frac{V_3}{l_p^3}\,. 
\end{align*}
If the timelike boundary is an edge of space of cosmological scale  then its volume $V_3$ is of order
$\frac{V_3}{l_p^3} \sim 10^{180}$ which dominates the EH term and so 
\begin{align*} 
\frac{1}{\hbar}S(C_{edge})& \sim 10^{180} \,.
\end{align*}
Now we have an indication of a mechanism for suppression of spacetime $M_{edge}$ compared to $M_{no\,\, edge}$. The large ratio of the magnitudes of their actions indicates that the measure of the neighbourhood of $C_{edge}$ over which the action is approximately constant is much smaller than that of $C_{no\,\, edge}$, leading to the suppression of $M_{edge}$ in the gravitational path integral. \par

The preceding argument is a rough sketch in which many issues are ignored. Not the least of these is the question of fluctuations around the mean of the causal set action and their implications for the sum-over-histories \cite{moradi2024fluctuationscorrelationscausalset}. It may be that the action in the sum-over-causal sets should be replaced by the smeared form of the action, which we used in this article as a device in our simulations exactly because its fluctuations are small for large $N$. In the meantime, the work in this article can be extended in many ways. We have already mentioned larger simulations and gaining more knowledge of the finite density and finite $\epsilon$ corrections. The conjectures should be tested on manifolds in curved spacetime, and the low curvature regime would be an obvious place to start \cite{10.1088/1361-6382/abc2fd, MachetWang_2021}. 
\par

Ultimately, if quantum gravity is a theory with no free parameters, one must hope that even the dimension of spacetime will emerge from the fundamental theory and not have to be entered by hand into the sum-over-histories for causal sets. One must hope then that the sum-over-histories given in (\ref{eq:causetsoh}) is  itself an effective theory that arises from a yet more fundamental dynamics. One proposal for such dynamics is the class of sequential growth models \cite{Rideout:1999ub, Dowker:2010qh,Surya:2020cfm} which hold the potential to be models of self-tuning universes. In such universes, what we now treat as fundamental constants such as $d=4$ arise through a process of ``cosmic renormalisation'' as cosmic epochs with different effective parameters succeed one another in an unceasing process \cite{Sorkin:1998hi, Martin:2000js, Dowker_2023}.

\section{Acknowledgements} We thank Josh Chevalier for useful discussions. FD is supported in part by STFC Grant ST/ST/X000575/1 and by STFC Grant ST/W006537/1. Research at Perimeter Institute
is supported by the Government of Canada through Industry
Canada and by the Province of Ontario through the Ministry
of Economic Development and Innovation. 

\appendix 
\section{Causal Set Actions for Specific Dimensions}\label{App:A}
The explicit forms of equation \eqref{action}, as examples, for $d=2,3,4$ are
\begin{equation} 
\frac{1}{\hbar}S^{(2)}(\mathcal{C}) = 2(N-2N_1+4N_2-2N_3) ,
\end{equation}
\begin{equation} 
\frac{1}{\hbar}S^{(3)}(\mathcal{C}) = \frac{1}{\Gamma(\frac{5}{3})}\left(\frac{\pi}{3\sqrt{2}}\right)^\frac{2}{3}\frac{l}{l_p}(N-N_1+\frac{27}{8}N_2-\frac{9}{4}N_3) ,
\end{equation}
\begin{equation} \label{eq:4daction}
\frac{1}{\hbar}S^{(4)}(\mathcal{C}) = \frac{4}{\sqrt{6}}\left(\frac{l}{l_p}\right)^2(N-N_1+9N_2-16N_3+8N_4).
\end{equation}
Their smeared versions, the explicit forms of equation \eqref{smeared_action} for $d=2,3,4$ are
\begin{equation} 
   \frac{1}{\hbar} S^{(2)}_\epsilon (\mathcal{C}) =2\left( \epsilon N-2 \epsilon^{2} \sum_{x\in \mathcal{C}}\sum_{y\prec x}f_2(n(x,y),\epsilon) \right)
\end{equation}
where
\begin{equation} 
    f_2(n,\epsilon)=(1-\epsilon)^n -2n\epsilon(1-\epsilon)^{n-1}+\binom{n}{2}\epsilon^2(1-\epsilon)^{n-2} ,
\end{equation}
\begin{equation} 
   \frac{1}{\hbar} S^{(3)}_\epsilon (\mathcal{C}) =\frac{1}{\Gamma(\frac{5}{3})}\left(\frac{\pi}{3\sqrt{2}}\right)^\frac{2}{3}\frac{l}{l_p}\left( \epsilon^{2/3} N-\epsilon^{5/3} \sum_{x\in \mathcal{C}}\sum_{y\prec x}f_3(n(x,y),\epsilon) \right)
\end{equation}
where
\begin{equation} 
    f_3(n,\epsilon)=(1-\epsilon)^n -\frac{27}{8}n\epsilon(1-\epsilon)^{n-1}+\frac{9}{4}\binom{n}{2}\epsilon^2(1-\epsilon)^{n-2},
\end{equation}
\begin{equation} 
   \frac{1}{\hbar} S^{(4)}_\epsilon (\mathcal{C}) =\frac{4}{\sqrt{6}}\left(\frac{l}{l_p}\right)^2\left( \epsilon^{1/2} N- \epsilon^{3/2} \sum_{x\in \mathcal{C}}\sum_{y\prec x}f_4(n(x,y),\epsilon) \right)
\end{equation}
where
\begin{equation} 
    f_4(n,\epsilon)=(1-\epsilon)^n -9n\epsilon(1-\epsilon)^{n-1}+16\binom{n}{2}\epsilon^2(1-\epsilon)^{n-2}-8\binom{n}{3}\epsilon^3(1-\epsilon)^{n-3}
\end{equation}
which we used in our simulations.
\section{The Cross Terms In the $d=2$ Diamond}\label{crossterm}
In this appendix we verify the  result from Section \ref{trianglesec} that the $d=2$ joint term for the corner is not present at subleading order in (\ref{eq:trianglemean}). A causal diamond has a mean action of 2 in the limit of infinite $\rho$. This is interpreted according to Conjecture 1 as being the contribution of the 2 points, $S^0$, of the joint. We can investigate the result for the triangle by considering it as one of two isometric halves of the diamond which gives us a chance to calculate an example of the bi-action defined above. 
Due to the bilocal nature of the action, we cannot simply sum the action of the left triangle and the right triangle to obtain the action of the diamond. Instead equation \eqref{bilocal_mean} must be used:
\begin{equation} 
\langle \boldsymbol{S}_{\text{diamond}}\rangle  
= \langle \boldsymbol{S}_{L}\rangle  + \langle \boldsymbol{S}_{R}\rangle  + \langle \boldsymbol{S}_{L,R}\rangle  + \langle \boldsymbol{S}_{R,L}\rangle 
\end{equation}
where $L$ refers to the left triangle and $R$ the right. $\langle \boldsymbol{S}_{L,R}\rangle $ is the mean action calculated by only considering intervals which begin in the right triangle and end in the left triangle. Similarly $\langle \boldsymbol{S}_{R,L}\rangle $, considers intervals which begin in the left and end in the right. To find $\langle \boldsymbol{S}_{R,L}\rangle $, cloning can again be used. We clone the left hand region and displace it by the defining vector $(u,v)$. The volume of realisation is then given by the overlapping area between the cloned left hand region and the right hand region.
\begin{figure} [H]
    \centering
    \includegraphics[width=80mm,scale=0.5]{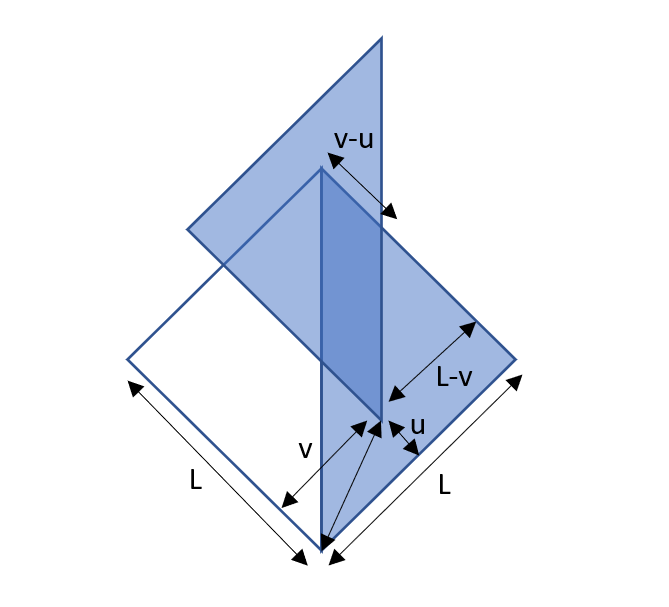}
    \caption{The volume of realisation (shaded) for the cross terms.}
    \label{bilocal_overlap}
\end{figure}
From Figure \ref{bilocal_overlap}, the overlapping area can be seen to be $(L-v)(v-u)$. In order for these two regions to intersect, $L> v> u$. Therefore, $X_\rho$ is given by
\begin{equation}
    \begin{split}
        X_\rho &=  \int_{0}^L \,du \int_{u}^L \,dv (L-v) (v-u) e^{-\rho  u v}\\
        & =\frac{-2 \sqrt{\pi } L \sqrt{\rho } \text{ Erf}\left(L \sqrt{\rho
   }\right)-\text{Ei}\left(-L^2 \rho \right)+L^2 \rho -2 e^{-L^2 \rho }+2 \log(L)+\log (\rho )+\gamma +2}{2 \rho ^2}.
    \end{split}
\end{equation}
The action can then be computed:
\begin{equation}
    \frac{1}{\hbar}\langle \boldsymbol{S}_{L,R}\rangle =-4\rho^2\hat{\mathcal O}_2 X_\rho=-\frac{1}{2} \sqrt{\pi } L \sqrt{\rho } \text{ Erf}\left(L \sqrt{\rho }\right)-e^{-L^2
   \rho }+1.
\end{equation}
From symmetry, $\langle \boldsymbol{S}_{L,R}\rangle  = \langle \boldsymbol{S}_{R,L}\rangle $ and $\langle \boldsymbol{S}_{L}\rangle  = \langle \boldsymbol{S}_{R}\rangle $. Therefore 
\begin{equation}
\begin{split}
     \frac{1}{\hbar}\langle \boldsymbol{S}_{\text{diamond}}\rangle  &= 2 \left(\frac{1}{2} \sqrt{\pi} L \sqrt{\rho} \text{Erf}(L \sqrt{\rho})\right) + 2 \left(-\frac{1}{2} \sqrt{\pi } L \sqrt{\rho } \text{ Erf}\left(L \sqrt{\rho }\right)-e^{-L^2 \rho }+1\right) \\
     &= 2(1-e^{-L^2 \rho })\\
\end{split}
\end{equation}
which is the known result \cite{Buck:2015oaa}. 
\section{$d=2$ Timelike-Spacelike (t-s) Corners in the Embedded Regime} \label{t-s_corners_section}
\subsection{Half Lozenge}\label{half_lozenge_sec}
\begin{figure} [H]
    \centering
    \includegraphics[width=50mm,scale=1]{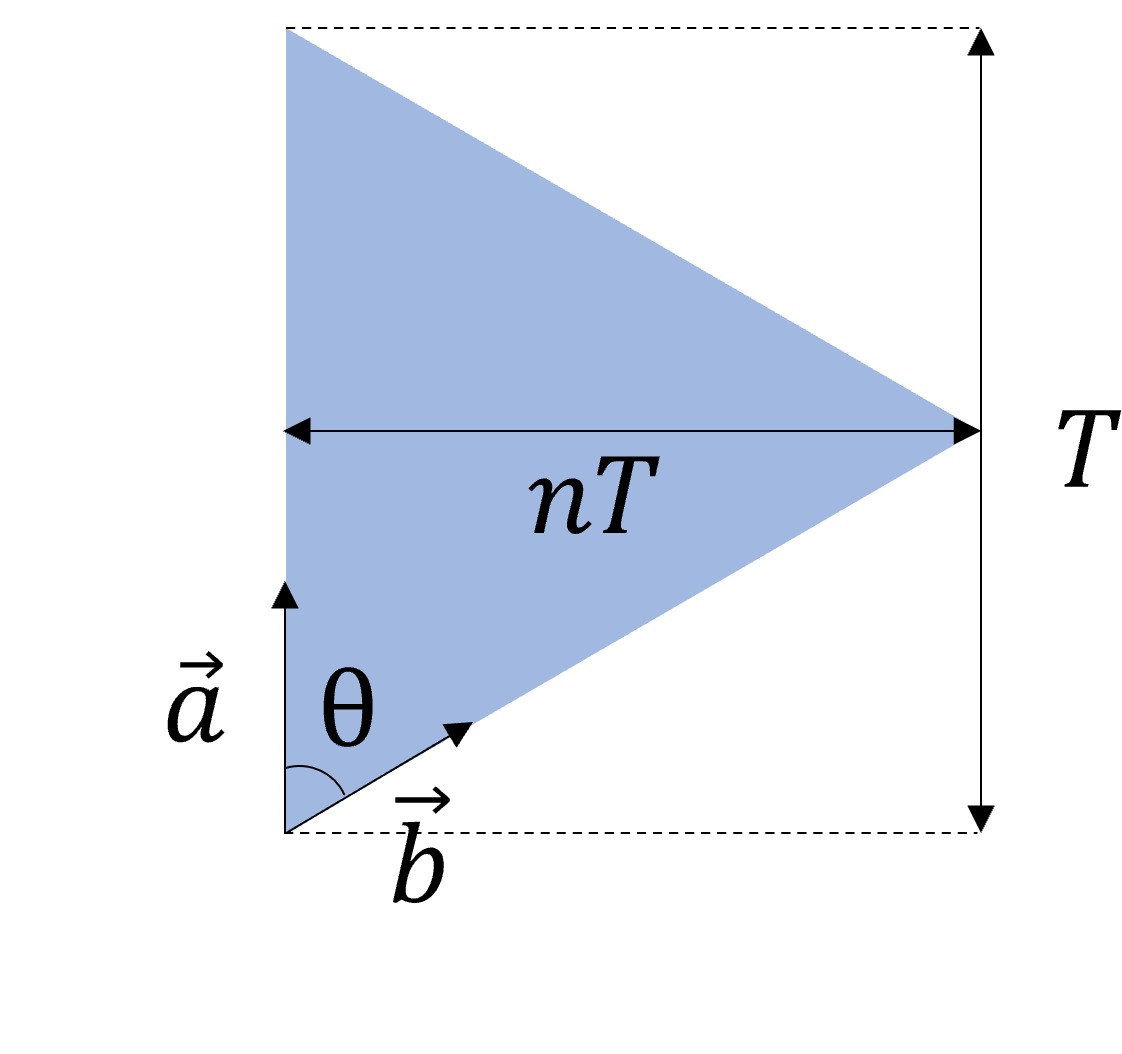}
    \caption{Half of the lozenge. The vectors are used to define the Lorentzian angle of the timelike-spacelike corner.}
    \label{half_lozenge}
\end{figure}

\begin{figure} [H]
    \centering
    \includegraphics[width=50mm,scale=1]{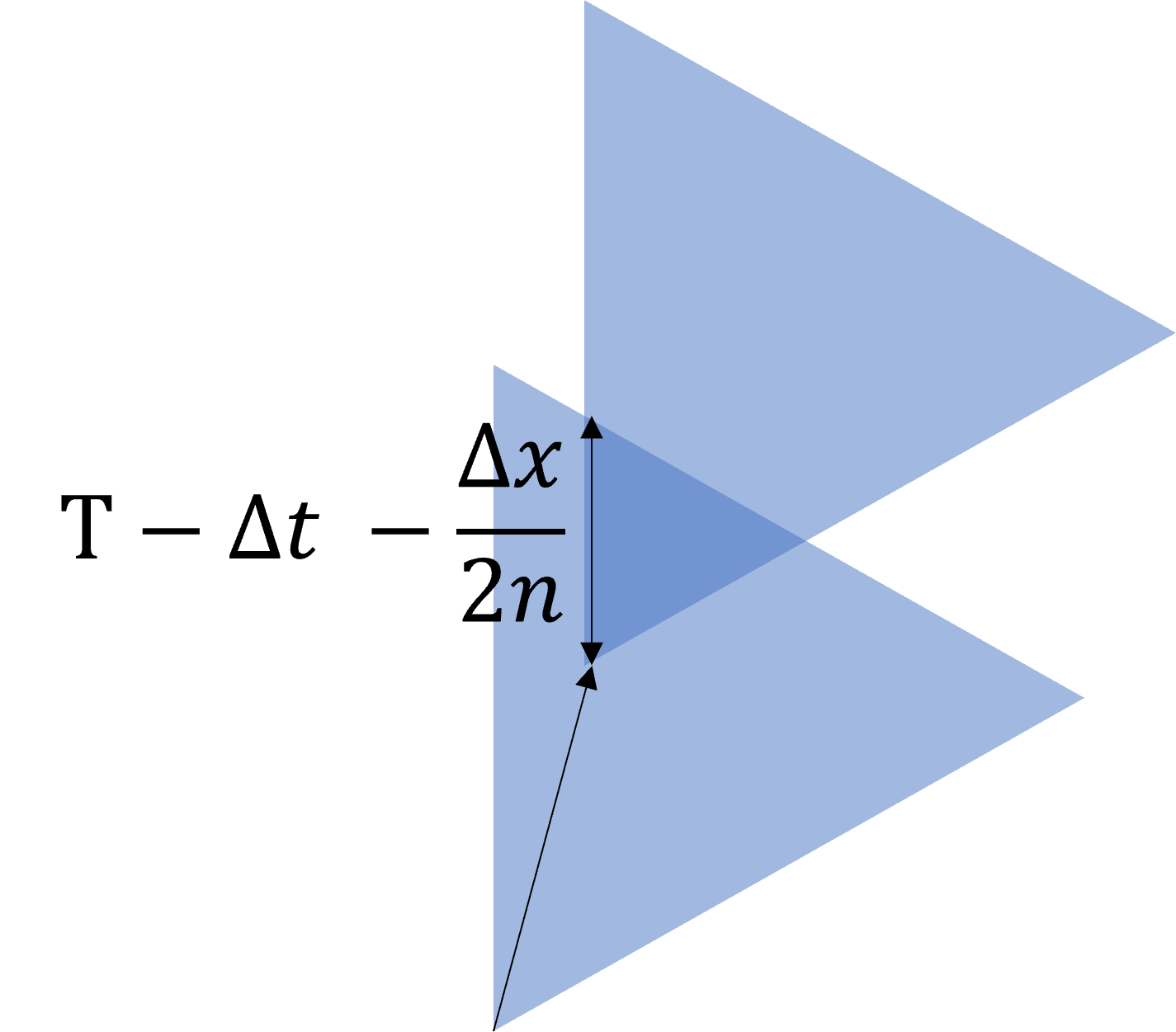}
    \caption{Finding the volume of realisation.}
    \label{half_lozenge_vol}
\end{figure}

To find the contribution from the intersection of timelike and spacelike boundaries, we can consider the action of half of the lozenge, shown in Fig. \ref{half_lozenge}. We want to conclude that due to the presence of the timelike boundary, any contributions from these corners will be dominated by the $\rho^\frac{1}{d}$ term, so they are not of great importance.
The volume of realisation is 
\begin{equation}
    \frac{1}{2} n \left(T-\Delta t-\frac{\Delta x}{2 n}\right)^2
\end{equation}
which can be seen from Fig. \ref{half_lozenge_vol}.
The limits of integration are the same as for the lozenge in Section \ref{lozenge subsection}. Therefore
\begin{equation}
    X_\rho = 2 \int_0^{\frac{\sqrt{2}T}{1+\frac{1}{2n}}} dv\int_0^v du \text{ } \frac{e^{-\rho  u v} \left(\sqrt{2} ((2 n-1) u+2 n v+v)-4 n T\right)^2}{32 n}.
\end{equation}
This gives
\begin{equation}
\begin{split}
    \frac{1}{\hbar}\langle\boldsymbol{S}^{(2)}_{\rho}(M)\rangle &= \rho nT^2-4\rho^2 \hat{\mathcal O}_2 X_\rho\\
    &=n-\frac{1}{4 n}+\frac{1}{2} \sqrt{\frac{\pi }{2}} \sqrt{\rho } T+\mathcal{O}\left(\left(\frac{1}{\rho}\right)^1\right)
\end{split}
\end{equation}
This has the expected $\sqrt{\rho}$ contribution from the timelike boundary. Since we know that the joint contributes $n+\frac{1}{4n}$ (from Section \ref{lorentzian_angle_1}), we expect that the 2 timelike-spacelike corners contribute $-\frac{1}{2n}$. Therefore 
\begin{equation}\label{eq:104}
     \lim_{{\rho \to \infty}}\frac{1}{\hbar}\langle\boldsymbol{S}^{(2)}_{\rho}(\text{t-s corner})\rangle=-\frac{1}{4n}.
\end{equation}
When $n=\frac{1}{2}$, the null case, this gives $-\frac{1}{2}$ and so is consistent with Section \ref{trianglesec}.
\subsection{Lorentzian Angle}
According to the definition of Lorentzian Angle $\theta$ in \cite{Sorkin2019LorentzianAA} for a timelike vector $\Vec{a}$ and a spacelike vector $\Vec{b}$, 
\begin{equation}
\begin{split}
    \cosh{\theta}=-i\frac{\Vec{a}\cdot\Vec{b}}{||\Vec{a}||||\Vec{b}||},
\end{split}
\end{equation}
where $||\Vec{a}||$ is the absolute value of $|\Vec{a}|$. From Section \ref{half_lozenge_sec}, we can find 
\begin{equation}
\begin{split}
    \cosh{\theta}=\frac{i}{2\sqrt{n^2-1/4}} \text{,    for    } \vec{a}=\begin{pmatrix} T \\ 0 \end{pmatrix}\ ,\vec{b}=\begin{pmatrix} T/2 \\ nT \end{pmatrix}\,
\end{split}
\end{equation}
where $\vec{a}$ and $\vec{b}$ come from Fig. \ref{half_lozenge_sec}. This can be rearranged to
\begin{equation}
     n=\frac{\sqrt{\cosh^2\theta-1}}{2\cosh{\theta}}=\frac{1}{2}\tanh{\theta}
\end{equation}
so we can substitute this result into equation \eqref{eq:104} which gives
\begin{equation}
     \lim_{{\rho \to \infty}}\frac{1}{\hbar}\langle\boldsymbol{S}^{(2)}_{\rho}(\text{t-s corner})\rangle=-\frac{1}{4n}=-\frac{1}{2}\coth{\theta}.
     \label{ts_conjecture}
\end{equation}
\subsection{Quarter Lozenge}
\begin{figure} [H]
    \centering
    \includegraphics[width=50mm,scale=1]{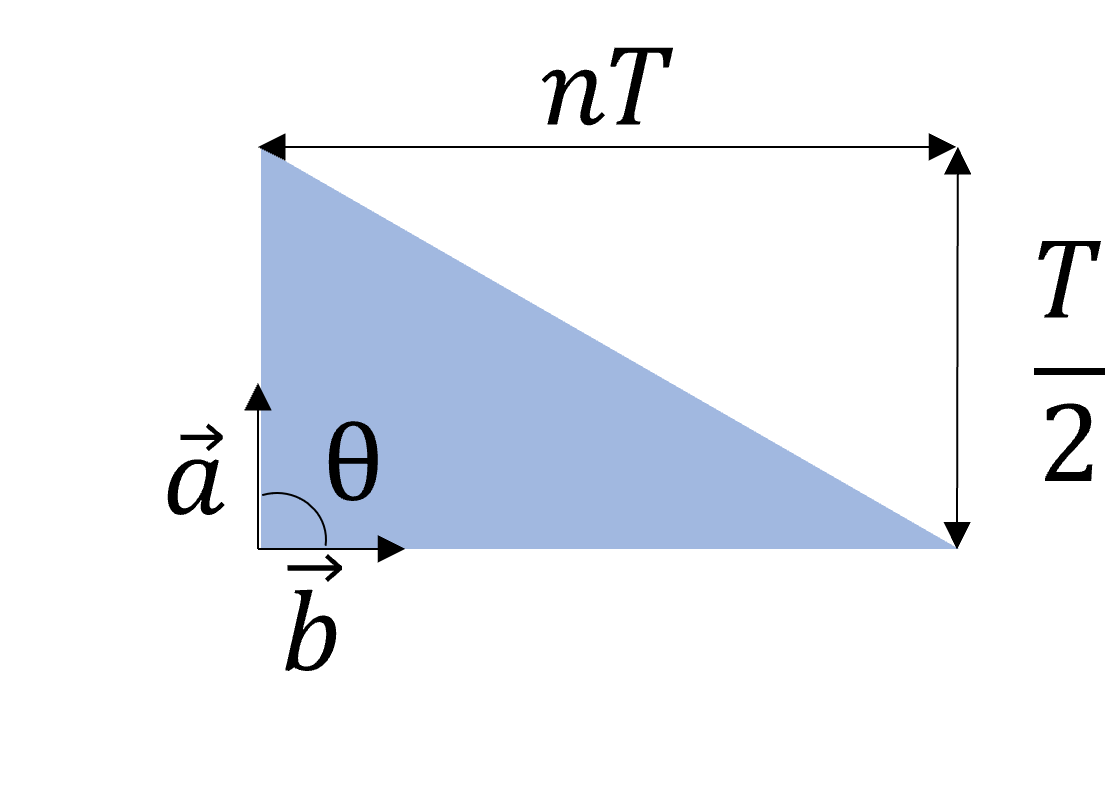}
    \caption{A quarter of the lozenge. This has 2 different spacelike-timelike corners.}
    \label{quarter_lozenge}
\end{figure}

\begin{figure} [H]
    \centering
    \includegraphics[width=120mm,scale=1]{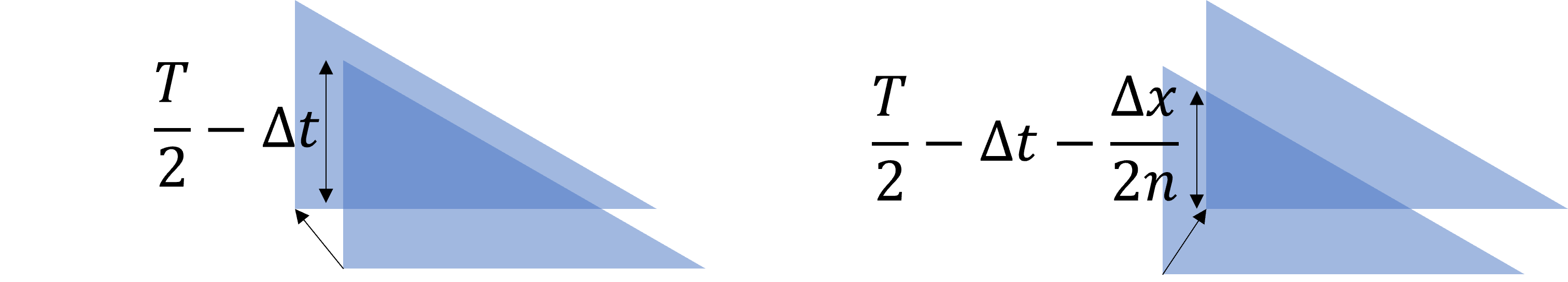}
    \caption{Finding the volume of realisation.}
    \label{quarter_lozenge_vol}
\end{figure}

\begin{figure} [H]
    \centering
    \includegraphics[width=160mm,scale=1]{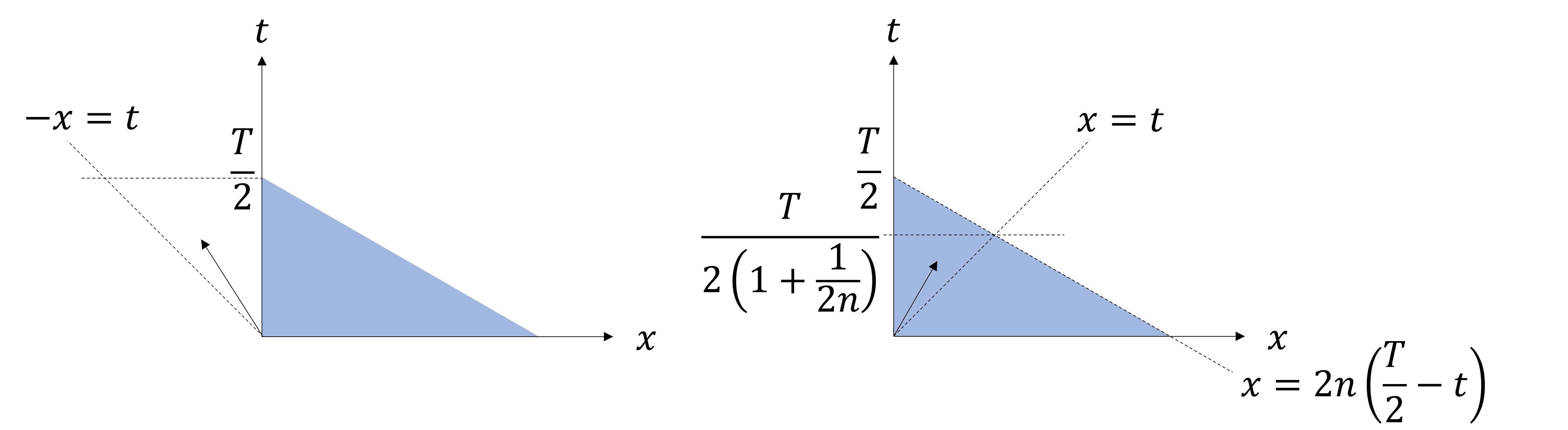}
    \caption{The limits of the integral can be found by considering the area over which the defining vector can range. This is different for $\Delta x <0$ (left) and $\Delta x>0$ (right).}
    \label{quarter_lozenge_limits}
\end{figure}

To check this result for different timelike-spacelike corners, we can compute the action of the quarter lozenge, shown in Fig. \ref{quarter_lozenge}.
For $\Delta x > 0$, this has volume of realisation 
\begin{equation}
    n \left(\frac{T}{2}-\Delta t-\frac{\Delta x}{2 n}\right)^2
\end{equation}
whereas for $\Delta x < 0$, the volume of realisation is 
\begin{equation}
    n \left(\frac{T}{2}-\Delta t\right)^2
\end{equation}
as can be seen from Fig. \ref{quarter_lozenge_vol}. 
To find the limits, we must consider when the defining vector gives a non-zero volume of resolution. This can be seen in Fig. \ref{quarter_lozenge_limits}.
Converting these limits to $(u,v)$ coordinates and using the approximation, we find
\begin{equation}
\begin{split}
    X_\rho &= \int_0^{\frac{\sqrt{2}T}{2(1+\frac{1}{2n})}} dv\int_0^v du \text{ }\frac{e^{-\rho  u v} \left(\sqrt{2} ((2 n-1) u+2 n v+v)-2 n T\right)^2}{16 n} \\
    &+ \int_0^{\frac{\sqrt{2}}{2}T} dv\int_0^v du \text{ }  \frac{1}{4} n \left(T-\sqrt{2} (u+v)\right)^2 e^{-\rho  u v}
\end{split}
\end{equation}
and so
\begin{equation}
\begin{split}
    \frac{1}{\hbar}\langle\boldsymbol{S}^{(2)}_{\rho}(M)\rangle &= \frac{1}{2} \rho nT^2-4\rho^2 \hat{\mathcal O}_2 X_\rho\\
    &=2 n-\frac{1}{4 n}+\frac{1}{4} \sqrt{\frac{\pi }{2}} \sqrt{\rho } T + \mathcal{O}\left(\left(\frac{1}{\rho}\right)^1\right).
\end{split}
\end{equation}
Again, there is the expected timelike boundary contribution. From Section \ref{half_lozenge_sec}, the upper t-s corner should contribute $-\frac{1}{4n}$. We also know from Section \ref{spacelike_triangle_sec} that the joint will contribute $2n$ (setting $m=2n$). Therefore the lower s-t corner should contribute 0. This is consistent with our conjecture in equation \eqref{ts_conjecture}. From Fig. \ref{quarter_lozenge}, we can work out
\begin{equation}
\begin{split}
    \cosh{\theta}=0 \text{,    for    } \vec{a}=\begin{pmatrix} T \\ 0 \end{pmatrix}\ ,\vec{b}=\begin{pmatrix} 0 \\ nT \end{pmatrix}
\end{split}
\end{equation}
so it is simple to see that the contribution to the action from this corner is indeed zero.

\section{Argument for Assumption} \label{proof}
Region 2 touches the $v$-axis at ($0,v_0$) in $u,v$ coordinates (where $v_0=\sqrt{2}T$ for the infinite slab case). $X_{\rho}^{(2)}$ therefore may contribute non-exponentially suppressed terms to the action.
\par
Recall from equation \eqref{mean_action_operator} that
\begin{equation}
    \frac{1}{\hbar}\langle \boldsymbol{S}_{\rho}^{(d)}(M)\rangle  = -\alpha_{d} \left(\frac{l}{l_{p}}\right)^{d-2}\left( \rho V 
    + \frac{\beta_{d}}{\alpha_{d}}
    \rho^{2} \hat{\mathcal O}_d X_{\rho}\right).
\end{equation}
Our aim is to show that $l^{d-2}\rho^2 \hat{\mathcal O}_d X_\rho^{(2)}=\rho^{1+2/d} \hat{\mathcal O}_d X_\rho^{(2)}$ does not contribute to $\mathcal{O}\left(\rho^{1/d}\right)$, the leading order of the action, which is the conjecture term.
\begin{figure}[H]
    \centering
    \includegraphics[width=60mm,scale=0.5]{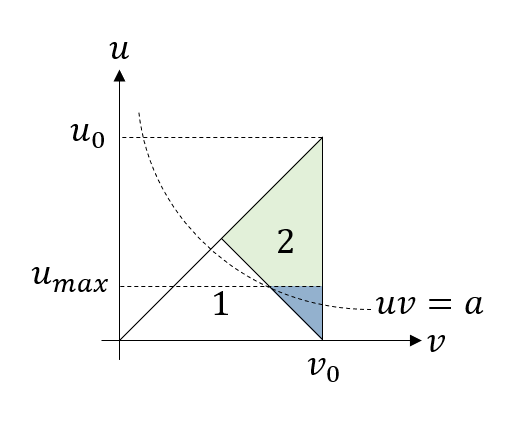}
    \caption{Region 1 here is the shaded region in Fig. \ref{6.1} (right). The shaded area in blue is labelled A(down), and the shaded area in green is labelled A(up).}
    \label{5.4}
\end{figure}
\begin{figure}[H]
    \centering
    \includegraphics[width=140mm,scale=0.5]{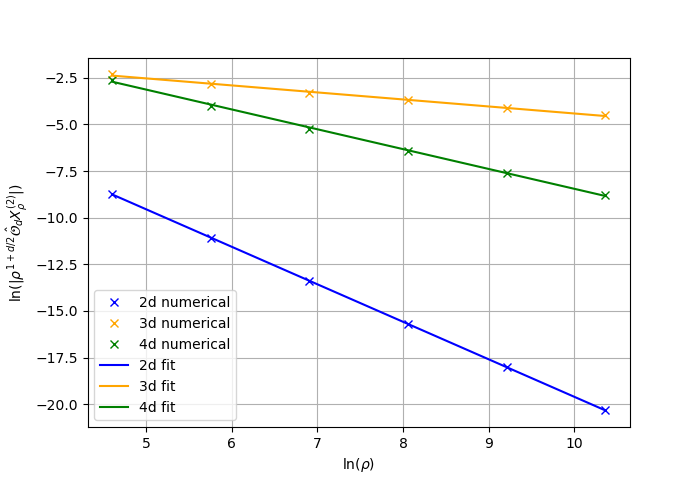}
    \caption{Plot of numerically integrated values $\rho^{1+2/d} \hat{\mathcal O}_d X_\rho^{(2)}$ against $\rho$ on a log scale. The data is fitted for each dimension with a straight line.}
    \label{numint}
\end{figure}
Firstly, we numerically calculate $\rho^{1+2/d} \hat{\mathcal O}_d X_\rho^{(2)}$ for $d=2,3,4$, and plotted the results for large $\rho$ values in Fig. \ref{numint}.
\par
By fitting each dimension with a straight line $\ln(|\rho^{1+2/d} \hat{\mathcal O}_d X_\rho^{(2)}|)=c_1\ln(\rho)+c_2$ on a log scale, we obtain $\rho^{c_1}$ as the leading order contribution to the action from region 2. For $d=2$, $c_1=-2.01\pm0.04,c_2=0.51\pm0.03$; for $d=3$, $c_1=-0.38\pm0.01,c_2=0.65\pm0.08$; for $d=4$, $c_1=-1.06\pm0.02,c_2=2.2\pm0.1$. It is noted that $c_1<1/d$ for these dimensions, so $l^{d-2}\rho^2 \hat{\mathcal O}_d X_\rho^{(2)}$ does not interfere with the conjecture term in these dimensions. These $c_1$ values are also close to -2, -1/3, and -1, which $c_1$ may approach as $\rho\rightarrow\infty$. 
\par
For an analytical argument, we can define 
\begin{equation}
\begin{split}
    &g(u,v) = \rho^{1+2/d} \hat{\mathcal O}_d \left(f(u,v)e^{-  c(d)\rho  (u v)^{d/2}}\right)
\end{split}
\end{equation}
where 
\begin{equation}
f(u,v)=\frac{(2 \pi )^{\frac{d}{2}-2} (v-u)^{d-2} \left(\sqrt{2} (u+v)-2 T\right) \left(2
   \pi  L \Gamma \left(\frac{d}{2}\right)+\sqrt{2 \pi } \Gamma
   \left(\frac{d-1}{2}\right) (u-v)\right)}{\Gamma (d-1)}
\end{equation}
from equation \eqref{eq:6.8}. Therefore
\begin{equation}\label{eq:5.9}
    \begin{split} 
    &\rho^{1+2/d} \hat{\mathcal O}_d X_\rho^{(2)}= \int_{-\infty}^{\infty} dx_2...\int_{-\infty}^{\infty} dx_{d-1} \int_{\frac{T}{\sqrt{2}}}^{\sqrt{2}T} dv \int_{\sqrt{2}T - v}^{v} du\text{  }g(u,v).\\
    \end{split}
\end{equation}
The strategy is to define a parameter $a=uv$, where $\rho a^{d/2} \gg 1$. Region 2 is then divided into 2 regions, with the area named $A$(up) and $A$(down) shown in Fig.  \ref{5.4}. We will separately argue that the contribution to the integral from $A$(up) and $A$(down) does not contribute to the conjecture term.
\subsection{Integration in $A$(up)}
We first argue that the contribution to the integral from $A$(up) can be arbitrarily small for a choice of large $\rho$. We define $g_{\text{max}}$ to be the maximum value of $g(u,v)$ in $A(\text{up})$. $A(\text{up})$ lies above the hyperbola, so $uv>a$ in this region. Therefore, 
\begin{equation}
    e^{-  c(d)\rho  (u v)^{d/2}} \leq e^{-  c(d)\rho  a^{d/2}},
\end{equation}
hence
\begin{equation}
    \int_{-\infty}^{\infty} dx_2...\int_{-\infty}^{\infty} dx_{d-1} \int \int_{A(\text{up})} du dv\text{  }g(u,v)\leq g_{\text{max}}A(\text{up}) \leq g_{\text{max}}A(2).
\end{equation}
where $A(2)$ is the area of region 2. Since $A(2)$ is finite, and $g_{\text{max}}$ can be arbitrarily small for large enough $\rho$, it is always possible to find
\begin{equation}
    \int_{-\infty}^{\infty} dx_2...\int_{-\infty}^{\infty} dx_{d-1} \int \int_{A(\text{up})} du dv\text{  }g(u,v) < \epsilon,
\end{equation}
for any $\epsilon$, hence this integral does not contribute to the leading order of the action (which diverges) in the large $\rho$ limit.
\subsection{Integration in $A$(down)}
To argue that
\begin{equation}
    \int_{-\infty}^{\infty} dx_2...\int_{-\infty}^{\infty} dx_{d-1} \int \int_{A(\text{down})} du dv\text{  }g(u,v)
\end{equation}
also does not contribute to the leading order of action, we choose
\begin{equation}\label{eq:5.15}
    a^\frac{d}{2}=\rho^{-1+\alpha}, \text{where } 0<\alpha<1.
\end{equation}
From this choice, as $\rho$ increases, $a=uv$ becomes closer to the $u=0$ line. The argument is to take a large $\rho$ such that $v$ is close to $v_0$ and so we can treat $g(u,v)$ as $g(u,v_0)$. We first notice that the operator $\hat{\mathcal O}_d$ is designed such that 
\begin{equation}\label{evendzero}
    \hat{\mathcal O}_d (\rho^{-2/d})=\hat{\mathcal O}_d (\rho^{-4/d})=...=\hat{\mathcal O}_d (\rho^{-(d+2)/d})=0
\end{equation}
for even $d$ and 
\begin{equation}\label{odddzero}
    \hat{\mathcal O}_d (\rho^{-2/d})=\hat{\mathcal O}_d (\rho^{-4/d})=...=\hat{\mathcal O}_d (\rho^{-(d+1)/d})=0
\end{equation}
for odd $d$ \cite{Sorkin:2007qi,Dowker:2013vba}. Then we rewrite
\begin{equation}\label{eq:59}
\begin{split}
    &\int_{-\infty}^{\infty} dx_2...\int_{-\infty}^{\infty} dx_{d-1} \int \int_{A(\text{down})} du dv\text{  }g(u,v)\sim\int_{-\infty}^{\infty} dx_2...\int_{-\infty}^{\infty} dx_{d-1}  \int_{0}^{u_{\text{max}}} du \text{  }g(u,v_0)\\
    &=\rho^{1+2/d}\hat{\mathcal O}_d\left(\int_{-\infty}^{\infty} dx_2...\int_{-\infty}^{\infty} dx_{d-1}  \int_{0}^{u_{\text{max}}} du \text{  }f(u,v_0)e^{-  c(d)\rho  (u v_0)^{d/2}}\right).
\end{split}
\end{equation}
We note that $f(u,v_0)$ can be written in the form of a polynomial of $u$, with $u^n$ where $n\in\mathcal{N}$. The integral 
\begin{equation}
    \int_{0}^{u_{\text{max}}} du \text{  }f(u,v_0)e^{-  c(d)\rho  (u v_0)^{d/2}}
\end{equation}
can then only produce polynomials of $\rho^m$, where $m=-2/d,-4/d...$ for $u_{\text{max}}>0$. Limited by equation \eqref{evendzero} and \eqref{odddzero}, the first non-vanishing orders according to $\hat{\mathcal O}_d$ are $\rho^{-(d+4)/d}$ for even $d$, and $\rho^{-(d+3)/d}$ for odd $d$. Together with the $\rho^{1+2/d}$ factor outside of the operator, the integral in $A$(down) should \textit{at most} be of order $\rho^{-2/d}$ for even $d$ and $\rho^{-1/d}$ for odd $d$, hence should always be smaller than the conjecture term. This agrees with the numerical work in Fig. \ref{numint}.

\newpage
\printbibliography

@article{Dowker_2023,
   title={Observables for cyclic causal set cosmologies},
   volume={40},
   ISSN={1361-6382},
   url={http://dx.doi.org/10.1088/1361-6382/ace149},
   DOI={10.1088/1361-6382/ace149},
   number={15},
   journal={Classical and Quantum Gravity},
   publisher={IOP Publishing},
   author={Dowker, Fay and Zalel, Stav},
   year={2023},
   month=jul, pages={155015} }

@mastersthesis{chevalier:2023,
    author = "Chevalier, Joshua ",
    title = "The Discrete Causal Action and Holes in
Spacetime",
    type  = "MSc Dissertation",
    school = "Imperial College London, Department of Physics",
    year = "2023"
}

@article{Saravani:2014gza,
    author = "Saravani, Mehdi and Aslanbeigi, Siavash",
    title = "{On the Causal Set-Continuum Correspondence}",
    eprint = "1403.6429",
    archivePrefix = "arXiv",
    primaryClass = "hep-th",
    doi = "10.1088/0264-9381/31/20/205013",
    journal = "Class. Quant. Grav.",
    volume = "31",
    number = "20",
    pages = "205013",
    year = "2014"
}

@article{Glaser:2013pca,
    author = "Glaser, Lisa and Surya, Sumati",
    title = "{Towards a Definition of Locality in a Manifoldlike Causal Set}",
    eprint = "1309.3403",
    archivePrefix = "arXiv",
    primaryClass = "gr-qc",
    doi = "10.1103/PhysRevD.88.124026",
    journal = "Phys. Rev. D",
    volume = "88",
    number = "12",
    pages = "124026",
    year = "2013"
}

@article{Loomis:2017jhn,
    author = "Loomis, S.P. and Carlip, S.",
    title = "{Suppression of non-manifold-like sets in the causal set path integral}",
    eprint = "1709.00064",
    archivePrefix = "arXiv",
    primaryClass = "gr-qc",
    doi = "10.1088/1361-6382/aa980b",
    journal = "Class. Quant. Grav.",
    volume = "35",
    number = "2",
    pages = "024002",
    year = "2018"
}

@article{Kleitman:1970,
 ISSN = {00029939, 10886826},
 URL = {http://www.jstor.org/stable/2037205},
 author = {D. Kleitman and B. Rothschild},
 journal = {Proceedings of the American Mathematical Society},
 number = {2},
 pages = {276--282},
 publisher = {American Mathematical Society},
 title = {The Number of Finite Topologies},
 urldate = {2024-10-27},
 volume = {25},
 year = {1970}
}

@misc{moradi2024fluctuationscorrelationscausalset,
      title={Fluctuations and Correlations in Causal Set Theory}, 
      author={Heidar Moradi and Yasaman K. Yazdi and Miguel Zilhão},
      year={2024},
      eprint={2407.03395},
      archivePrefix={arXiv},
      primaryClass={gr-qc},
      url={https://arxiv.org/abs/2407.03395}, 
}

@misc{carlip2024causalsetsemergingcontinuum,
      title={Causal Sets and an Emerging Continuum}, 
      author={Steven Carlip},
      year={2024},
      eprint={2405.14059},
      archivePrefix={arXiv},
      primaryClass={gr-qc},
      url={https://arxiv.org/abs/2405.14059}, 
}

@misc{Benincasa:thesis,
	Author = {Benincasa, Dionigi},
	Note = {PhD thesis, Imperial College, London https://spiral.imperial.ac.uk/bitstream/10044/1/14170/1/Benincasa-DMT-2013-PhD-Thesis.pdf},
	School = {Imperial College London},
	Title = {The Action of a Causal Set},
	Year = {2013}
	}

@article{Surya:2011du,
   author = "Surya, Sumati",
   title = "{Evidence for a Phase Transition in 2D Causal Set Quantum Gravity}",
   eprint = "1110.6244",
   archivePrefix = "arXiv",
   primaryClass = "gr-qc",
   doi = "10.1088/0264-9381/29/13/132001",
   journal = "Class. Quant. Grav.",
   volume = "29",
   pages = "132001",
   year ="2012"
}

@article{Surya_2019,
   title={The causal set approach to quantum gravity},
   volume={22},
   ISSN={1433-8351},
   url={http://dx.doi.org/10.1007/s41114-019-0023-1},
   DOI={10.1007/s41114-019-0023-1},
   number={1},
   journal={Living Reviews in Relativity},
   publisher={Springer Science and Business Media LLC},
   author={Surya, Sumati},
   year={2019},
   month={Sep}
}

@article{Sorkin:1998hi,
      author         = "Sorkin, Rafael D.",
      title          = "{Indications of causal set cosmology}",
      booktitle      = "{3rd Peyresq Workshop on Quantum and Stochastic Gravity,
                        String Cosmology and Inflation Peyresq, France, June
                        28-July 3, 1998}",
      journal        = "Int. J. Theor. Phys.",
      volume         = "39",
      year           = "2000",
      pages          = "1731-1736",
      doi            = "10.1023/A:1003629312096",
      eprint         = "gr-qc/0003043",
      archivePrefix  = "arXiv",
      primaryClass   = "gr-qc",
      reportNumber   = "SU-GP-00-02-1"
}

@article{dowker2021recovering,
      title={Recovering General Relativity from a Planck scale discrete theory of quantum gravity}, 
   author = {Butterfield, Jeremy and Dowker, Fay}, 
      year={2021},
      eprint={2106.01297},
      archivePrefix={arXiv},
      primaryClass={gr-qc}
}

@article{Benincasa:2010as,
      author         = "Benincasa, Dionigi M.T. and Dowker, Fay and Schmitzer,
                        Bernhard",
      title          = "{The Random Discrete Action for 2-Dimensional Spacetime}",
      journal        = "Class.Quant.Grav.",
      volume         = "28",
      pages          = "105018",
      doi            = "10.1088/0264-9381/28/10/105018",
      year           = "2011",
      eprint         = "1011.5191",
      archivePrefix  = "arXiv",
      primaryClass   = "gr-qc",
      SLACcitation   = "%%CITATION = ARXIV:1011.5191;%%",
}

@article{Bombelli:1987aa,
  title = {Space-time as a causal set},
  author = {Bombelli, Luca and Lee, Joohan and Meyer, David and Sorkin, Rafael D.},
  journal = {Phys. Rev. Lett.},
  volume = {59},
  issue = {5},
  pages = {521--524},
  numpages = {0},
  year = {1987},
  month = {Aug},
  publisher = {American Physical Society},
  doi = {10.1103/PhysRevLett.59.521},
  url = {https://link.aps.org/doi/10.1103/PhysRevLett.59.521}
}

@incollection{Sorkin:2007qi,
	Archiveprefix = {arXiv},
	Author = {Sorkin, Rafael D.},
	Booktitle = {{Approaches to Quantum Gravity: Towards a New Understanding of Space and Time}},
	Editor = {Oriti, D.},
	Eprint = {gr-qc/0703099},
	Publisher = {Cambridge University Press},
	Title = {{Does locality fail at intermediate length-scales?}},
	Year = {2006}}

@article{MachetWang_2021,
doi = {10.1088/1361-6382/abc274},
url = {https://dx.doi.org/10.1088/1361-6382/abc274},
year = {2020},
month = {dec},
publisher = {IOP Publishing},
volume = {38},
number = {1},
pages = {015010},
author = {Ludovico Machet and Jinzhao Wang},
title = {On the continuum limit of Benincasa–Dowker–Glaser causal set action},
journal = {Classical and Quantum Gravity}
}

@article{Dowker:2010qh,
	Archiveprefix = {arXiv},
	Author = {Dowker, Fay and Johnston, Steven and Surya, Sumati},
	Doi = {10.1088/1751-8113/43/50/505305},
	Eprint = {1007.2725},
	Journal = {J.Phys.A},
	Pages = {505305},
	Primaryclass = {gr-qc},
	Slaccitation = {%%CITATION = ARXIV:1007.2725;%%},
	Title = {{On extending the Quantum Measure}},
	Volume = {A43},
	Year = {2010}}

@article{Martin:2000js,
	Author = {Martin, Xavier and O'Connor, Denjoe and Rideout, David P. and Sorkin, Rafael D.},
	Eprint = {gr-qc/0009063},
	Journal = {Phys. Rev.},
	Pages = {084026},
	Slaccitation = {%%CITATION = GR-QC 0009063;%%},
	Title = {On the `renormalization' transformations induced by cycles of expansion and contraction in causal set cosmology},
	Volume = {D63},
	Year = {2001}}

@article{Kleitman:1975,
	Author = {Kleitman, D.J. and Rothschild, B.L.},
	Journal = {Trans. Amer. Math. Society},
	Pages = {205-220},
	Title = {Asymptotic enumeration of partial orders on a finite set},
	Volume = {205},
	Year = {1975}}

@article{Buck:2015oaa,
    author = "Buck, Michel and Dowker, Fay and Jubb, Ian and Surya, Sumati",
    title = "{Boundary Terms for Causal Sets}",
    eprint = "1502.05388",
    archivePrefix = "arXiv",
    primaryClass = "gr-qc",
    doi = "10.1088/0264-9381/32/20/205004",
    journal = "Class. Quant. Grav.",
    volume = "32",
    number = "20",
    pages = "205004",
    year = "2015"
}

@article{10.1088/1361-6382/abc2fd,
	author={Fay Dowker},
	title={Boundary contributions in the causal set action},
	journal={Classical and Quantum Gravity},
	url={http://iopscience.iop.org/article/10.1088/1361-6382/abc2fd},
	year={2020}
}

@article{Dowker:2013vba,
    author = "Dowker, Fay and Glaser, Lisa",
    title = "{Causal set d'Alembertians for various dimensions}",
    eprint = "1305.2588",
    archivePrefix = "arXiv",
    primaryClass = "gr-qc",
    doi = "10.1088/0264-9381/30/19/195016",
    journal = "Class. Quant. Grav.",
    volume = "30",
    pages = "195016",
    year = "2013"
}

@article{Glaser:2013xha,
    author = "Glaser, Lisa",
    title = "{A closed form expression for the causal set d'Alembertian}",
    eprint = "1311.1701",
    archivePrefix = "arXiv",
    primaryClass = "math-ph",
    doi = "10.1088/0264-9381/31/9/095007",
    journal = "Class. Quant. Grav.",
    volume = "31",
    pages = "095007",
    year = "2014"
}

@article{Cunningham:2019rob,
    author = "Cunningham, William J. and Surya, Sumati",
    title = "{Dimensionally Restricted Causal Set Quantum Gravity: Examples in Two and Three Dimensions}",
    eprint = "1908.11647",
    archivePrefix = "arXiv",
    primaryClass = "gr-qc",
    doi = "10.1088/1361-6382/ab60b7",
    journal = "Class. Quant. Grav.",
    volume = "37",
    number = "5",
    pages = "054002",
    year = "2020"
}

@article{Glaser:2017sbe,
    author = "Glaser, Lisa and O'Connor, Denjoe and Surya, Sumati",
    title = "{Finite Size Scaling in 2d Causal Set Quantum Gravity}",
    eprint = "1706.06432",
    archivePrefix = "arXiv",
    primaryClass = "gr-qc",
    reportNumber = "DIAS-STP-17-03",
    doi = "10.1088/1361-6382/aa9540",
    journal = "Class. Quant. Grav.",
    volume = "35",
    number = "4",
    pages = "045006",
    year = "2018"
}

@article{Rideout:1999ub,
      author         = "Rideout, D. P. and Sorkin, R. D.",
      title          = "{A Classical sequential growth dynamics for causal sets}",
      journal        = "Phys. Rev.",
      volume         = "D61",
      year           = "2000",
      pages          = "024002",
      doi            = "10.1103/PhysRevD.61.024002",
      eprint         = "gr-qc/9904062",
      archivePrefix  = "arXiv",
      primaryClass   = "gr-qc",
      reportNumber   = "SU-GP-99-4-1",
      SLACcitation   = "%%CITATION = GR-QC/9904062;%%"
}

@article{Benincasa:2010ac,
    author = "Benincasa, Dionigi M.T. and Dowker, Fay",
    title = "{The Scalar Curvature of a Causal Set}",
    eprint = "1001.2725",
    archivePrefix = "arXiv",
    primaryClass = "gr-qc",
    doi = "10.1103/PhysRevLett.104.181301",
    journal = "Phys. Rev. Lett.",
    volume = "104",
    pages = "181301",
    year = "2010"
}

@article{Surya:2020cfm,
    author = "Surya, Sumati and Zalel, Stav",
    title = "{A Criterion for Covariance in Complex Sequential Growth Models}",
    eprint = "2003.11311",
    archivePrefix = "arXiv",
    primaryClass = "math-ph",
    doi = "10.1088/1361-6382/ab987f",
    journal = "Class. Quant. Grav.",
    volume = "37",
    number = "19",
    pages = "195030",
    year = "2020"
}

@article{Mathur:2009,
    author = "Mathur, Abhishek and Singh, Anup Anand and Surya, Sumati",
    title = "{Entropy and the Link Action in the Causal Set Path-Sum}",
    eprint = "2009.07623",
    archivePrefix = "arXiv",
    primaryClass = "gr-qc",
    doi = "10.1088/1361-6382/abd300",
    journal = "Class. Quant. Grav.",
    volume = "38",
    number = "4",
    pages = "045017",
    year = "2021"
}

@article{Sorkin2019LorentzianAA,
  title={Lorentzian angles and trigonometry including lightlike vectors},
  author={Rafael D. Sorkin},
  journal={arXiv: General Relativity and Quantum Cosmology},
  year={2019},
  url={https://api.semanticscholar.org/CorpusID:201646393}
}

@online{feynman,
    author = {Richard P. Feynman and Robert B. Leighton and Matthew Sands},
    title = {The Feynman Lectures on Physics, Vol. II. Chapter 19: The principle of least action},
    year = {1964},
    url = {https://www.feynmanlectures.caltech.edu/II_19.html},
    urldate = {2024-04-02}
}

@article{action_constants,
   title={A closed form expression for the causal set d’Alembertian},
   volume={31},
   ISSN={1361-6382},
   url={http://dx.doi.org/10.1088/0264-9381/31/9/095007},
   DOI={10.1088/0264-9381/31/9/095007},
   number={9},
   journal={Classical and Quantum Gravity},
   publisher={IOP Publishing},
   author={Glaser, Lisa},
   year={2014},
   month=apr, pages={095007} }

@article{dalembertians_various_dimensions,
   title={Causal set d’Alembertians for various dimensions},
   volume={30},
   ISSN={1361-6382},
   url={http://dx.doi.org/10.1088/0264-9381/30/19/195016},
   DOI={10.1088/0264-9381/30/19/195016},
   number={19},
   journal={Classical and Quantum Gravity},
   publisher={IOP Publishing},
   author={Dowker, Fay and Glaser, Lisa},
   year={2013},
   month=sep, pages={195016} }

@article{dowker2021recovering_published,
      title={Recovering General Relativity from a Planck scale discrete theory of quantum gravity}, 
      author={Fay Dowker and Jeremy Butterfield},
      year={2024},
      journal={Philosophy of Physics},
      volume={2},
      issue={1},
doi={10.31389/pop.17}
}

@article{CarlipCarlipSurya2024,
author = {Carlip, Peter and Carlip, Steve and Surya, Sumati},
year = {2024},
month = {06},
pages = {},
title = {The Einstein–Hilbert action for entropically dominant causal sets},
volume = {41},
journal = {Classical and Quantum Gravity},
doi = {10.1088/1361-6382/ad506e}
}
\end{document}